\documentclass[apj]{emulateapj}
\slugcomment{{\sc Accepted to Icarus:} 1 November 2012} 

\usepackage{tabularx}

\usepackage{graphics}
\usepackage{natbib}
\usepackage{graphicx}
\usepackage{amssymb}
\usepackage{amsmath}
\usepackage{epstopdf}
\usepackage{float}
\usepackage{url}

\newcommand{\OH}{\mathrm{OH}}
\newcommand{\CO}{\mathrm{CO}}
\newcommand{\CH}{\mathrm{CH}}
\newcommand{\HH}{\mathrm{H}}
\newcommand{\OO}{\mathrm{O}}
\newcommand{\M}{\mathrm{M}}
\newcommand{\cm}{\mathrm{cm}}
\newcommand{\s}{\mathrm{s}}

\begin{document}
\title{Constraining the Origins of Neptune's Carbon Monoxide Abundance with CARMA Millimeter-wave Observations}
\shortauthors{S. H. Luszcz-Cook and I. de Pater}

\author{
  S. H. Luszcz-Cook\altaffilmark{1},
  I. de Pater\altaffilmark{1}}

\altaffiltext{1}{Astronomy Department, University of California, 
Berkeley, CA, USA}

\begin{abstract}
We present observations of Neptune's 1- and 3-mm spectrum from the Combined Array for Research in Millimeter-wave Astronomy (CARMA). Radiative transfer analysis of the CO (2--1) and (1--0) rotation lines was performed to constrain the CO vertical abundance profile. We find that the data are well matched by a CO mole fraction of $0.1^{+0.2}_{-0.1}$ parts per million (ppm) in the troposphere, and $1.1^{+0.2}_{-0.3}$ ppm in the stratosphere. A flux of  $0.5$--$20 \times 10^8$ CO molecules cm$^{-2}$ s$^{-1}$ to the upper stratosphere is implied. Using the \cite{zahnle03} estimate for cometary impact rates at Neptune, we calculate the CO flux that could be formed from (sub)kilometer-sized comets; we find that if the diffusion rate near the tropopause is small (200 cm$^{2}$ s$^{-1}$), these impacts could produce a flux as high as $0.5^{+0.8}_{-0.4}\times 10^8$ CO molecules cm$^{-2}$ s$^{-1}$. We also revisit the calculation of Neptune's internal CO contribution using revised calculations for the CO$\rightarrow$CH$_4$ conversion timescale in the deep atmosphere \citep{visscher11}. We find that an upwelled CO mole fraction of 0.1 ppm implies a global O/H enrichment of at least 400, and likely more than 650, times the protosolar value.



\vspace{\baselineskip}
\small{NOTICE: this is the authors' version of a work that was accepted for publication in Icarus. Changes resulting from the publishing process may not be reflected in this document. A definitive version was subsequently published in {\it Icarus} 222 (2013) 379--400.}

\end{abstract}
 
\section{Introduction}\label{sec:introduction}
In equilibrium, carbon monoxide (CO) should be confined to the warm interiors of the Solar System giant planets. Therefore, its detection in the upper atmospheres of all four of these planets \citep{beer75,noll86,encrenaz04,marten91} indicates disequilibrium processes at work. Two pathways exist for enriching the atmosphere in CO: vertical mixing from the deep atmosphere and external supply from the environment. 

CO  production occurs via the net thermochemical reaction
\begin{eqnarray}
\CH_4 +\HH_2\OO=\CO+3\HH_2.
\end{eqnarray}
At the temperatures and pressures of Neptune's atmosphere, the left-hand side of this equation dominates, with nearly all of the carbon present in the form of CH$_4$.  Indeed, CH$_4$ has been observed in Neptune's troposphere at an abundance of 2.2\% \citep{baines95}. CO is more stable and therefore more abundant at the warmer temperatures of the deep atmosphere. Consequently, CO abundances in the upper atmosphere can exceed their equilibrium value if convective transport from the warm interior is more rapid than the CO destruction rate \citep{prinn77,fegley86}. The CO abundance originating from the interior therefore depends on the internal water abundance (Eq. (1)) as well as the speed of vertical mixing, and thereby acts as a chemical probe of the deep atmosphere \citep{fegley94,lodders94,lodders02, visscher10}.

An alternative source of CO in giant planet atmospheres is the planetary environment. Observations of stratospheric water and CO$_2$ \citep{feucht97, degraauw97, lellouch97b} indicate an external supply of oxygen to the giant planets, which could originate from rings and icy satellites or interplanetary dust (meteoroids) \citep{feucht97, encrenaz99}. Oxygen then forms CO by combining in the stratosphere with the byproducts of methane photolysis \citep{rosenqvist92, moses00}.  However,  the CO/H$_2$O ratios observed for Jupiter and Neptune seem to be inconsistent with oxygen supply by these mechanisms \citep{lellouch02,bezard02,lellouch05}.  Shock chemistry due to infall of comets also can produce significant amounts of CO in the upper atmosphere, as demonstrated by the 1994 impact of comet Shoemaker-Levy 9 with Jupiter \citep{lellouch95}. Infall of (sub)kilometer-sized comets has been shown to be sufficient for supplying Jupiter's observed stratospheric CO abundance if the eddy mixing coefficient near the tropopause is smaller than 300 cm$^2$ s$^{-1}$ \citep{bezard02}.  Once it is produced, CO is stable and is removed from the stratosphere by downward transport.

Atmospheric CO enrichment by these two very different pathways can be distinguished by measuring the vertical CO profile: a uniform distribution of CO throughout the upper atmosphere indicates that CO is being mixed up to observable levels from the deep atmosphere. If instead CO is being produced in the stratosphere of the planet, then downward transport will act as a sink, and there will be a higher CO abundance in the stratosphere than the troposphere. Early measurements of Neptune's stratospheric \citep{marten93,rosenqvist92,marten05} and tropospheric \citep{guilloteau93,naylor94,courtin96, encrenaz96} CO mole fractions were all roughly consistent with a CO abundance of 1 part per million (ppm), which led several authors to tentatively conclude that Neptune's CO was internal in origin \citep[e.g.][] {courtin96,marten05}. However, the estimates of Neptune's CO abundance from these observations were highly divergent. More recent observations \citep{lellouch05,hesman07} suggest that both internal and external sources play a role in supplying CO to Neptune: these experiments simultaneously determine the stratospheric and tropospheric abundances from the shapes of CO rotational lines. Figure 1 illustrates the contributions of the various atmospheric levels to the line intensity at a range of frequency offsets from line center, for the three lowest CO rotational transitions.  Peaks in the contribution functions range from above 0.1 mbar at line center, down to several bars in the far wings indicating  that characterization of the full CO vertical profile requires both high frequency resolution to measure the shape of the narrow ($\sim$10 MHz wide) central line peak, and broad frequency coverage (of order 6-10 GHz) to probe the CO mole fraction below the tropopause. From their analyses of the CO (2--1) and CO (3--2) lines, respectively,  \cite{lellouch05} and \cite{hesman07} conclude that Neptune's CO mixing ratio is measurably higher in the stratosphere, indicating that CO has both an internal and external origin. Such a dual origin is also indicated for Jupiter and Saturn \citep{bezard02,cavalie09}. Quantitatively, however, the \cite{lellouch05} and \cite{hesman07} Neptunian CO profiles are inconsistent to within their quoted uncertainties, particularly in the upper stratosphere. Uncertainties in calibration and in Neptune's thermal profile may be responsible for discrepancies between these published CO values. Recent observations at infrared wavelengths have not resolved the discrepancy:   \cite{fletcher10} found a CO mixing ratio of $2.5\times10^{-6}$ (mole fraction of 2.1 ppm) at altitudes above 10 mbar, which is consistent with the \cite{hesman07} result, whereas \cite{lellouch10} favor a 1 ppm stratospheric mole fraction in agreement with \cite{lellouch05}.

\begin{figure}
\begin{center}$
\begin{array}{ccc}
\includegraphics[width=0.45\textwidth]{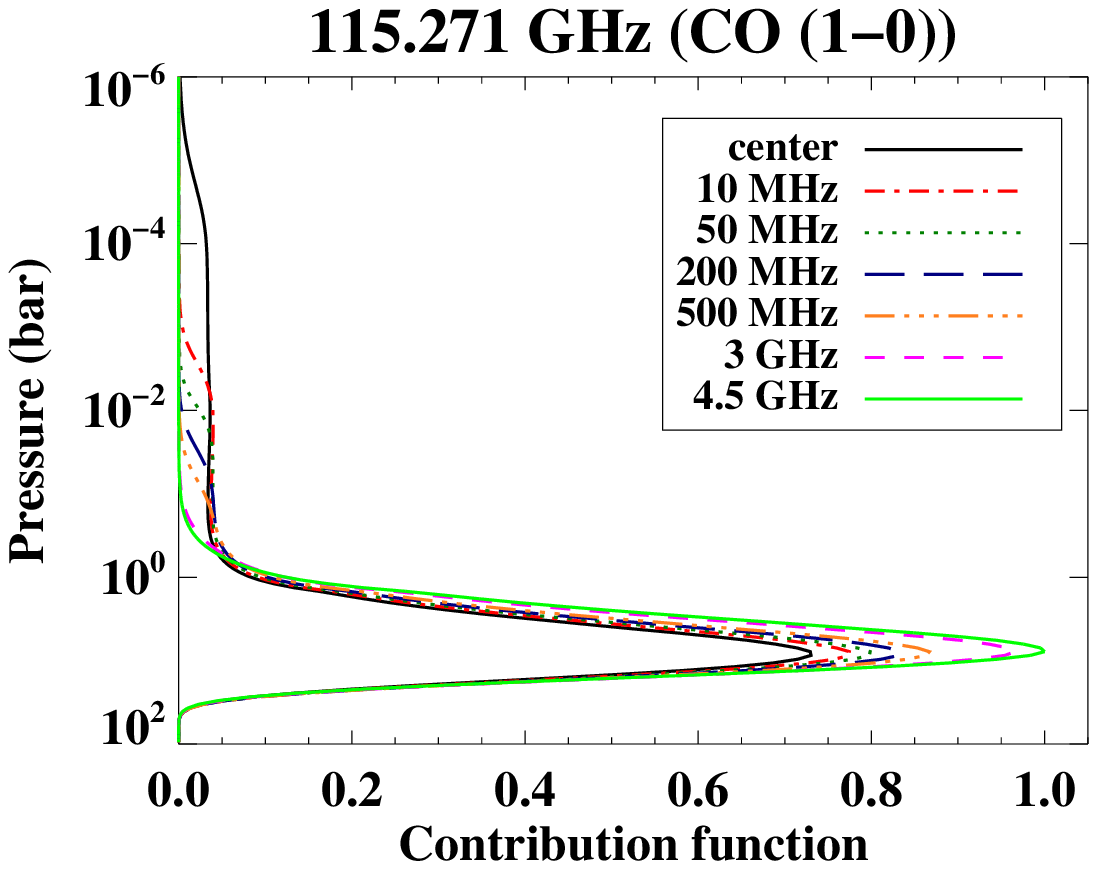}\\
\includegraphics[width=0.45\textwidth]{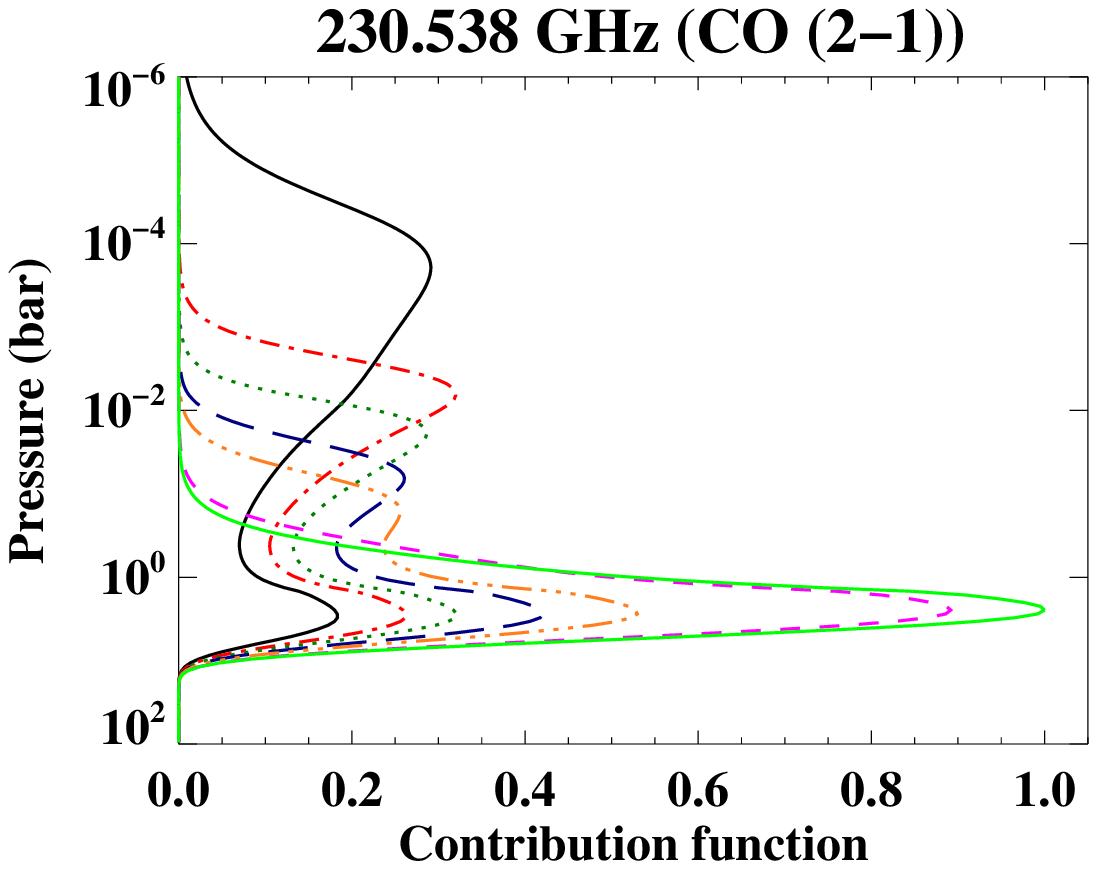}\\
\includegraphics[width=0.45\textwidth]{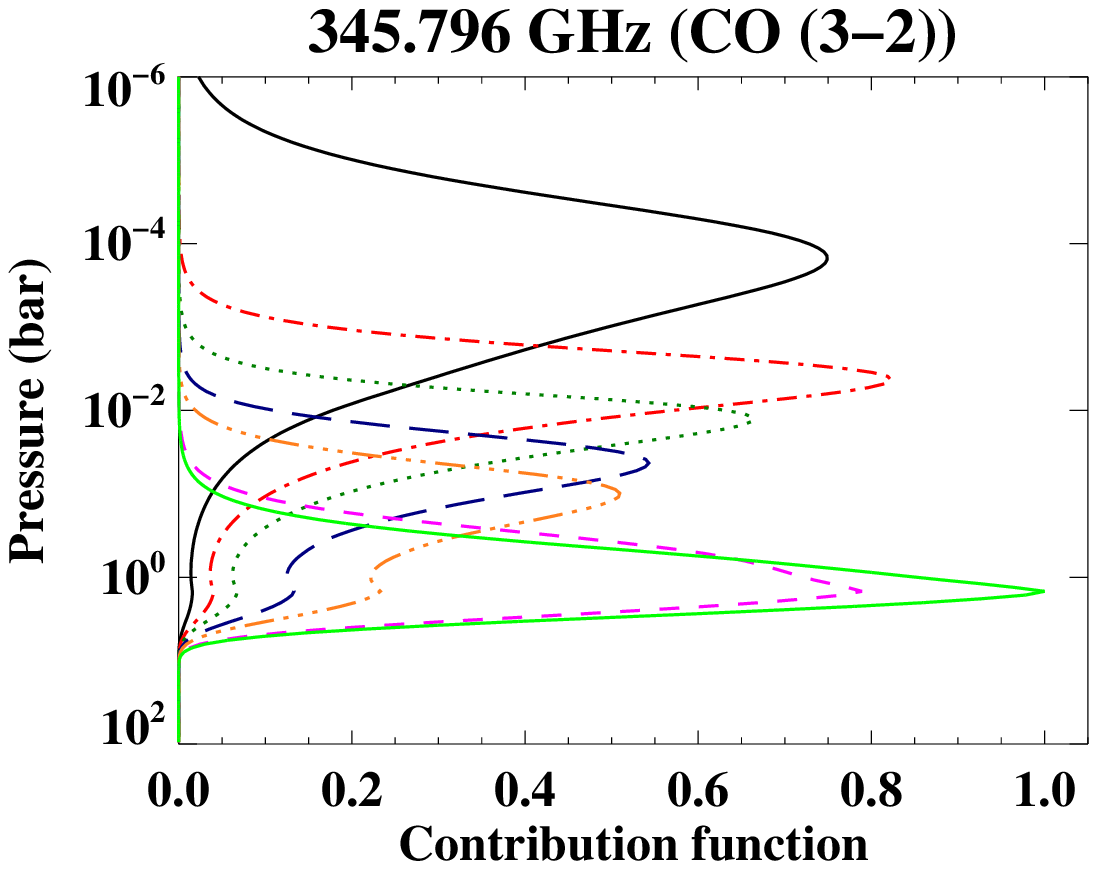}
\end{array}$
\end{center}
\caption[Contribution functions for the CO (1--0), (2--1), and (3--2) lines]{\label{fig:contrib} \small Contribution functions for the CO (1--0), (2--1), and (3--2) lines, illustrating the altitudes contributing to the line at offsets from 0 to 4.5 GHz from line center. The contribution functions include H$_2$-H$_2$, H$_2$-He and H$_2$-CH$_4$ collision-induced absorption, and absorption due to a constant 1 ppm abundance of CO. Close to line center, the emission originates in the stratosphere; at larger offsets from line center, deeper levels contribute to the line. The CO (1--0) and (2--1) lines are the subject of this work; the CO (3--2) line was observed most recently by \cite{hesman07}.  
}

\end{figure}

In this paper we present new observations and modeling of the CO (2--1) and (1--0) rotational lines in Neptune's millimeter spectrum in order to better constrain the vertical CO profile. We investigate how the uncertainty in Neptune's atmospheric thermal profile affects these constraints. This is followed by a discussion of the implications of our derived CO profile on Neptune's global oxygen abundance: updated laboratory measurements of reaction rates have shown that the chemical scheme proposed by \cite{prinn77} and adopted by \cite{lodders94} is actually too slow to be relevant for CO quenching kinetics \citep{griffith99}.  \cite{visscher11} have further revised the kinetic scheme for CO$\rightarrow$CH$_4$ conversion using new values for the reaction rate coefficients. Using this new rate-limiting step, we calculate the CO mole fraction that is transported upwards from Neptune's deep atmosphere, assuming an effective mixing length scale as determined by \cite{smith98}. We also use the \cite{zahnle03} impact rates to determine the effectiveness of cometary impacts in supplying oxygen to Neptune's atmosphere for stratospheric CO production.

\section{Data}\label{sec:data}
\subsection{Observations}
Disk-integrated observations of Neptune in the J = 1--0 and J = 2--1 transitions of CO were performed with the Combined Array for Research in Millimeter-wave Astronomy (CARMA) in March and April 2009. CARMA is a 23 element interferometer  that combines six 10-meter antennas, nine 6-meter antennas, and  eight 3.5-meter antennas. Our observations were performed with the 6- and 10-meter antennas only, for a total of 105 baselines and 2900 m$^2$ of total collecting area. CARMA rotates between 5 different standard antenna configurations; our CO (2--1) data were taken in CARMA's D array, which consists of baselines of 11--148 meters for a synthesized beam of 3.0$\times$1.9 arcseconds at 230 GHz. Our CO (1--0) data were observed in CARMA's C array, with baselines of 26--370 meters for a synthesized beam of 2.1$\times$1.8 arcseconds at 115 GHz. In both sets of observations, the beam size is comparable to the size of the planet; therefore we do not obtain information about spatial variations in the CO distribution. 
\begin {table*}[]
 \footnotesize
\begin{center}
 \caption{} {\small Correlator setups used in the observations. At 1 mm, each track uses one of setups `a'--`d', while all 3-mm tracks use correlator setup `e'. Bands (``windows") 1--3 are located in the same sideband as the CO line center frequency: Band 2 is always a 62 MHz band centered at the CO line rest frequency, and Bands 1--3 are wide (500 MHz) bands offset from line center. Bands 4--6 correspond to Bands 1--3 in the opposite sideband; these data are also recorded, and the wide-band data from both sidebands are used in our analysis. }
 \vspace{\baselineskip}
 
\label{tab:config} 
\begin{tabular}{l l l l l l | l}
\vspace{\baselineskip}
Configuration/setup 	&				&a 		&b   	 	&c  		&d			&e \\
\hline \hline
Band 1			&sideband 		& LSB	&USB	&LSB	&USB		&USB\\
			&bandwidth (MHz)   	& 500	&500 	&500	&500	&500\\
			&center frequency (GHz)    & 231.288	&230.318 	&230.758	&229.538	&114.521\\
\hline
Band 2\footnote{centered on the CO line}			&sideband 		& LSB	&USB	&LSB	&USB		&USB\\
			&bandwidth (MHz)   & 62		&62 		&62		&62			&62\\
				&center frequency   (GHz)   & 230.538	&230.538 	&230.538	&230.538	&115.271\\
\hline
Band 3			&sideband 		& LSB	&USB	&LSB	&USB		&USB\\
				&bandwidth (MHz)   & 500	&500 	&500	&500	&500\\
				&center frequency (GHz)    & 230.288	&229.218 	&232.538	&230.688	&115.521\\
\hline
\hline
Band 4			&sideband                & USB	&LSB	&USB	&LSB		&LSB\\
				&bandwidth (MHz)   & 500	&500 	&500	&500	&500\\
				&center frequency (GHz)   &  235.788	&221.818 	&239.258	&225.538	& 110.021\\
\hline		
Band 5\footnote{not used}&sideband      & USB	&LSB	&USB	&LSB		&LSB\\
				&bandwidth (MHz)   & 62	&62 	&62	&62	&62\\
				&center frequency (GHz)   & 236.538	&221.598 	&239.478	&224.538	&109.271\\

\hline
\nopagebreak Band 6			&sideband 		& USB	&LSB	&USB	&LSB		&LSB\\
				&bandwidth (MHz)   & 500	&500 	&500	&500	&500\\
				&center frequency (GHz)   & 236.788	&222.918 	&237.478	&224.388	&109.021\\
\hline
\end{tabular}

\end{center}
\end{table*}

\normalsize

At the time of our observations the CARMA correlator had three dual bands (or ``windows") with configurable bandwidth of either 500, 62, 31, 8 or 2 MHz. Each band could be placed independently anywhere within the 4 GHz IF bandwidth, and appears symmetrically in the upper and lower sidebands of the first local oscillator. The sideband, total bandwidth, and central frequency of the 6 windows (3 in each sideband) for each of our correlator setups are given in Table \ref{tab:config}. In each setup, we tuned the receivers to the CO (1--0) or (2--1) rest frequency in either the lower (setup a,c)  or upper (setup b,d,e) sideband.  Band 2 was always configured with a bandwidth of 62 MHz across 63 channels, to observe the CO line center at a resolution of 0.98 MHz per channel.  Bands 1 and 3 were each configured for maximum bandwidth (500 MHz across 15 channels) and positioned at a frequency offset from the CO line center. Bands 4--6 correspond to Bands 1--3 in the opposite sideband.  For the CO (1--0) line observations, we configured Band 1 at an offset of $-0.75$ GHz and Band 3 at an offset of +0.30 GHz (correlator setup `e' in Table \ref{tab:config}). For the CO (2--1) line, we alternated between 4 different correlator setups of the 500 MHz bands (Table \ref{tab:config}, setups `a-d') to extend our frequency coverage of the broad CO line wings. Frequency offsets of the 500 MHz bands from line center vary from $0.15-9.3$ GHz. 

Each individual observation consisted of a $3.6-6.8$ hour track, with a 15 minute observation of a passband calibrator, followed by a series of observing cycles of 15 minutes on source and 3 minutes on a phase calibrator. Optical pointing was performed every hour. No planets were available for primary flux calibration, so we observed MWC349 when it was available. The quality of the data from tracks shorter than 3 hours and tracks with very poor weather were too low, and hence the data were rejected (2 of 8 tracks at  both 1 mm and 3 mm). Weather conditions during the remaining observations were generally fair to good, with typical root mean square (rms) path errors of 150-400 $\mu$m on a 100-m baseline,  and zenith optical depths of $0.1--0.2$. The total time on source was about 14.9 hours at 1 mm and 18.6 hours at 3 mm. These observations are summarized in Table \ref{tab:data}.

\subsection{Calibration}
The data are reduced and calibrated using the MIRIAD software package \citep{sault11}. Prior to any calibration, flagging is performed: we flag 10 edge channels for the narrow bands and 3 edge channels for the wide bands; the narrow window that is not centered on the emission core of the CO line (Band 5: see Table \ref{tab:config}); and any bad data.  After performing passband calibration using our bright quasar observations, time-dependent gain solutions are derived using the wide-band data, and then applied to the full data set. We do an initial self calibration using the phase calibrator. This first calibration is performed in two steps: a record-by-record phase solution is found for the phase calibrator, to remove short-term variations. Then, a phase and amplitude self calibration is performed using an 18 minute interval and applied to the Neptune data. Finally,  a record-by-record phase-only self calibration is performed on Neptune itself to remove short-term phase variations in the Neptune data.

\vspace{\baselineskip}
\begin {table*}[]
\begin{minipage}{\textwidth}
 \footnotesize
\begin{center}
 \footnotesize
\setlength{\tabcolsep}{0.05in}

\caption{}{\small Summary of observations (all in 2009) \label{tab:data}}

\vspace{\baselineskip}
\begin{tabular}{l l l l l l l l}

\hline
Line& Frequency (GHz) &Date 	& Correlator setup\footnote{as specified in Table 1}	&T$_{int}$ (hours)\footnote{time on source}	& Calibrators & rms path ($\mu$m) & $\tau_{zen}$ \tabularnewline
\hline
CO (2--1) 	&	230.538	&	07 Mar  & a & 2.33 & 3C454.3\footnote{Passband calibrator},2229-085\footnote{Phase calibrator}&208&0.22 \\
 		&			&	08 Mar  & b &2.08 & 3C454.3$^c$, 2229-085$^d$&165&0.24\\
  		&			&	10 Mar  & c & 2.83 & 3C454.3$^c$, 2229-085$^d$&157&0.19\\
  		&			&	11 Mar  & d & 2.33 &  3C454.3$^c$, 2229-085$^d$&133&0.23\\
  		&			&	27 Mar  & a & 2.57 & 3C454.3$^c$, 2229-085$^d$&162&0.17\\
  		&			&	28 Mar  & a & 2.71 & 3C454.3$^c$, 2229-085$^d$, MWC349\footnote{Flux calibrator}&147&0.23\\
 CO (1--0) 	&115.271		&	16 Apr  & e & 4.25 &1751+096$^c$, 3C446$^d$, MWC349$^e$&413&0.12\\
       		&			&	17 Apr  & e & 3.40 & 3C446$^{c,d}$, MWC349$^e$&306&0.12\\
      		&			&	23 Apr  & e & 3.03 &3C454.3$^c$, 3C446$^d$, MWC349$^e$&335&0.16\\
     		&			&	24 Apr  & e & 2.25 &3C454.3$^c$, 3C446$^d$, MWC349$^e$&424&0.18\\
   	 	&			&	28 Apr  & e & 3.15 & 3C454.3$^c$, 3C446$^d$, MWC349$^e$&200&0.11\\
     		&			&	29 Apr  & e & 2.50 & 3C454.3$^c$, 3C446$^d$&180&0.10\\

\hline

\end{tabular}
\end{center}
\end{minipage}
\end{table*}

\subsubsection{CO (J=2--1)}
For this set of observations, only one track (28 March 2009) contains observations of the primary flux calibrator, MWC349. We flux calibrate this track using an assumed value of 1.86 Jy for MWC349 at 230 GHz, based on flux density measurements obtained with CARMA between 2007 and 2011\footnote{\url{http://cedarflat.mmarray.org/fluxcal/primary\_sp\_index.htm}}. We estimate the flux calibration to be accurate to better than 20\%. For a first pass at the flux calibration of the other tracks, we use our phase calibrator as a secondary flux calibrator. We then bin the Neptune visibility data into 8 ($u,v$) bins between 0 and 80 $k\lambda$, and adjust the flux of the phase calibrator to align the binned ($u,v$) data as much as possible: these adjustments are about $2$--$5$\%. From the deviation of these fits, we estimate the remaining error in day-to-day gains to be less than 3\%. 

\subsubsection{CO (J=1--0)}
All but one of our 3-mm tracks contains primary flux calibrator data (Table \ref{tab:data}). However, we find that in general, the flux determination from the primary flux calibrator is very sensitive to bad/noisy data. We are able to get more stable fluxes by using our best primary flux calibrator data from 28 April 2009, and applying the same procedure as for the 1-mm data. We use a flux density of 1.29 Jy for MWC349 at 3 mm$^1$.

\subsection{Imaging}\label{sec:maps}
\begin{figure*}[htb!]
\begin{center}$
\begin{array}{cccccc}
\includegraphics[width=0.33\textwidth]{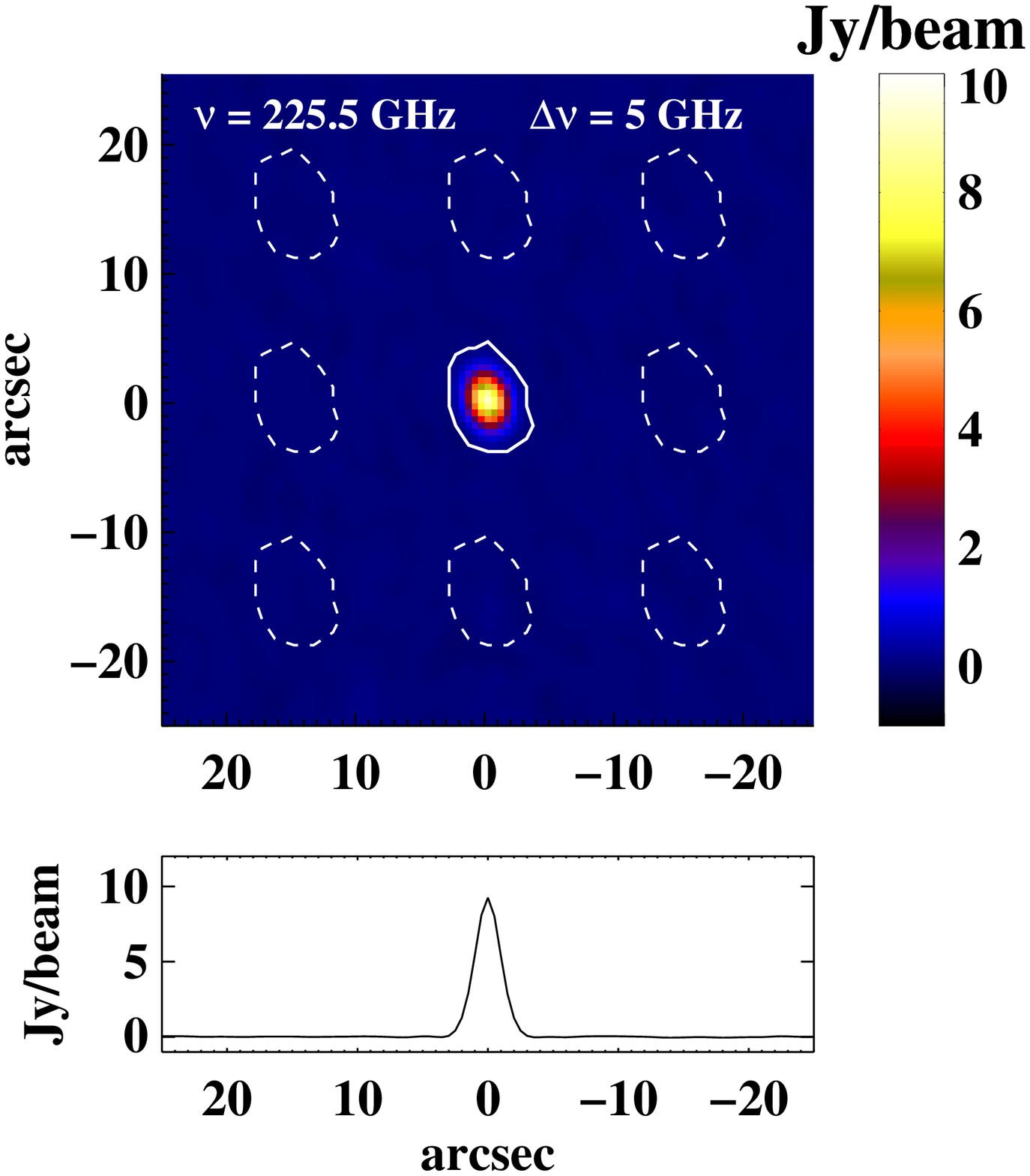}
\includegraphics[width=0.33\textwidth]{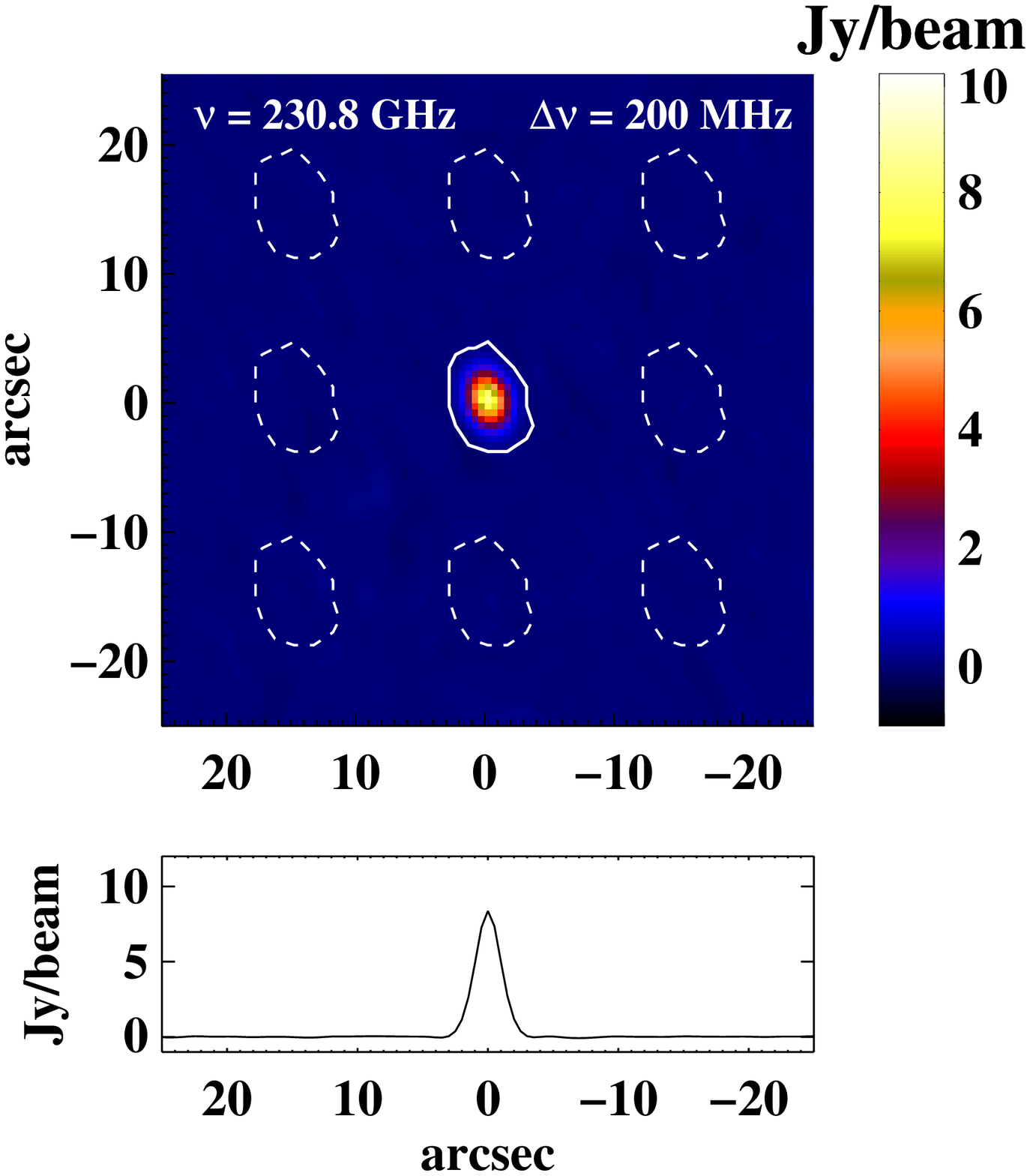}
\includegraphics[width=0.33\textwidth]{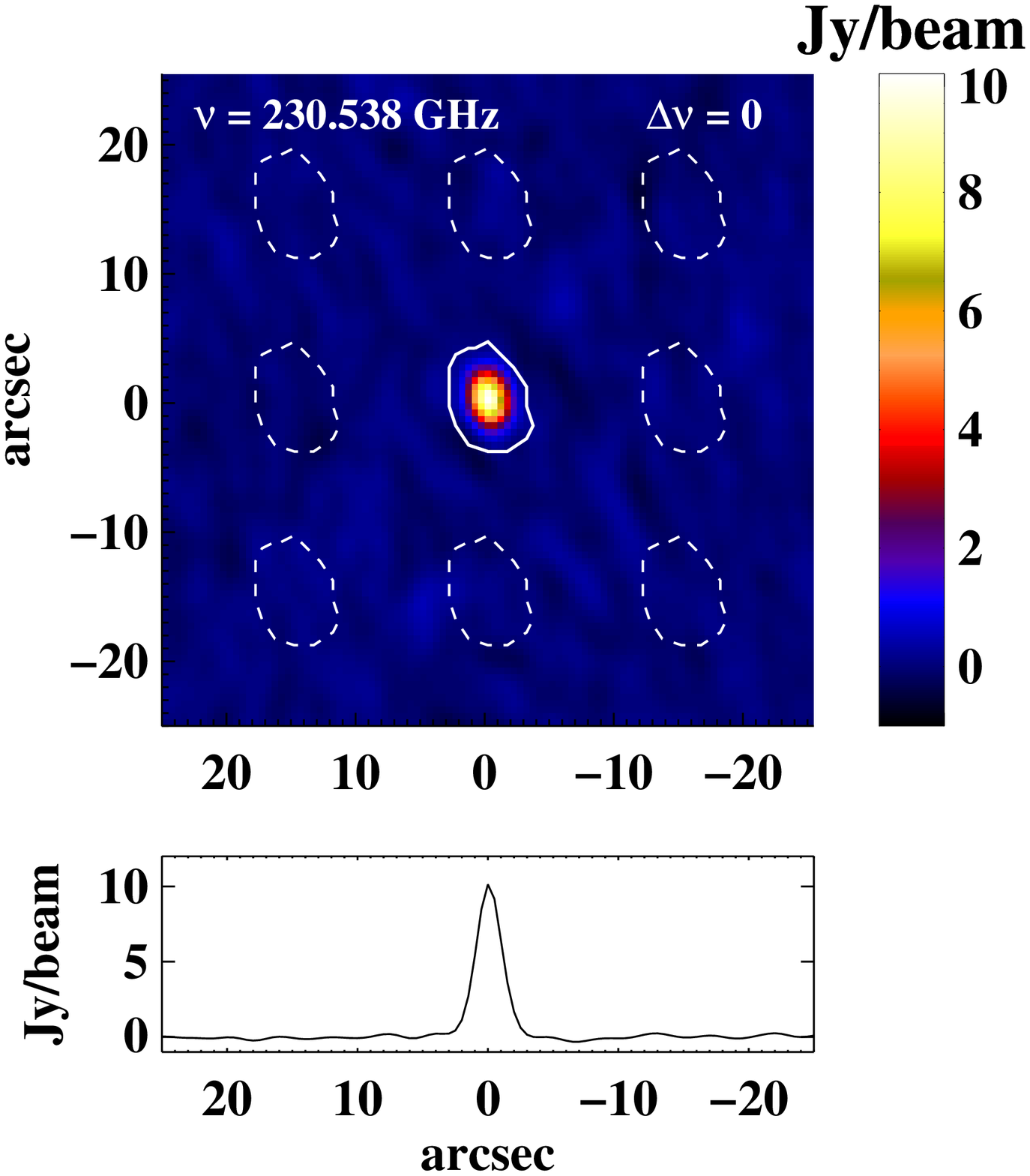}\\\
\includegraphics[width=0.33\textwidth]{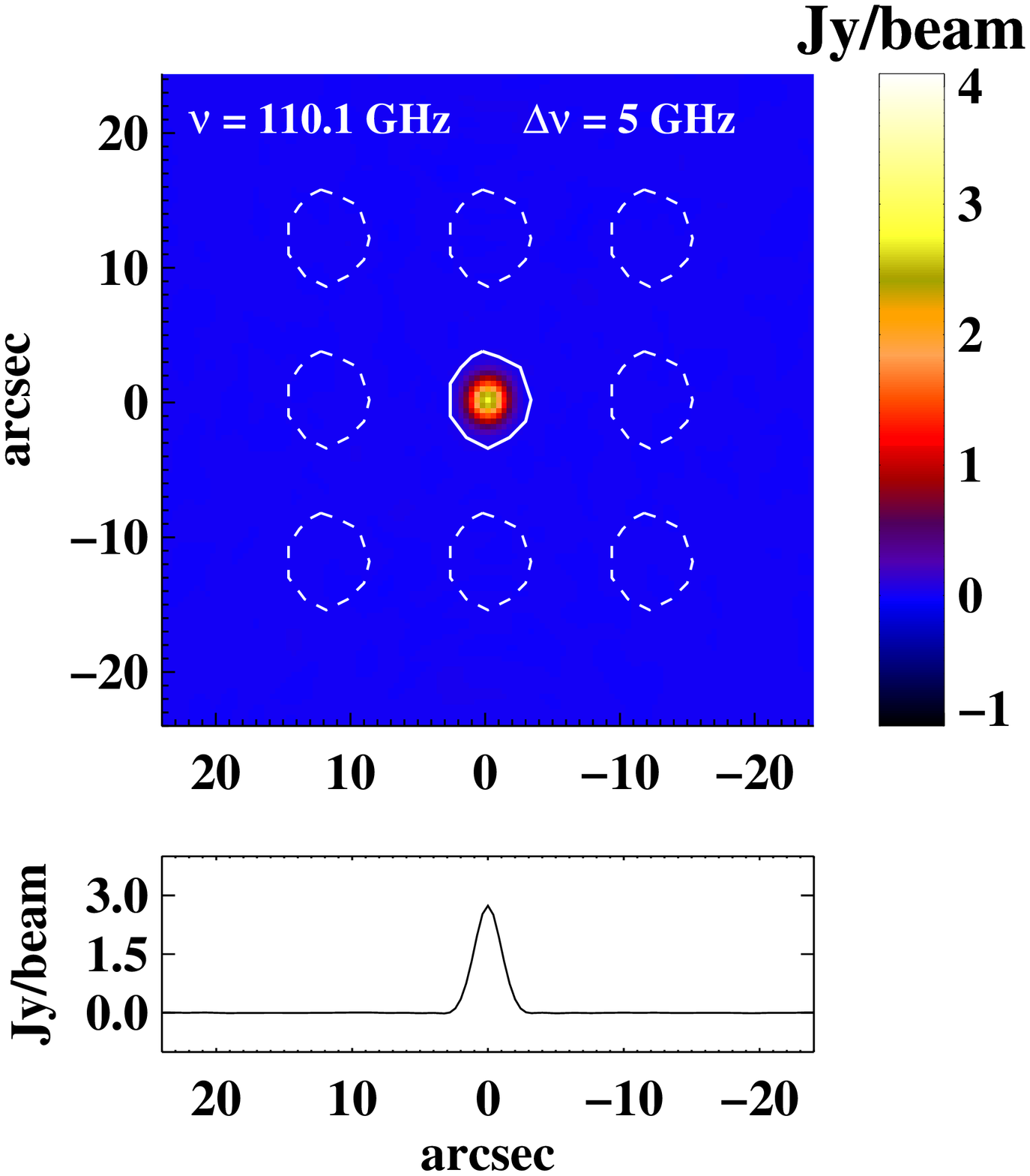}
\includegraphics[width=0.33\textwidth]{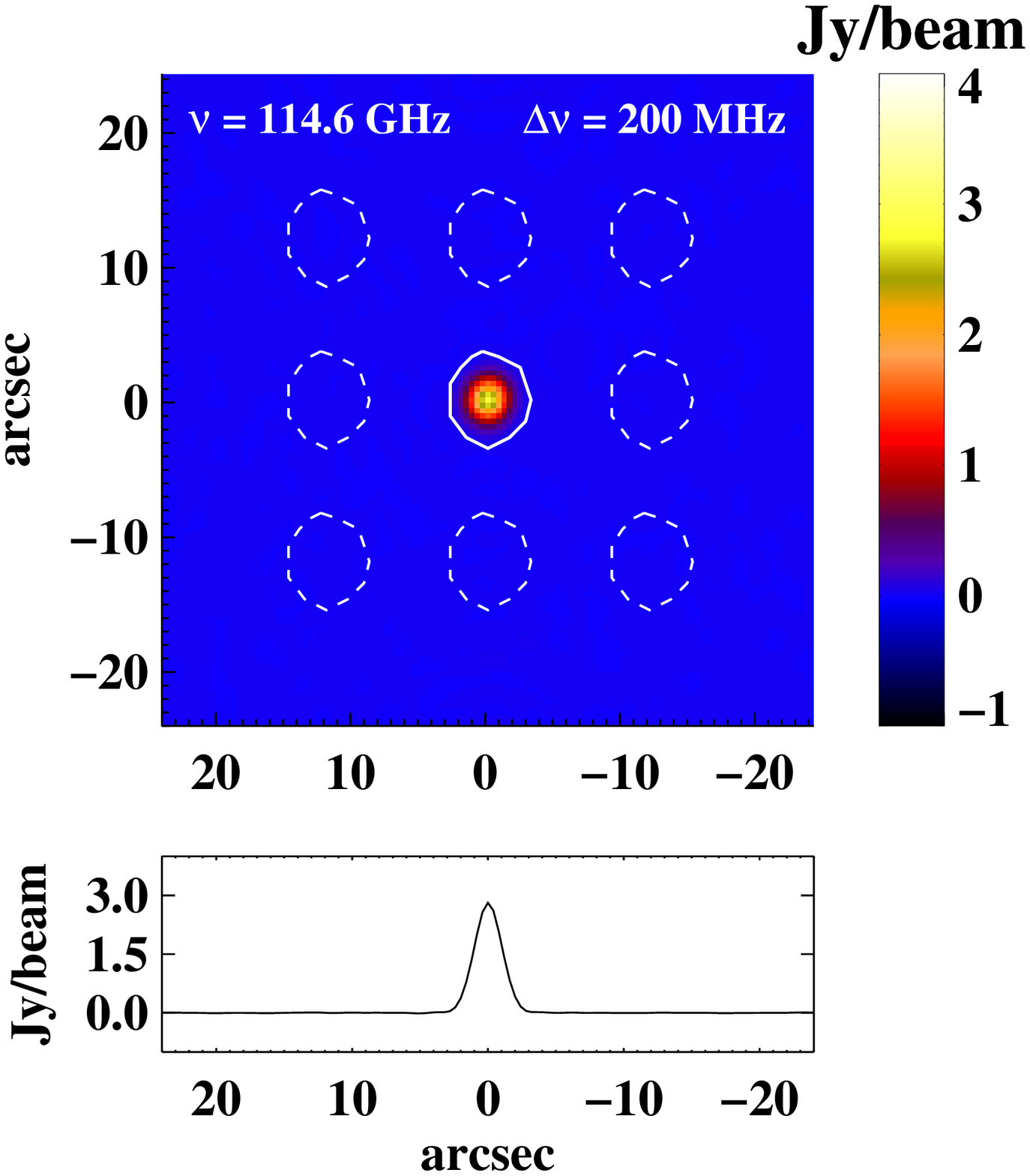}
\includegraphics[width=0.33\textwidth]{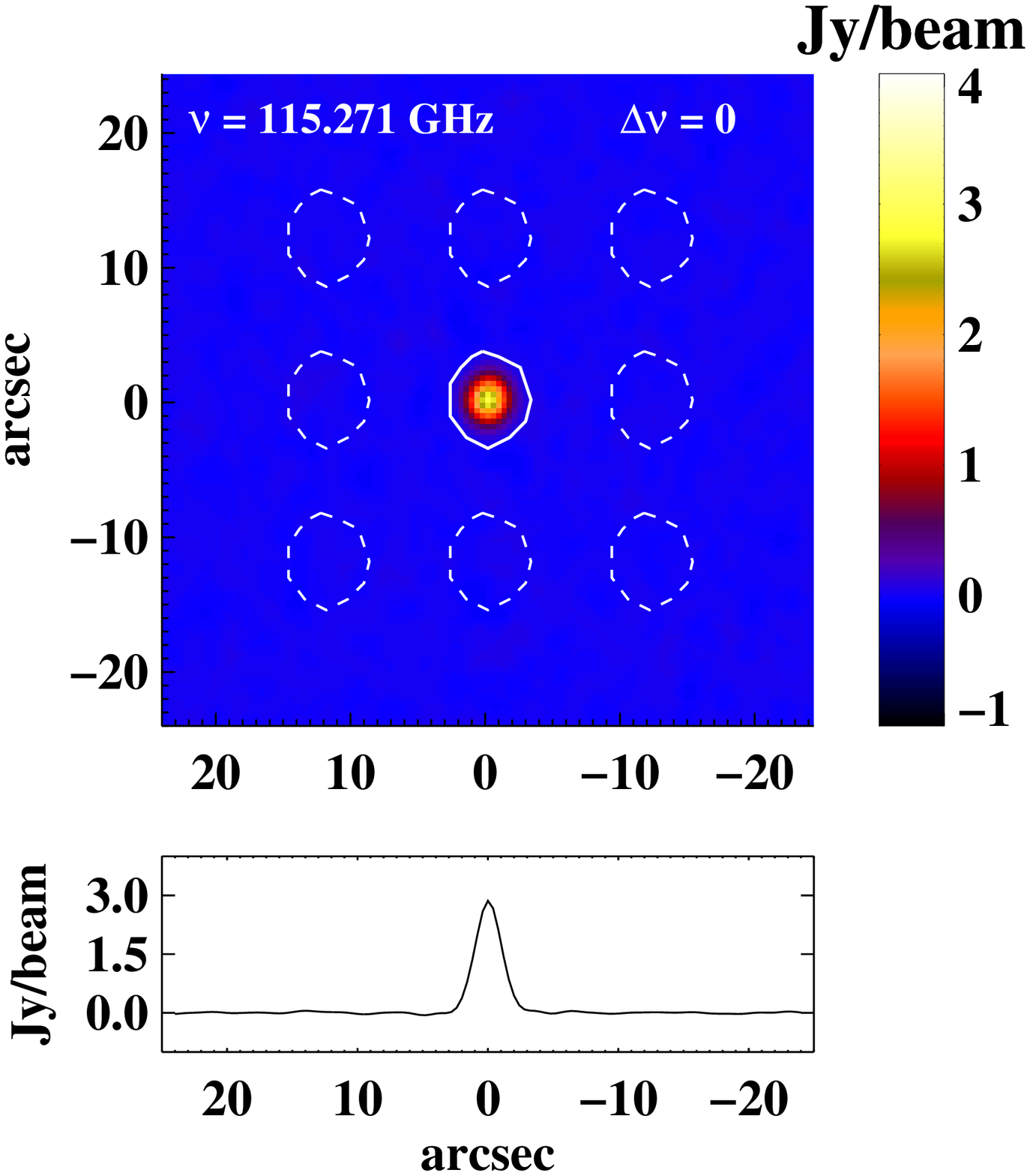}\\\
\end{array}$
\end{center}
\caption[Representative channel maps]{\label{fig:maps} \small Representative channel maps; color scale indicates intensity in Jy/beam. The top row are maps from a typical 1-mm day (11 March); the bottom row are the channel maps created from all days of 3-mm data. The left-most figures are wideband channels, roughly 5 GHz from line center. In the middle column are maps from wide channels in the absorption, about 500 MHz from line center. Maps on the right are from narrow channels at line center. Below each map is a scan through the map Right Ascension (a value of 0 on the y- axis). Neptune is unresolved in the maps; the total flux in each channel is determined by summing over the region indicated by the solid white line. The error is estimated by the regions indicated by dashed lines- see Section \ref{sec:maps} for discussion. 
}
\end{figure*}

Our 3-mm data sets are all observed with the same correlator setup; therefore, we combine all the 3-mm data prior to imaging. In contrast, the 1-mm tracks are observed at different frequencies and we image each track separately. After imaging each channel, we deconvolve the dirty maps using the CLEAN algorithm.  Figure \ref{fig:maps} shows several of our final maps. Each map represents a single channel of data: the top row are 1-mm channel maps from a day with typical weather conditions (11 March), and the bottom row are examples of 3-mm channel maps. Below each map is a scan in Right Ascension through the center of Neptune.

\subsection{Flux determination and error estimate}
\begin{figure*}
\begin{center}$
\begin{array}{cc}
\includegraphics[width=0.73\textwidth,clip,trim=.4in .1in 0 0]{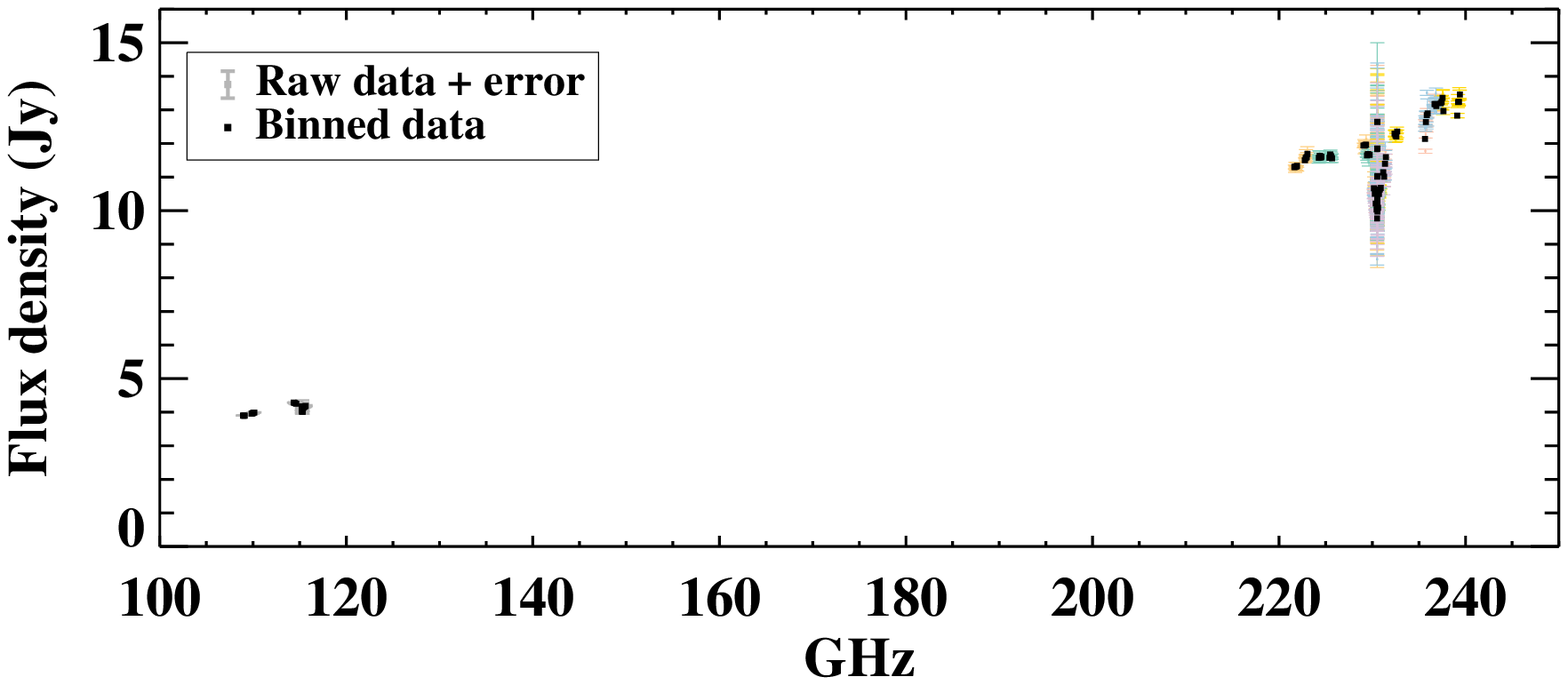}\\
\includegraphics[width=0.75\textwidth]{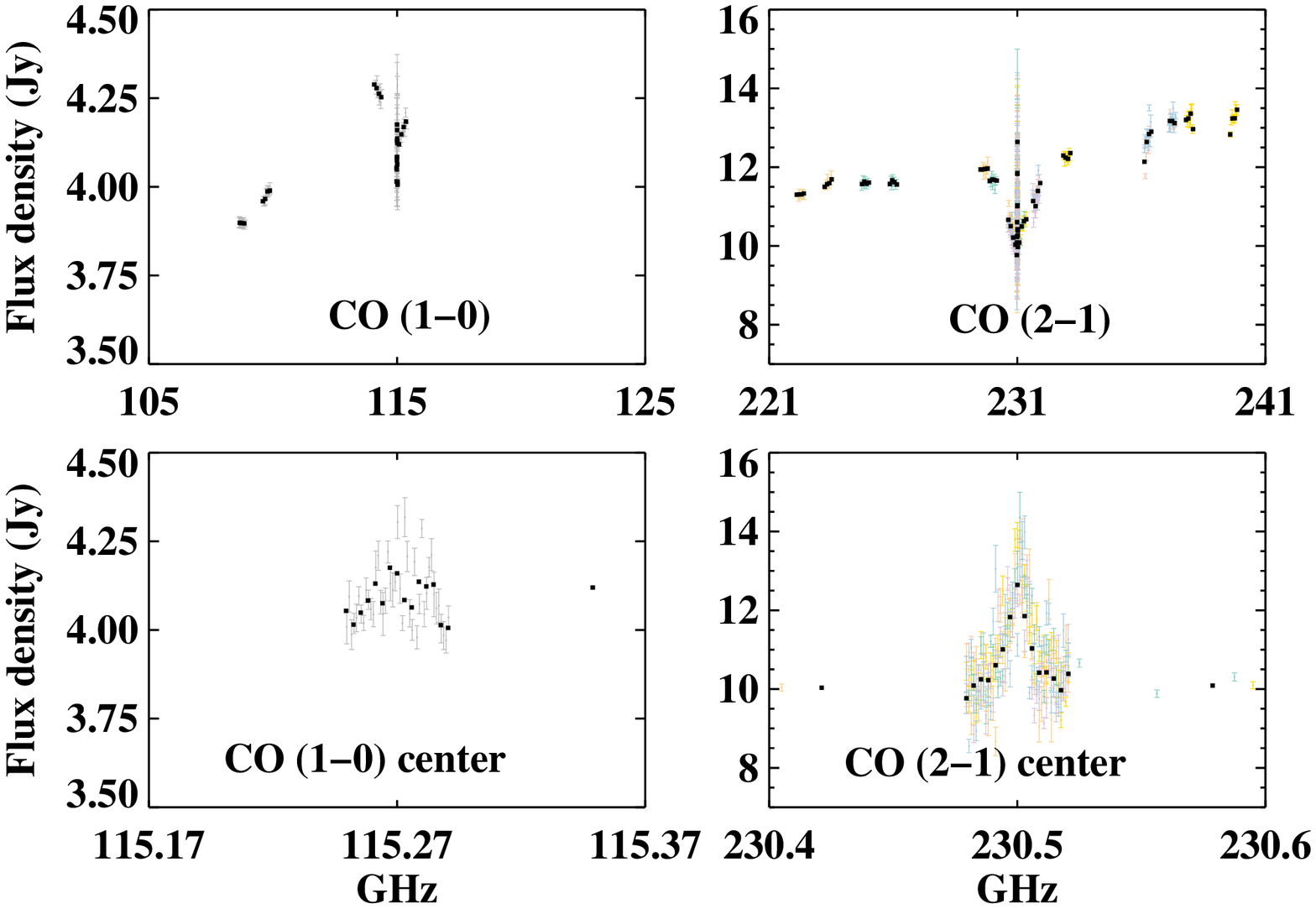}
\end{array}$
\end{center}
\caption[Raw CO (1--0) and (2--1) spectra]{\label{fig:coraw} \small Raw CO (1--0) and (2--1) spectra. Each grey point corresponds to the flux density calculated from a single  3-mm channel map; colors are used to delineate between individual tracks at 1 mm. Error bars are calculated using the method described in Section \ref{sec:maps}. Black points are binned to 3 times the width of the individual channels.  
 }
\end{figure*}

The flux density in each channel is determined as the integrated intensity over a hand-selected region, chosen to include all of the signal from Neptune, as indicated by the solid line in each map in Fig.  \ref{fig:maps}. To quantify the errors in the determined flux density, we then shift this region to 8 different positions outside of the source (indicated by dashed lines in  Fig.  \ref{fig:maps}). The error in each channel is estimated as the standard deviation of the integrated fluxes of these regions. We also calculate the flux errors using the rms of the pixel values in the residual maps. We find that the two methods of error estimation are in good agreement. These error estimates do not include systematic effects, such as errors in the bandpass calibration or in the assumed flux of the calibrator; however, they are useful weights for fitting the data to models (see Sections \ref{sec:analysis} and \ref{sec:errors}).  Raw spectra, with uncertainties,  are shown in Fig. \ref{fig:coraw}; for reference, the individual 1-mm tracks are differentiated by color. A frequency-binned version of the data is plotted in black, using a bin interval of three times the channel width.

\section{Model}\label{sec:model}

To model Neptune's millimeter spectrum, we developed a line-by-line radiative transfer code that integrates the equation of radiative transfer, assuming local thermodynamic equilibrium (LTE)

\begin{eqnarray}
B_\nu(T_D,\mu) =\int_0^\infty B_\nu(T)e^{-\tau/\mu} d\tau/\mu
\end{eqnarray}
over a model atmosphere consisting of 2000 plane-parallel layers extending from 200 bar to $5\ \mu$bar. In this equation, $T_D$ is the disk brightness temperature, $\mu=\cos \theta$ with $\theta$ defined as the angle between the line of sight and local vertical, $B_\nu(T)$ is the Planck function for temperature $T$, and $\tau$ is the optical depth.  The code is optimized for least-squares fitting of the CO altitude profile (Section \ref{sec:analysis}), and has been thoroughly tested against the microwave radiative transfer code described in \cite{depater91}, after the latter was updated as described by \cite{depater05a}. The \cite{depater91} code has been further revised to include the \cite{orton07c} H$_2$ absorption coefficients (Section \ref{sec:opacity}), and absorption due to CO using subroutines from \cite{depater91v} that were updated with the parameters described in Section \ref{sec:opacity}. We find that the two radiative transfer codes produce consistent model spectra for the same input parameters, except at the CO line center, because the new code includes  the effect of Doppler broadening on the shape of the emission peak (Section \ref{sec:disk}) and the revised \cite{depater91}  code does not. The details of the new model are described below.

\subsection{Composition}\label{sec:composition}

Neptune's upper atmosphere is dominated by H$_2$ and He.  \cite{conrath91} estimated the relative abundances of these two species using constraints on the atmospheric mean molecular weight from Voyager infrared and radio occultation measurements; they found a best-fit helium mole fraction of 0.19$\pm 0.032$. The detection of HCN on Neptune \citep{marten91} led \cite{marten93} to suggest a scenario \citep[originally proposed by][]{romani89} in which Neptune's nitrogen is predominantly present as N$_2$, rather than NH$_3$. Such a high N$_2$ abundance is greater than expected from thermochemical equilibrium arguments \citep[e.g.][for Uranus]{fegley91}, but could be the source of atomic nitrogen for the production of stratospheric HCN \citep{marten93}, and would help explain the observed atmospheric NH$_3$ deficit \citep[][discussed below]{romani89}. Reanalysis of the \cite{conrath91} data \citep{conrath93} showed that a mole fraction of 0.003 of N$_2$ and 0.15 of He is consistent with the Voyager measurements. Using spectra from the Infrared Space Observatory Long-Wavelength Spectrometer (ISO-LWS),  \cite{burgdorf03} found a He/(H$_2$+He) mass ratio of 26.4$^{+2.6}_{-3.5}\%$, and an N$_2$ mixing ratio of less than 0.7\%. They determined that the correlation of their value of the He mole fraction with the \cite{conrath93} results gives an N$_2$ mole fraction of $0.3\pm 0.2\%$.  Accordingly, for our atmospheric models we maintain a He/H$_2$ ratio of 0.15/0.847 and an N$_2$/H$_2$ ratio of 0.003/0.847 by number throughout our model atmosphere; these numbers are consistent with the solutions of both \cite{conrath93} and \cite{burgdorf03} within the 1$\sigma$ uncertainties. Since the presence of N$_2$ in Neptune's atmosphere is not certain, we also tested models with the same He/H$_2$ ratio and no N$_2$; the difference was negligible.  

Previous observations have shown that CH$_4$ is supersaturated in Neptune's stratosphere \citep{orton87,orton90,yelle93}. We adopt the stratospheric CH$_4$ profile recently derived by \cite{fletcher10} from AKARI infrared data, which has a mole fraction of $9 \times 10^{-3}$ at 50 mbar. Below the methane condensation level, we adopt a mole fraction of 0.022 as measured by \cite{baines95}, which implies an enrichment factor of $\sim50$ over the protosolar C/H ratio \citep{asplund09}.

While the composition of Neptune's deep troposphere has yet to be uniquely determined, data from centimeter wavelengths suggest that NH$_3$ is depleted in Neptune's atmosphere  \citep{romani89,depater91}.  Good fits to the cm wavelength range are obtained by enhancing H$_2$S 30-50 times over the solar S/H value \citep{depater91,deboer96} and using a solar abundance of nitrogen in NH$_3$. \cite{hoffman01} suggest that PH$_3$ may be an important source of microwave absorption as well; however, \cite{moreno09} determined an upper limit of phosphorous in Neptune's upper atmosphere of 0.1 times the solar abundance, from observations of the PH$_3$ (1--0) transition. Due to the uncertainty in our absolute flux calibration (as much as 20\%, see Section \ref{sec:data}), we do not attempt to constrain the abundances of species other than CO from our data. However, we do investigate the potential effect of these absorbers on the 1- and 3-mm spectra: using the code described by \cite{depater91,depater05a}, we model the deep atmosphere with a 10-50 times solar enrichment of H$_2$O and H$_2$S, a solar abundance of NH$_3$, and 0.1 times solar PH$_3$.  Trace species are removed from the model atmosphere at higher altitudes by condensation, when the partial pressures of the trace gases exceed the saturation vapor pressure (see \cite{depater05a} for a detailed description). Figure \ref{fig:microwave} shows the effects on the microwave spectrum of absorption due to each of these gases individually, as well as that from all gases combined.

\subsection{Opacity}\label{sec:opacity}

To determine the optical depth $\tau_\nu(z)$ we consider the following sources of opacity:

\subsubsection{Collision-induced H$_2$ absorption}
For Neptune, the dominant millimeter-wave opacity source is collision-induced absorption by H$_2$ with H$_2$, He, and CH$_4$.  We use  the absorption coefficients  calculated from revised ab-initio models of \cite{orton07c}, assuming an equilibrium distribution of hydrogen. These authors incorporate a correction to the Borysow models \citep{borysow85,borysow88,borysow91,borysow92,borysow93}, and show that the new coefficients are an improvement at low temperatures. In practice, we find that spectra produced using the \cite{orton07c} coefficients are 1.6-2 K higher than those made using the \cite{borysow85} models.

\subsubsection{CO}
The absorption due to the CO (1--0) and (2--1) rotation lines is calculated assuming a Voigt line shape profile. The H$_2$ broadened line half-width is determined from a fit to the data of \cite{mengal00} and is $\sim$2.8 MHz/Torr at 300 K for the CO (1--0) and (2--1) lines. The remaining line parameters are taken from the HITRAN 2008 database \citep{hitran09}. A Van Vleck-Weisskopf line shape was tested using the updated \cite{depater91} code and found to be a nearly identical match to the Voigt line shape in the atmospheric region probed.

\subsubsection{Additional microwave absorbers}

\begin{figure*}[t!]
\begin{center}$
\begin{array}{cc}
\includegraphics[width=0.5\textwidth,clip,trim=.4in 0 0 0]{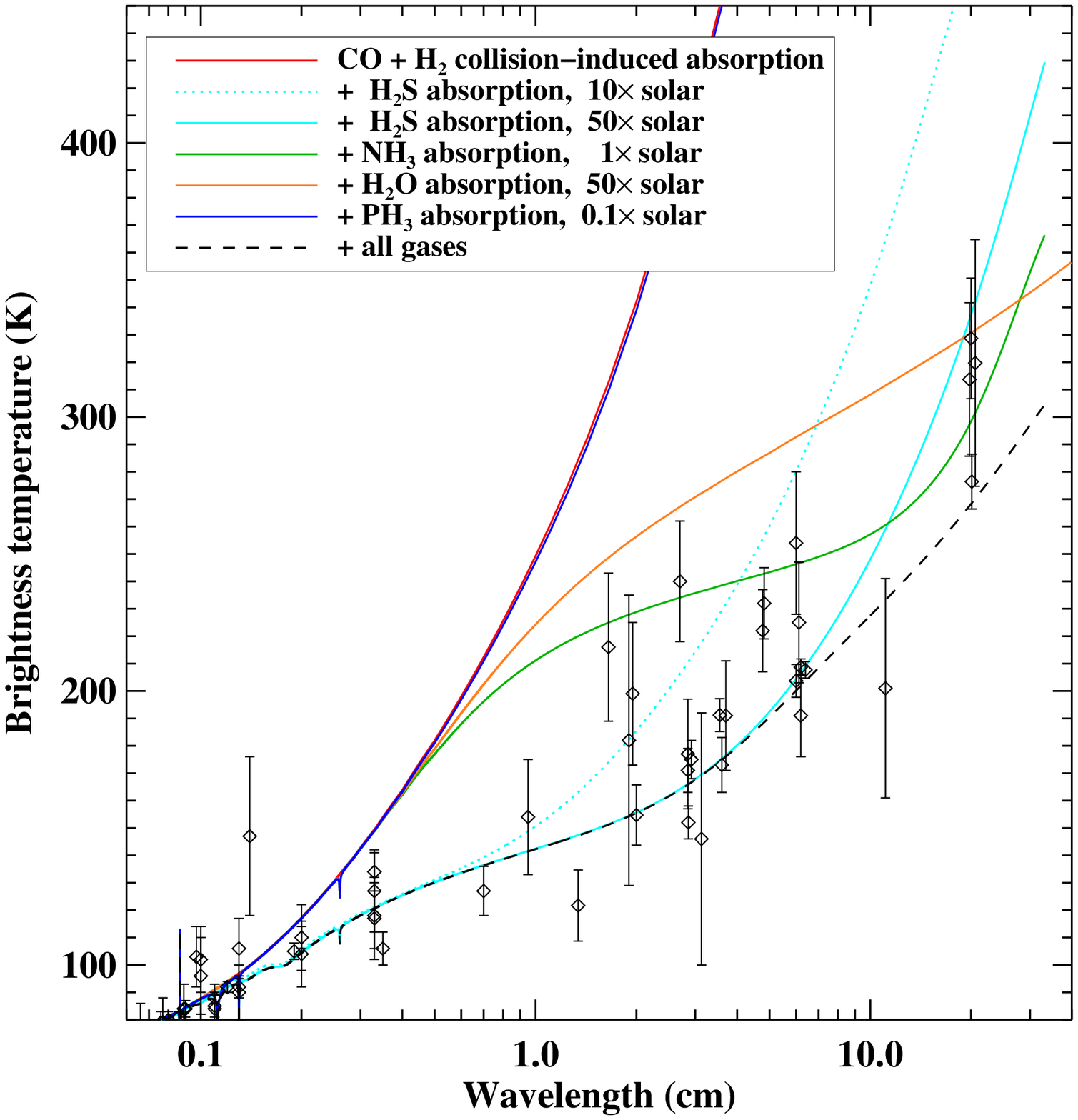}
\includegraphics[width=0.5\textwidth,clip,trim=.4in 0 0 0]{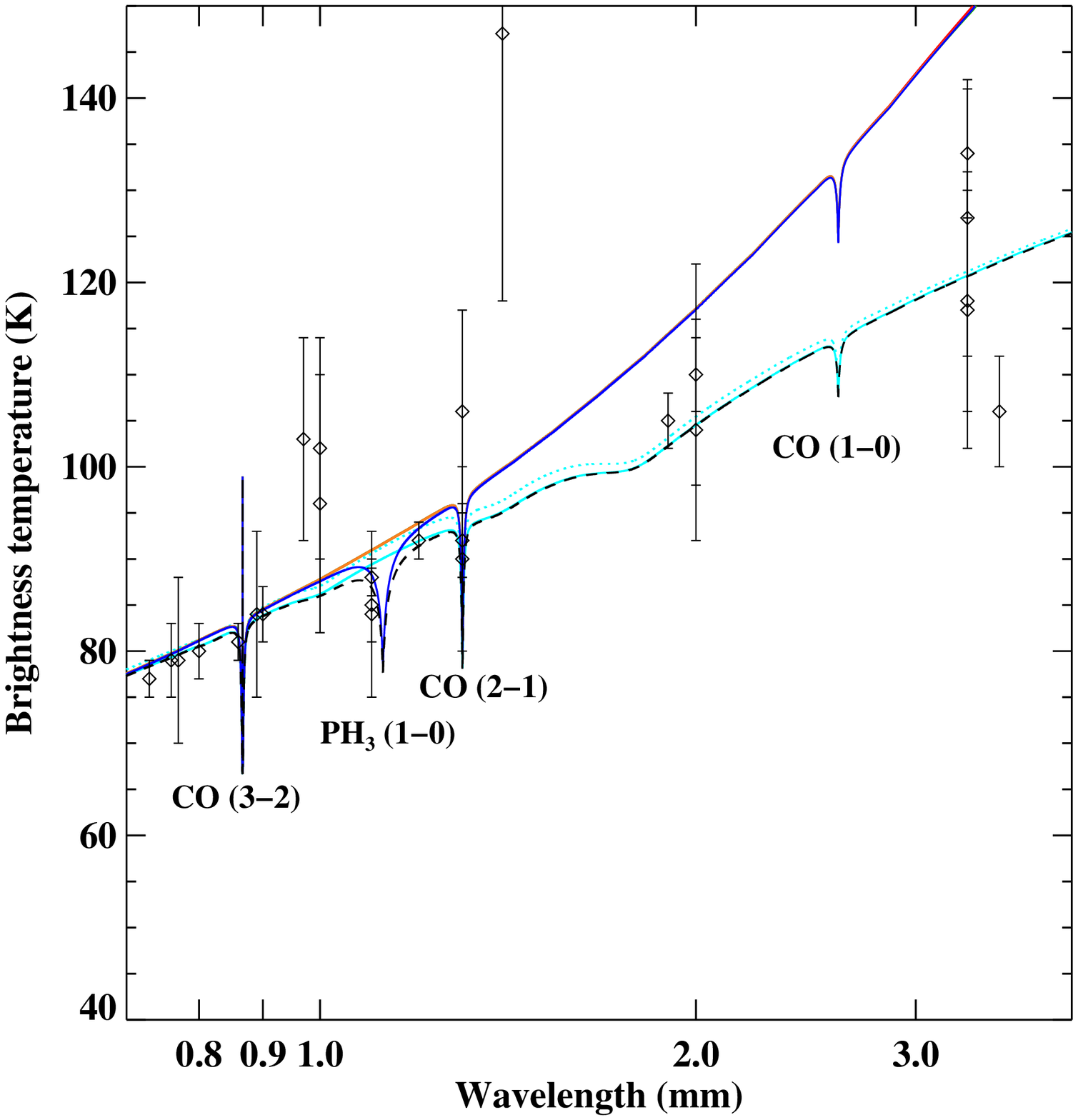}

\end{array}$
\end{center}
\caption[Model atmosphere calculations for mm--cm wavelengths]{\label{fig:microwave} \small Model atmosphere calculations. All models include CO and H$_2$ collision-induced absorption (see Section \ref{sec:opacity}); the red line represents these two opacity sources only.  Opacity due to a 10 and 50 times protosolar abundance of H$_2$S (cyan, dashed and solid lines, respectively); a 50 times protosolar abundance of H$_2$O (orange); a protosolar abundance of NH$_3$ (green) and a 0.1 times protosolar abundance of PH$_3$ (blue) are included. The black dashed model contains all of these absorptions (note that a 50 times protosolar abundance was used for H$_2$S, rather than the 10 times solar value). A compilation of microwave data taken before 1991 is plotted for comparison (diamonds); these data are described in \cite{depater91} and references therein. The right plot zooms in on millimeter wavelengths, and shows the importance of  H$_2$S opacity at these wavelengths. 
}
\end{figure*}

We model several other potential sources of radio-wavelength opacity (H$_2$O, NH$_3$, H$_2$S, PH$_3$) using the updated \cite{depater91} code and the abundances from Section \ref{sec:composition}. We find that H$_2$O and NH$_3$ do not affect the spectrum at 1-3 mm wavelengths (Fig. \ref{fig:microwave}). The far wings of the PH$_3$ (1--0) line at 266.9  GHz could influence the high frequency wing of the CO (2--1) line; however, for a 0.1 times solar abundance of PH$_3$ the line is too weak to significantly affect the CO profile.  We note that a 0.1 times solar abundance of PH$_3$ is an upper limit \citep{moreno09}; larger values would also be inconsistent with the available data according to our own models, as shown. We find that absorption due to H$_2$S is important at  wavelengths longer than 1 mm; at 3 mm the addition of $>10$ times solar abundance of H$_2$S decreases the continuum level by $\sim$20 K.

We therefore incorporate H$_2$S absorption into our new radiative transfer code following the formalism described by \cite{deboer94}; line parameters are taken from the JPL catalog \citep{jpl}, while parameters for the Ben Reuven line shape come from \cite{deboer94}. We choose as our nominal atmospheric model no H$_2$S absorption, and additionally investigate models with absorption due to 10 and 50 times solar  H$_2$S abundances.  We note that high H$_2$S abundances give a best fit to the cm wavelength region (Section \ref{sec:composition}; Fig. \ref{fig:microwave}).

\subsection{Thermal profile}\label{sec:TP}

The observed shapes of the CO absorption lines depend strongly on both the CO abundance and the temperature-pressure (TP) profile in the atmosphere. Therefore, we look at a range of TP profiles to understand the effect of uncertainties in the TP profile on our derived CO altitude profile.  Our nominal TP profile is that of \cite{fletcher10}; they adopt the TP profile of \cite{moses05} below the tropopause and retrieve the stratospheric temperature profile from mid-infrared AKARI spectra. Fits to HD lines in Neptune's 51-220 $\mu$m spectrum from Herchel/PACS \citep{lellouch10} are sensitive to the 10-500 mbar levels; these authors favor a thermal profile much like that of \cite{fletcher10} in the stratosphere as well.

 \begin{figure}[h]
\begin{center}$
\includegraphics[width=.45\textwidth,clip,trim=.3in 0 0 0]{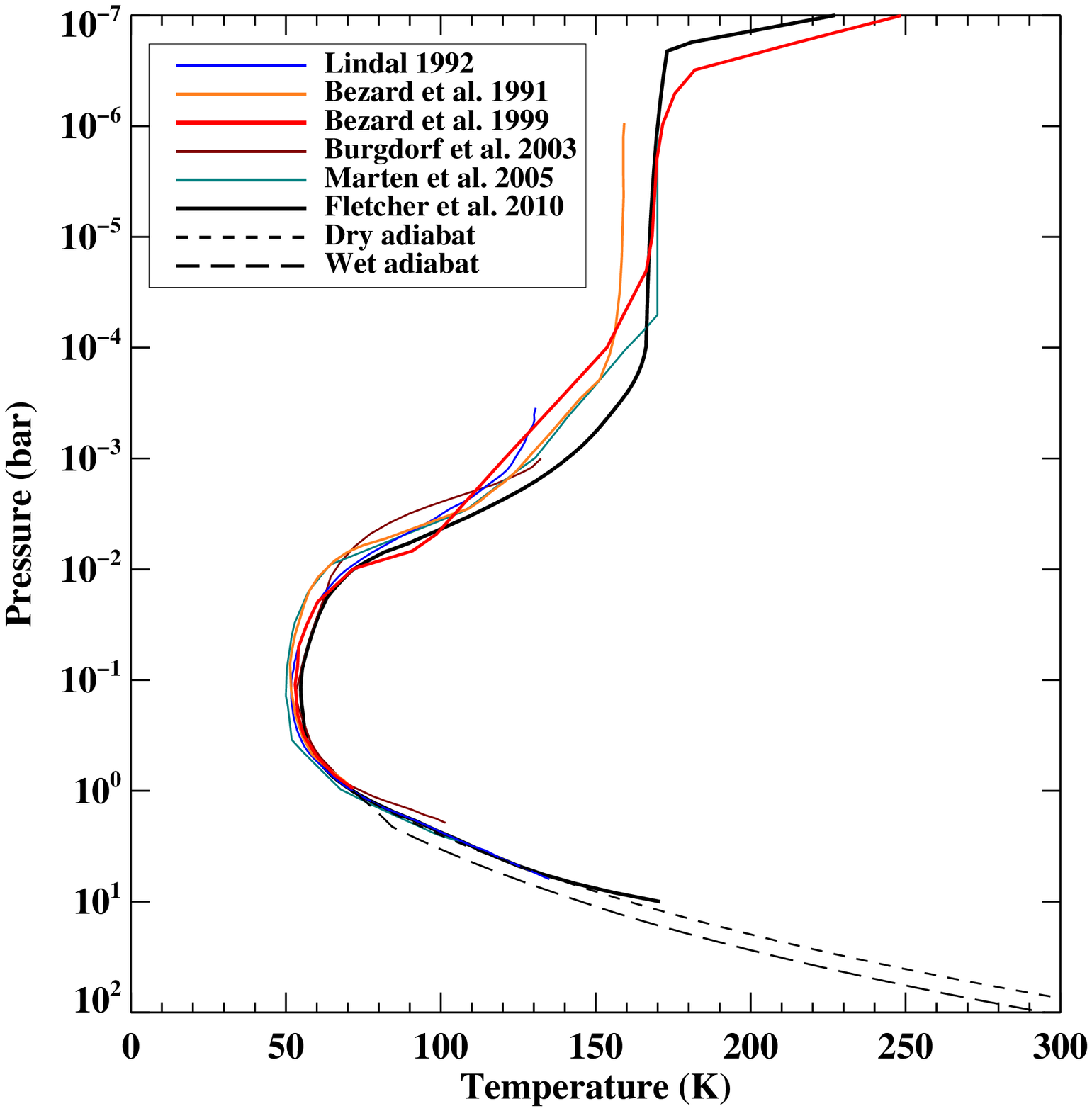}
$
\end{center}
\caption[Thermal profiles for Neptune]{\label{fig:tpprof} \small Selected temperature-pressure (TP) profiles for Neptune from the literature. The \cite{marten05} thermal profile (teal) was derived from their submillimeter data concurrently with CO abundance. \cite{lellouch05} adopted the \cite{bezard91} TP profile (orange) in their study; the thermal profile used by \cite{hesman07} was based on the \cite{burgdorf03} profile (brown). Our nominal choice for this work is the \cite{fletcher10} profile (thick black line), and we also consider the \cite{bezard99} profile (thick red line). At pressures greater than 1 bar, we follow a dry adiabat (black, short-dashed line).}
\end{figure}

 \begin{figure}[h]
\begin{center}$
\includegraphics[width=.45\textwidth,clip,trim=.3in 0 0 0]{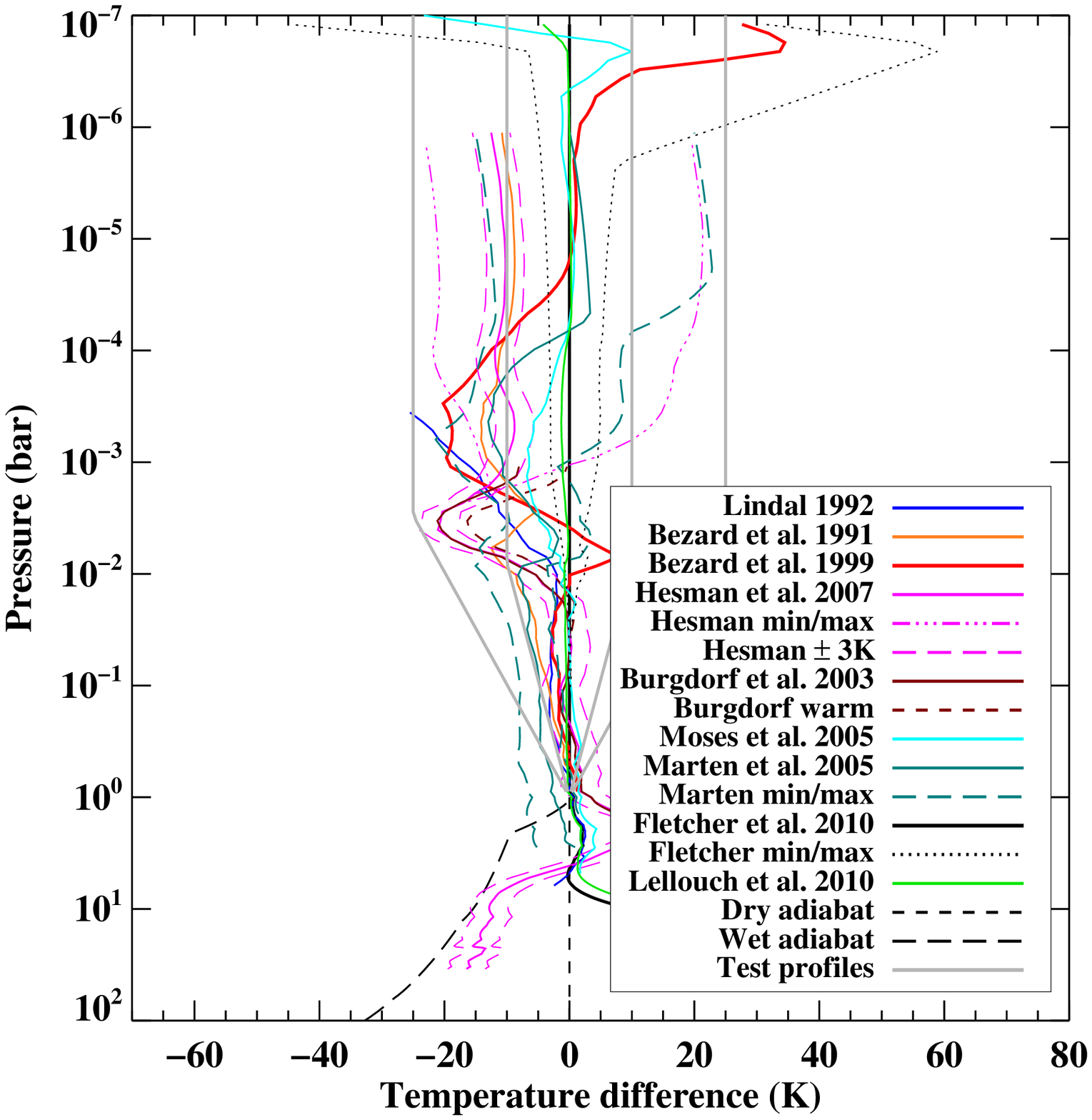}
$
\end{center}
\caption[Differences between thermal profiles and \cite{fletcher10}]{\label{fig:tpdiff}  \small Difference between selected TP profiles and our nominal TP profile. Line colors are the same as in Fig. \ref{fig:tpprof}; several additional thermal profiles are also shown. Published alternate profiles/uncertainties are included as broken lines. The light grey lines indicate our nominal test profiles, chosen to span the range of published profiles above 1 bar (see Section \ref{sec:model} for details). From left to right, these are the `extreme cool', `moderate cool', `moderate warm' and `extreme warm' TP profiles.}
\end{figure}

Figure \ref{fig:tpprof} shows a selection of published TP profiles, including our nominal profile and the profiles used in the recent CO studies of \cite{lellouch05}, \cite{marten05} and \cite{hesman07}. Published profiles generally match the Voyager RSS occultation profile \citep{lindal92} at $\sim$1 bar ($T$ = 71.5 K), and assume an adiabatic extrapolation down to deeper levels. Our nominal profile is slightly warmer than the \cite{lindal92} profile at the tropopause, which  could be due to seasonal changes in the atmospheric temperature since the Voyager era \citep{orton07,hammel07}. At altitudes above 1 bar, the published profiles diverge greatly. We consider four additional TP profiles that span the range of published profiles; each of these test profiles matches the nominal profile at altitudes below 1 bar.  The `extreme low' and `extreme high' profiles decrease/increase steadily to $\pm 25$ K from the nominal profile, respectively, between 1 and $0.003$ bar and then remain at $\pm 25$ K from nominal as the pressure continues to decrease. The `moderate low' and `moderate high' profiles decrease/increase steadily to $\pm 10$K from the nominal profile, respectively, between 1 and $0.008$ bar and then remain at $\pm 10$ K from nominal as the pressure continues to decrease. These offsets are illustrated in Fig. \ref{fig:tpdiff}. While these test profiles envelope the published profiles and their error bars, they do not represent all of the possible TP profile shapes. For example, the profile of \cite{bezard99} is $\sim$20 K cooler than our nominal profile around $10^{-3}$ bar, and warmer than our nominal profile at higher and lower pressures. The provisional temperature profile from the analysis of 2005 Spitzer Infrared Spectrometer data, reported in \cite{line08}, is similar to the \cite{bezard99} profile. We do not test all of the published TP profiles, but we do consider the \cite{bezard99} profile in addition to our nominal profile and four test profiles. The profiles used in the analyses of \cite{lellouch05} and \cite{hesman07} are most similar to our 'moderate cool' profile. 

At pressures greater than 1 bar, where the temperature profile is unknown, we assume the atmosphere is convective and extrapolate adiabatically. We consider two possibilities: a dry adiabat (nominal) and a wet adiabat (see \cite{depater91,depater05a} for a detailed description). In the adiabatic extrapolation, we assume that the specific heat of hydrogen is near that of normal hydrogen (though the opacity for equilibrium hydrogen is used- this is the ``intermediate'' hydrogen case \citep{massie82,depater93}).

\subsection{Disk averaging}\label{sec:disk}
To account for variations in viewing geometry on Neptune's disk-averaged spectrum, we calculate $T_B(\nu,\mu)$ for 25 different viewing angles $\mu$, which represent the average spectra within 25 concentric rings.  Doppler broadening due to the planet's rotation, which has a velocity of $V_{eq}= 2.7$ km s$^{-1}$ at Neptune's equator, affects the shape of the emission peak in the center 2 MHz of the line. Following \cite{moreno01}, the velocity at a given distance $x$ from the central meridian is:
\begin{eqnarray}
V = \frac{x}{R} V_{eq}
\end{eqnarray}
where 
\[
x= R \cos(lat)\sin(\Delta long)\cos(SEPL)
\]
We define \textit{R} as the planetary radius, \textit{lat} as planetary latitude, $\Delta long$ as longitude from the central meridian and $SEPL$ as the latitude of the sub-earth point. We divide the disk into 100 values of $x$. At each $x$, we calculate the average spectrum given the relative contribution of each representative viewing angle $\mu$. We then shift the spectrum in frequency given the value of $V(x)$. Finally we coadd the Doppler-shifted spectra to get our final disk-averaged spectrum. Once the disk-averaged spectrum is computed, models are convolved to the instrumental resolution ($\sim$1 MHz near the peak, $\sim$33 MHz in the wings). 


\section{Analysis}\label{sec:analysis}

To determine Neptune's vertical CO profile, we perform robust non-linear least-squares fitting to the data weighted by the data errors using the MPFIT IDL package \citep{markwardt09}. We first consider models in which the CO mole fraction  $n_{\CO}/n_\mathrm{TOT}$ is held constant throughout the atmosphere. This corresponds to the case where Neptune's observed CO primarily comes from vertical mixing from deep levels in the atmosphere. In addition to the value of $n_{\CO}/n_\mathrm{TOT}$, these fits have two additional free parameters. These are amplitude correction factors for the 1- and 3-mm data, which account for the uncertainty in the absolute flux calibration of the data. In practice, we find the values of these flux density correction parameters to be in the range of 5--10\% at both wavelengths. 

The least-squares fitting procedure is performed for models that use our nominal TP profile as well as each of the test thermal profiles described in Section \ref{sec:TP}. We also perform the fit on the 1-mm and 3-mm datasets separately. We report the fitted values for the CO mole fraction, 1- and 3-mm amplitude correction factors, and the statistical errors from the covariance matrices of the fits in Table \ref{tab:1lev}. We also report the reduced chi-squared ($\widehat{\chi^2}$) of the best fit: 

\begin{equation}
\widehat{\chi^2}=\frac{1}{M-N} \sum^{M-1}_{m=0} \frac{\delta y_m^2}{\sigma_{meas,m}^2}
\end{equation}
where M is the $\widehat{\chi^2}$number of data points, and N is the number of fit parameters (in this case, three), so that $M-N$ is the number of degrees of freedom (DOF) for the fit. The parameter $\delta y_m$ is the difference between the data and model value for point $m$, and $\sigma_{meas,m}$ is the measurement error for point $m$. Our values of $\widehat{\chi^2}$ are greater than one: this is discussed in detail in Section \ref{sec:errors}. 


\begin{table*}[htb!]
\footnotesize
\begin{center}
 \caption{}{ \small Best-fit one-level CO profiles. The CO mole fraction is held constant as a function of altitude (Section \ref{sec:results1lev}). The 1- and 3-mm flux density factors are the factors by which the data must be multiplied in the best-fit solution. These factors account for errors in the gain calibration of the data, which may be as much as 20\% (see Section \ref{sec:data}).  $\widehat{\chi^2}$ is the reduced $\chi^2$ of the best-fit model, as defined in Eq. (4). The thermal profile used in each model is specified. }
 
 \vspace{\baselineskip}
 
\begin{tabular}{ll  l p{2.8cm} p{2.8cm} l }
 \hline
Data&TP profile		&CO (ppm) &1-mm flux density factor & 3-mm flux density factor & $\widehat{\chi^2}$ \\
\hline
1 mm&&&&\\
&nominal&$0.50\pm0.02$ &$1.050\pm0.003$ & &15.3\\
3 mm&&&&\\
&nominal&$0.37\pm0.05$ & &$1.09\pm0.004$ &14.6\\

1 mm + 3 mm & & & &\\
&nominal				& $0.49 \pm 0.02$	&$ 1.052 \pm 0.003 $ & $1.081\pm 0.002$ &15.3\\

&\cite{bezard99}	& $ 0.45\pm 0.01 $	&$ 1.055\pm 0.003$ & $1.083\pm 0.002$ & 15.4 \\
&extreme cool		&  $0.264 \pm0.008 $	&$ 1.061\pm 0.003 $ & $1.090\pm 0.002 $ & 18.3\\
&moderate cool		&  $ 0.37\pm 0.01$	&$ 1.055 \pm0.003 $ & $1.085\pm 0.002$ &16.2 \\
&moderate warm	& $ 0.60\pm0.02$	&$ 1.056\pm 0.003 $ & $1.079\pm 0.002$ &16.5\\
&extreme warm		&  $ 0.67\pm 0.03 $	&$ 1.069\pm 0.004 $ & $1.080\pm 0.003$ & 21.5\\

\hline
\end{tabular}
\label{tab:1lev}

\end{center}
\end{table*}
\normalsize

\begin{figure}[]
\begin{center}$
\includegraphics[width=0.45\textwidth,clip,trim=.3in .4in .2in .2in]{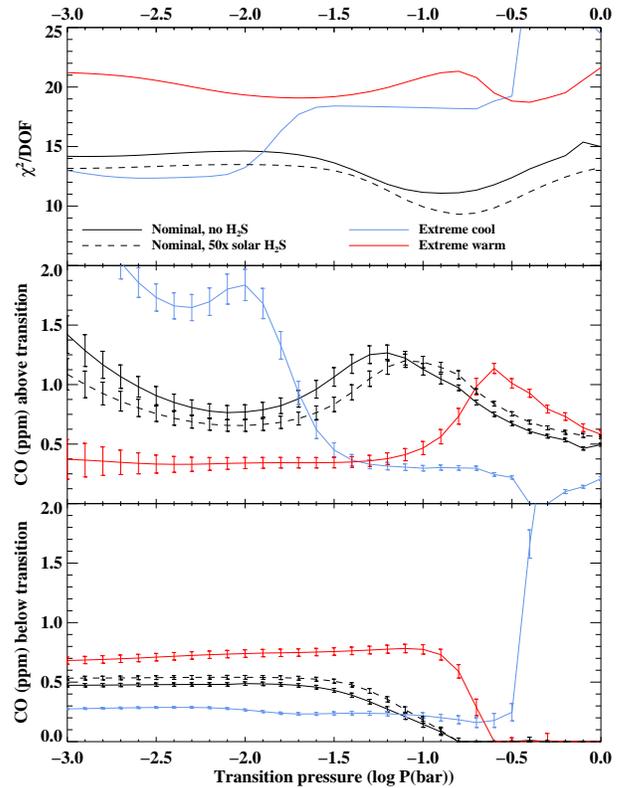}
$
\end{center}
\caption[Best-fit parameters as a function of transition pressure]{\label{fig:vsplev}  \small Best-fit parameters as a function of transition pressure level for the nominal thermal profile (black), both with (dashed) and without (solid) H$_2$S opacity; and for the extreme warm (red) and extreme cool (blue) thermal profiles. Pressure is given in units of $\log(P(\mathrm{bar}))$, from $\log(P(\mathrm{bar})) = -3.0$ (1 mbar) to $\log(P(\mathrm{bar}))=0.0$ (1 bar). 
}
\end{figure}

Following previous authors \citep{lellouch05,hesman07} we repeat the fitting process using a two-level CO profile. In this case the CO abundance is assumed to be constant within each of two levels, above and below some transition pressure. Fits of this form are representative of the scenario in which CO is produced in the stratosphere and diffuses downward, along with being mixed upward from the interior. In addition to the CO mole fractions and the 1- and 3- mm amplitude correction factors, the pressure that defines the transition between the two atmospheric levels must also be determined. We find that our fitting procedure does not perform well when we allow this transition pressure to be a free parameter; the fitting program will rarely vary the pressure level from its start value. Therefore, for each model thermal profile, we perform a series of fits: in each run we fix the pressure level, testing transition pressures in steps of 0.1 in $\log{P(\mathrm{bar})}$ from -3.0 (1 mbar) to 0.0 (1 bar). The best-fit solutions as a function of transition pressure are shown in Fig. \ref{fig:vsplev} for a selection of thermal profiles. We find that for a given TP profile, the $\widehat{\chi^2}$ value has a global minimum, and all the parameters are smooth functions of the $\log{P(\mathrm{bar})}$ of the transition pressure. We therefore take the solution with the lowest $\widehat{\chi^2}$ value as our overall best-fit model for each test TP profile; for example, for our nominal profile (no H$_2$S), the minimum $\widehat{\chi^2}$ reveals a transition pressure of $\log{P(\mathrm{bar})}=0.9$, or 0.13 bar (Fig. \ref{fig:vsplev}, top panel). This corresponds to a CO abundance of 1.1 ppm at altitudes above 0.13 bar (middle panel) and 0.1 ppm at altitudes below 0.13 bar (lower panel).  These best-fit values are reported in Table \ref{tab:2lev}. As when modeling a constant CO mole fraction, we repeat the two-level fit for  the 1- and 3-mm datasets separately.  We also test the effect of  H$_2$S absorption on the best-fit two-level CO profile, by repeating the two-level fits assuming a 10 and 50 times solar abundance of H$_2$S. These results are reported in Table \ref{tab:2lev} as well.

\begin{table*}[]
\begin{center}
 \scriptsize
\setlength{\tabcolsep}{0.05in}
 \caption{}{\small Best-fit two-level CO profiles. The CO mole fraction transitions between a high altitude value (``CO above") and a deep atmosphere value (``CO below") at the best-fit transition pressure $P$ which is given in units of $\log{P(\mathrm{bar})}$ as well as in bars (Section \ref{sec:results2lev}-\ref{sec:resultsh2s}).  As in Table \ref{tab:1lev}, the 1- and 3-mm flux density factors are the factors by which the data must be multiplied to match the model in the best-fit solution, and $\widehat{\chi^2}$ is the reduced $\chi^2$ of the best-fit model, as defined in Eq. (4). \label{tab:2lev}} 
 
 \vspace{\baselineskip}
 
\begin{tabularx}{1\linewidth}{l l l l l l p{2.3cm} p{2.3cm} l }

\hline
Data &TP profile				& CO below (ppm)	& CO above (ppm) & $\log{P(\mathrm{bar})}$& $P$(bar)& 1-mm flux density factor & 3-mm flux density factor & $\widehat{\chi^2}$ \\
\hline
1 mm& & & & &\\
&nominal				& $0.09 \pm 0.03$	& $1.12 \pm .04$& -0.9& 0.13& $1.079 \pm 0.003$ & &10.7\\
3 mm& & & & &\\
&nominal				& $0.00 \pm 0.01$	& $0.89 \pm .08$& -1.0& 0.100 & & $1.100 \pm .0002$ &9.3\\
all & & & &\\
&nominal			& $0.08 \pm 0.03$	& $1.06 \pm0 .04$ & -0.9&0.13&$1.082 \pm 0.003$ & $1.088 \pm 0.002$ &11.1 \\
&\cite{bezard99}	& $0.12 \pm 0.03$	& $1.05 \pm 0.05 $& -1.0&0.10&$1.080 \pm 0.003 $& $1.088 \pm 0.002$& 11.4 \\
&extreme cool		& $0.285\pm 0.007$& $1.9 \pm 0.1 $     & -2.6&0.0025  &$1.062 \pm 0.003$ & $1.089 \pm 0.002$ & 12.3  \\
&moderate cool	& $0.30 \pm 0.01$	&$1.37 \pm 0.09$ & -1.6&  0.025       & $1.066 \pm 0.003 $& $1.085 \pm 0.002$& 12.7 \\
&moderate warm	&$0.001\pm 0.03$	&$1.13 \pm 0.04$ & -0.7& 0.20          &$1.094 \pm 0.003 $& $1.090 \pm 0.002$&12.3 \\
&extreme  warm	& $0.00 \pm 0$	& $0.93\pm 0.03$          & -0.4&   0.40        &$1.100 \pm 0.003$ & $1.092\pm 0.002$& 18.7 \\
\hline
+10$\times$ solar H$_2$S\footnote{Opacity due to  10$\times$ solar enrichment in H$_2$S included } &nominal&$0.00\pm 0$&$1.08\pm0.02$&-0.8& 0.16   &$1.036\pm0.002$&$0.904\pm0.001$&9.3\\
+50$\times$ solar H$_2$S\footnote{Opacity due to  50$\times$ solar enrichment in H$_2$S included } &nominal&$0.00\pm 0$&$1.08\pm0.02$&-0.8 &  0.16&$1.039\pm0.002$&$0.909\pm0.001$&9.3\\

\hline
\end{tabularx}
\end{center}
\end{table*}
\normalsize

 \begin{figure}[]
\begin{center}$
\includegraphics[width=0.45\textwidth,clip,trim=.4in .2in .3in .4in]{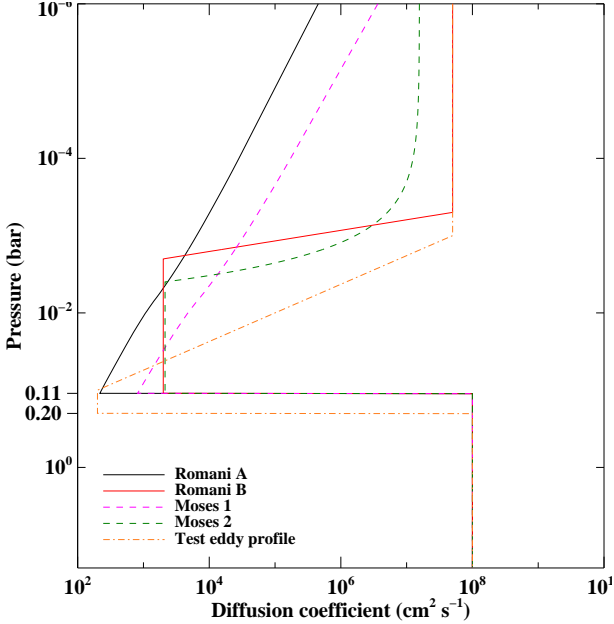}
$
\end{center}
\caption[Eddy diffusion profiles]{\label{fig:eddy}\small Eddy diffusion coefficient profiles from \cite{moses92} and \cite{romani93}. Also included is one of our `test profiles', designed to produce a CO profile more similar to our best-fit two-level CO profile (see Section \ref{sec:resultsphys} for discussion). The tropopause pressure values of  0.11/0.20 bar are indicated. }
\end{figure}

\begin{figure}[]
\begin{center}$
\includegraphics[width=0.45\textwidth,clip,trim=.4in .2in .3in 0in]{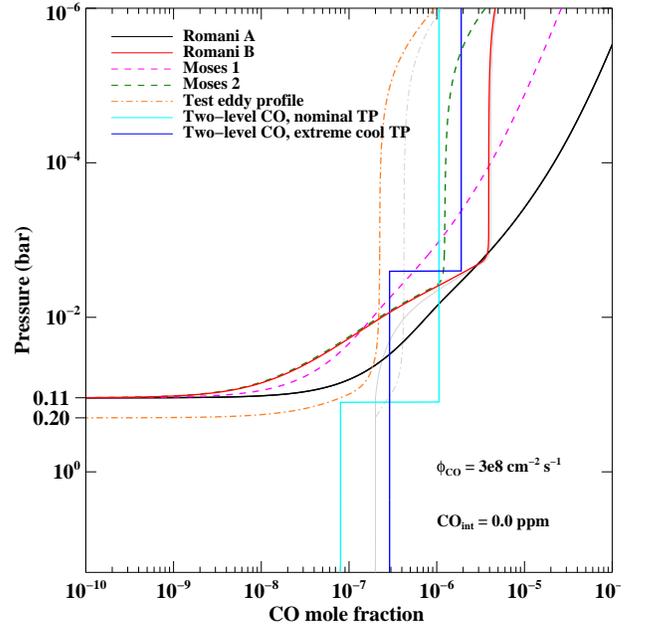}
$
\end{center}
\caption[Sample physical CO profiles]{\label{fig:cophys}\small Sample CO profiles that result from eddy profiles in Fig. \ref{fig:eddy}, assuming $\phi_\CO=3\times 10^8$ cm$^{-2}$s$^{-1}$ and no internal contribution to the CO.  To illustrate the effect of internal CO on the CO profile, we also show the CO profile for the Romani B eddy profile (solid red line) with the addition of 0.2 ppm of internal CO (solid grey line); and the `test eddy profile' (orange dot-dashed line) with 0.2 ppm of internal CO included (grey dot-dashed line).  The best-fit parameters for physical CO profiles depend on the TP profile in addition to the eddy profile; fits are summarized in Table \ref{tab:physco}. For reference, the best-fit two-level profiles from our nominal (no H$_2$S, cyan line) and extreme cool (blue line) TP profile fits are shown.}
\end{figure}

Finally, we use a model to derive several `physical' CO profiles. To model the vertical diffusion rate,  the eddy diffusion coefficient ($K$) profiles of \cite{moses92} and \cite{romani93}, which were developed to match photochemical models of observed hydrocarbon distributions, are used (Fig. \ref{fig:eddy}). Molecular diffusion is included using \cite{marrero72}:
\begin{multline}
D_{CO} =\\  \frac{1}{P(atm)} \frac{1.539\times 10^{-2} T^{1.548}}{\left(\ln{[T/3.16\times10^7]}\right)^2e^{-2.8/T}e^{1067/T^2}} \ \ \ \cm^2\ \s^{-1}
\end{multline}
 Using these profiles, we solve the diffusion equation:
 \begin{multline}
 \phi_\CO= -K\left(\frac{\partial n_\CO}{\partial z} + n_{\CO}\left[ \frac{1}{H} +\frac{1}{T}\frac{\partial T}{\partial z}\right]\right) \\
 -D_\CO\left(\frac{\partial n_\CO}{\partial z} + n_{\CO}\left[ \frac{1}{H_\CO} +\frac{1}{T}\frac{\partial T}{\partial z}\right]\right)
 \end{multline}
where $n_\CO$ is the number density of CO molecules in cm$^{-3}$, $z$ is the altitude in cm, $H$ is the total atmospheric pressure scale height, and $H_\CO$ is the scale height of CO.  We then find the best-fit value for the external flux $\phi_{\CO}$ from the upper atmosphere, allowing for an additional CO contribution from the planet's interior ($n_{\CO}/n_{\mathrm{TOT}}|_{(z=0)}$).  Sample physical CO profiles, assuming an influx rate of $\phi_\CO= 10^8\ \cm^{-2}\s^{-1}$  and no internal contribution of CO, are illustrated in Fig. \ref{fig:cophys}. To demonstrate the effect of internal CO on these physical CO profiles, we show the same profiles with 0.2 ppm of internal CO added for two of the eddy profile cases.  Best-fit values of  $\phi_\CO$ and  $n_\CO/n_{\mathrm{TOT}}|_{(z=0)}$ from our least-squares modeling are summarized in Table \ref{tab:physco}.

\begin{table*}[]
 \begin{minipage}{1.0\textwidth}
\begin{center}
 \small
 \setlength{\tabcolsep}{0.05in}

 \caption{} {\small Best-fit physical CO profiles. The external flux $\phi_\CO$ and the mole fraction of CO brought up from the deep atmosphere (``internal CO") that produce the best match between the spectrum and model are given, for a selection of thermal and eddy diffusion coefficient profiles (see Section \ref{sec:resultsphys}). $\widehat{\chi^2}$ is the reduced $\chi^2$ of the best-fit model, as defined in Eq. (4). The flux density scale factors for these fits are not shown. }

\vspace{\baselineskip}

\begin{tabular}{l  l l l l l l}

\hline
TP profile					& eddy diffusion profile	&$\log\phi_\CO($cm$^{-2}$s$^{-1})$ & internal CO (ppm)& $\widehat{\chi^2}$ \\
\hline
nominal			& Romani `A' 		&$7.37\pm0.07$ &$0.48\pm0.01$	&14.2	\\
moderate low		& Romani `A' 		&$7.71\pm0.05$ &$0.37\pm0.01 $	&13.2\\
extreme low		& Romani `A' 		&$8.07\pm0.04$ &$0.260\pm0.007 $&13.0	\\
\hline
nominal			& Romani `B' 		&$7.55\pm 0.08$	&$0.48\pm0.01$&14.4\\
moderate cool		& Romani `B' 		&$7.93\pm0.04$ & $0.38\pm0.01$	&13.0\\
extreme cool		& Romani `B' 		&$8.33\pm0.03$ & $0.279\pm0.007 $ & 12.3\\
\hline
nominal			& Moses `1' 		&$7.94\pm0.07$ &$0.48\pm0.01$	&14.2	\\
moderate cool		& Moses `1' 		&$8.27\pm0.05 $ &$0.37\pm0.01$	&13.1\\
extreme cool		& Moses `1' 		&$8.60\pm0.04 $ &$0.254\pm0.007 $&12.9	\\
\hline
nominal			& Moses `2' 		&$7.93\pm0.08$ &$0.48\pm0.01$& 14.5		\\
moderate cool 		& Moses `2' 		&$8.38\pm 0.04$ &$0.37\pm0.01 $	& 13.0 \\
extreme cool		& Moses `2' 		&$8.85\pm 0.03$ &$0.259\pm 0.008$& 12.2	\\
\hline
nominal					& test profile\footnote{see description in Section \ref{sec:resultsphys}}; p$_{trop}$ = 0.11 bar&$9.33\pm0.04$ &$0.26\pm0.02$	&11.8	\\
nominal			& test profile$^a$; p$_{trop}$ = 0.2 bar&$9.17\pm0.03$ &$0.15\pm0.03$	&11.2	\\
moderate cool		& test profile$^a$; p$_{trop}$ = 0.2  bar&$9.06\pm0.04 $	&$0.12\pm0.03 $&13.5\\
extreme cool		& test profile$^a$; p$_{trop}$ = 0.2 bar&$8.4\pm0.2$	&$0.21\pm0.03 $&18.2	\\
\hline
\end{tabular}
\label{tab:physco}
\end{center}
\end{minipage}
\end{table*}
\normalsize

\section{Errors and uncertainty}\label{sec:errors}
As described in Section \ref{sec:data}, the least-squares fitting utilizes the residual map rms scatter to weight the data for calculating  $\widehat{\chi^2}$. This weighting scheme compensates for the difference in signal-to-noise between the narrow ($\sim1$ MHz) and wide ($\sim33$ MHz) frequency channels, which we expect to differ by a factor of $\sim$6. Noisy channels that are not flagged during data reduction are down-weighted.  Day-to-day variations due to weather or array performance, and differences in the noise between 1 and 3 mm, are also accounted for.  However, we expect that these weights underestimate the total data uncertainties. While the overall calibration offset is a free parameter in the least-squares fits, additional systematic errors in the flux densities, for example due to errors in the bandpass calibration, could affect the relative flux between frequency windows. A visual comparison of individual 1-mm tracks (see Fig. \ref{fig:coraw} ) shows small day-to-day offsets in the average flux density in a given frequency window, for example at 230 GHz and 235 GHz, that are likely due to this type of systematic error. 

Additionally, gain calibration offsets between each of our 1-mm tracks could also cause systematic, frequency-dependent errors in the final spectrum. Since we expect a residual uncertainty of 3\% in the relative gain calibration between our 1-mm datasets, we tested the effect of such errors using a Monte Carlo method. We assume 3\% random errors in the relative gains between data sets and perform 300 trials of our constant-value $n_{\CO}/n_\mathrm{TOT}$ fit. We find that the fit solution is stable to the inclusion of these errors, but the statistical uncertainty of the best-fit CO abundance increases by a factor of 2.  While this experiment is too time intensive to repeat for the other model cases, this result suggests that 1) in general we are underestimating the errors in the fits by ignoring the error in the relative gains between 1-mm tracks, and 2) at least in this case, including the errors in the relative gains between 1-mm tracks leads to an increase in the uncertainties, but does not change the best-fit solution.

 Because of the systematic errors, we consider our values of $\sigma_{meas,m}$ to be weights rather than true estimates of the uncertainties in our data points. This is reflected in the values of $\widehat{\chi^2}$ for our fits, which we would expect to be 1.0 for a model that fully captures the data \textit{if} the values of  $\sigma_{meas,m}$ were reflective of the true data uncertainties. A $\widehat{\chi^2}$ of 11.1 for our best-fit, two-level model implies that if this model were a perfect representation of the data, then the data weights $\sigma_{meas,m}$ underestimate the total data uncertainty by a factor of 3.3. However, the high value of $\widehat{\chi^2}$ is likely due to deficiencies in the model in addition to an underestimate in the data errors. Factors that contribute to this include errors in the continuum opacity: either in the mixing ratios of the continuum species or in the coefficients/line profiles used to derive continuum absorption. Our tests with H$_2$S, a species which we observe to affect our millimeter spectrum but whose abundance is poorly constrained, provide one example of this. Errors in the shape of the TP profile and CO profile can also contribute, since we only investigate a finite number of TP  and CO profile shapes. 

In summary, the $\widehat{\chi^2}$ parameter does not reach the theoretical expectation of $\widehat{\chi^2} \approx 1$ for a perfect fit because 1) while the values of $\sigma_{meas,m}$ adopted from the rms scatter in the maps are good estimates of the relative dispersions of the data points,  they do \textit{not} characterize the total intrinsic uncertainties; and 2) the models (even our best models) are not perfect. Regardless, $\widehat{\chi^2}$ is a useful measure of the variance of the data point residuals ($\delta y_m^2$, see above) for our different fits. We encourage the reader to use both the reported $\widehat{\chi^2}$ values and the plots when evaluating the success of various models in reproducing the data.

\begin{figure*}[t!]
\begin{center}$
\begin{array}{cc}
\includegraphics[width=0.6\textwidth,clip,trim=0.5in 0 0 0]{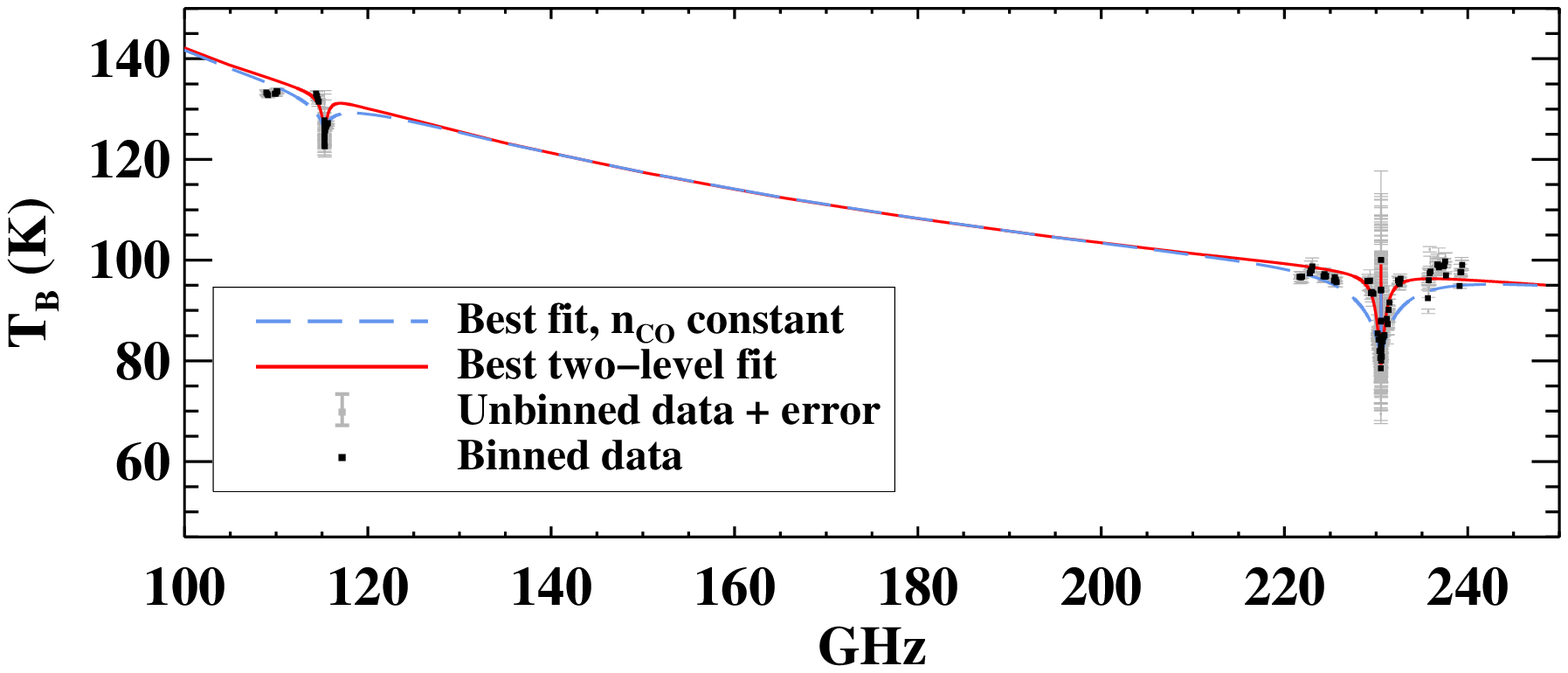}\\
\includegraphics[width=0.6\textwidth,clip,trim=0.5in 0 0 0]{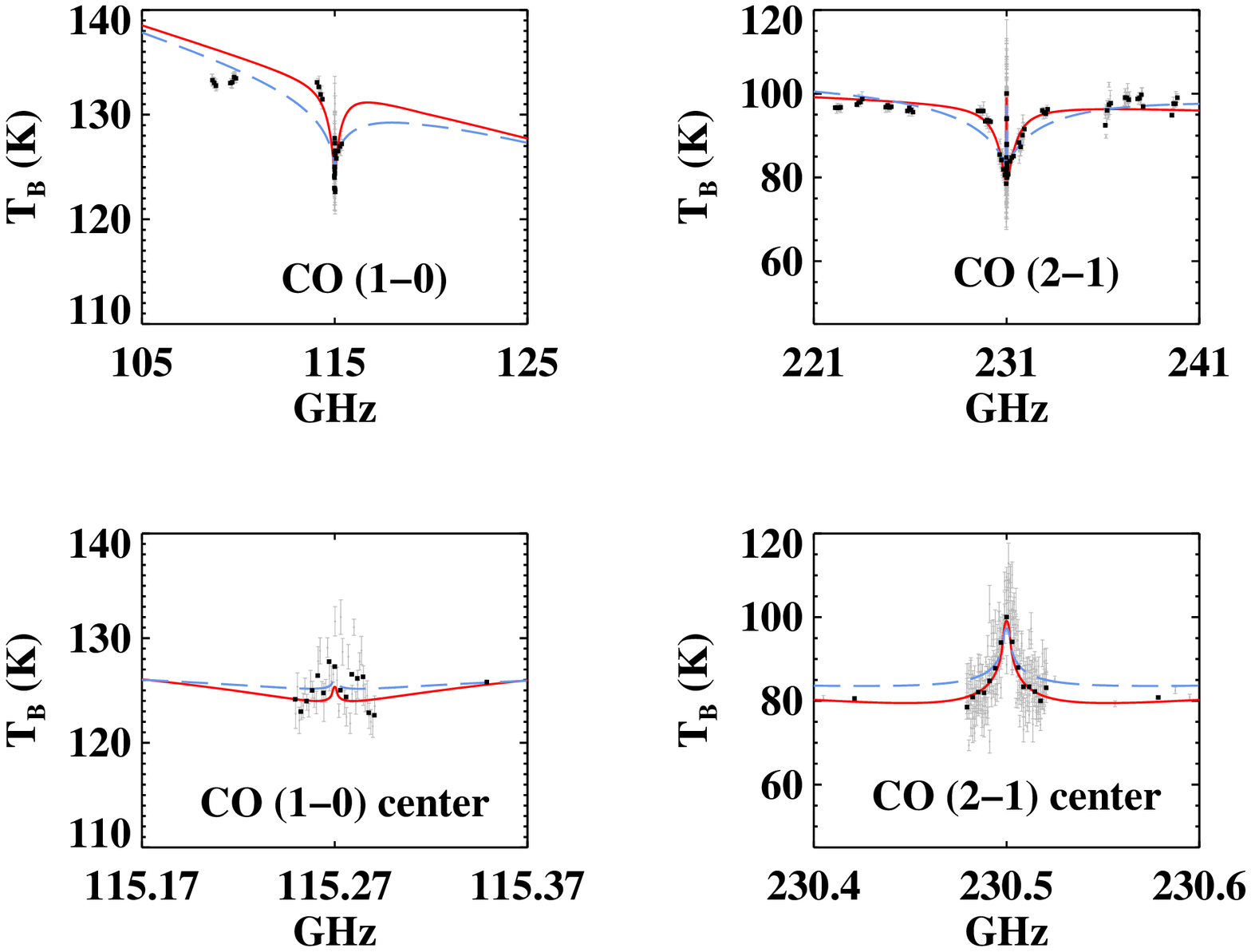}
\end{array}$
\end{center}
\caption[One- and two-level CO profile data fits]{\label{fig:co1vs2lay}  \small Models of the millimeter spectrum corresponding to the best-fit constant (0.5 ppm) CO profile in blue; and two-level (1.1 ppm above the 0.13 bar level, 0.1 ppm deeper than 0.13 bar) CO profile in red.  H$_2$S absorption is omitted here. For plotting purposes, the 1- and 3-mm data were scaled up by 8.2\% and 8.8\% respectively, to match nominal TP, best-fit two-level model; likewise, subsequent models are all scaled by the relative flux correction factors to permit comparison. This plot shows that a two-level profile fits the data significantly better than a one-level profile. }
\end{figure*}
\section{Results}\label{sec:results}

The results of our model fits, which are summarized in Tables 3-6, are illustrated in Figs. 10-21 and discussed below. Plots of the CO spectra show both the unbinned single-channel data with uncertainties $\sigma_{meas,m}$, and the data binned to 3 times the channel width; 1- and 3-mm flux densities are typically scaled by the amplitude correction factors from the nominal TP profile best two-level fit (not including H$_2$S), before being converted to brightness temperature. For ease of comparison, additional model spectra are also scaled to align with this nominal two-level model.
 
\subsection{Constant CO models}\label{sec:results1lev}

Our best-fit model for a constant CO distribution uses the nominal TP profile and a CO mole fraction of 0.5 ppm. The best-fit CO abundance decreases for cooler TP profiles, and increases for warmer TP profiles, within a range of $0.3-0.7$ ppm (Table \ref{tab:1lev}). Figure \ref{fig:co1vs2lay} illustrates that a constant CO profile that has the appropriate line depth is generally too broad in the wings. The emission peak at line center is also a poor match to the data.

\subsection{Two-level CO models, no H$_2$S}\label{sec:results2lev}

\begin {table*}[]
\begin{center}
\footnotesize
\setlength{\tabcolsep}{0.05in}
 \caption{}{\label{tab:plev} \small Best-fit two-level CO profiles for a series of transition pressure levels, for models using our nominal thermal profile \citep{fletcher10}. The CO mole fraction transitions between a high-altitude mole fraction (``CO above") and a deep-atmosphere mole fraction (``CO below") at a transition pressure $P$ which is given in units of $\log{P(\mathrm{bar})}$ as well as in bars (Section \ref{sec:results2lev}). Transition pressures in the range of $\log{P(\mathrm{bar})}=-3.0$ to 0.0 were tested with the model, and a subset of these tests is shown here. As in Table \ref{tab:1lev}, the 1- and 3-mm flux density factors are the factors by which the data must be multiplied to match the model in the best-fit solution, and $\widehat{\chi^2}$ is the reduced $\chi^2$ of the best-fit model, as defined in Eq. (4).}
 
 \vspace{\baselineskip}
 
\begin{tabular}{ll ll l p{2.8cm} p{2.8cm}}

 \hline
 $\log P($bar$)$& $P$(bar)	&$\widehat{\chi^2}$ & CO below (ppm) & CO above (ppm) & 1-mm flux density factor 	& 3-mm flux density factor\\
 \hline
    -1.2&0.063&	11.8& 	$0.27  \pm 0.02 $&	$1.27 \pm 0.07$& $1.074 \pm 0.003$&	$1.084 \pm 0.002$\\
    -1.1& 0.079&    11.4&	$0.21  \pm 0.02 $&	$1.22 \pm 0.06$& $1.077 \pm 0.003$&     $1.085\pm 0.002$\\
    -1.0&0.10&	11.2&	$0.15  \pm 0.03 $&	$1.15 \pm 0.05$& $1.079 \pm 0.003$&     $1.087\pm 0.002$\\
\bf{-0.9}&\bf{0.13}&     \bf{11.1}&	\boldmath{$0.08\pm0.03$}&	\boldmath{$1.07 \pm 0.04$}&	\boldmath{$1.082 \pm 0.003$}& \boldmath{$1.088\pm 0.002$}\\
    -0.8& 0.16&    11.1&	$0.01  \pm0.02  $&	$0.98 \pm 0.03$& $1.084\pm 0.003$& $1.090\pm 0.002$\\
    -0.7&  0.20&   11.3&	$0.000\pm0.003$&	$0.86 \pm 0.02$& $1.082\pm 0.002$& $1.090\pm 0.002$\\
    -0.6& 0.25&	11.8&	$0.004\pm0.005$&	$0.75 \pm 0.02$& $1.079\pm 0.002$& $1.090\pm 0.002$\\
\hline
\end{tabular}
\end{center}
\end{table*}
\normalsize

\begin{figure}[]
\begin{center}$
\includegraphics[width=0.45\textwidth]{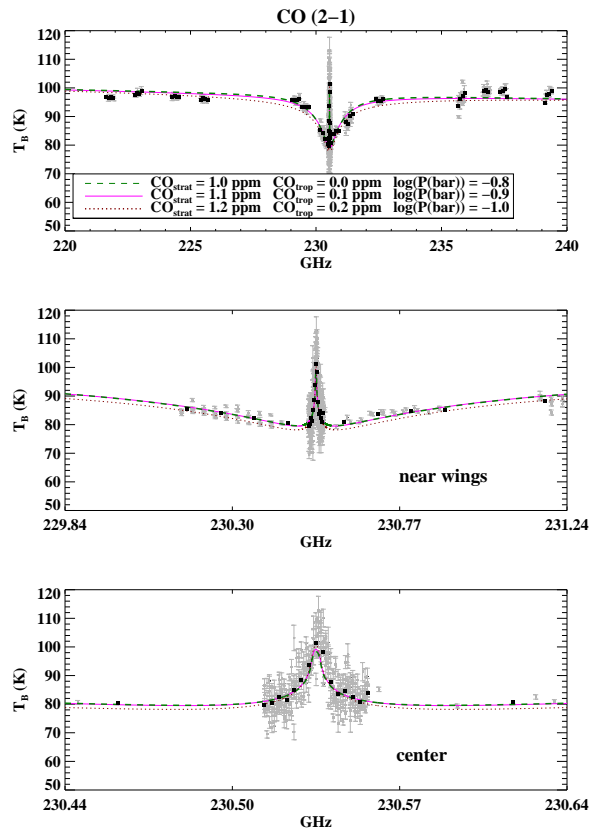}
$
\end{center}
\caption[Spectral models for varying CO profile transition pressures]{\label{fig:testplev}  \small Spectral models for transition pressures of $\log(P(\mathrm{bar}))= -0.8, -0.9$ and $-1.0$ (0.16, 0.13, and 0.10  bar, respectively), and our nominal TP profile, shown for the CO (2--1) line (see Table \ref{tab:plev}). Solutions for $\log{P(bar)}=-0.8$ (green dashed line) and  $-0.9$ (solid magenta line) are somewhat more favorable in the far wings (2-3 GHz from line center), but all three models match the data quite well. }
\end{figure}

Model spectra produced with our two-level CO profiles offer an improved fit to the data; this is also illustrated in Figure \ref{fig:co1vs2lay}. Overall, our nominal TP profile allows the best fit to the data, in a $\widehat{\chi^2}$ sense, with the \cite{bezard99} TP profile solution a close second. The best two-level model solution has a $\widehat{\chi^2}$  of 11.1, with 1.1 ppm of CO in the upper atmosphere and 0.1 ppm CO in the lower atmosphere. The transition pressure for this model is located near the tropopause, at $\log{P(bar)} = -0.9$ ($P = 0.13$ bar). The best-fit parameters for several transition pressures near the $\widehat{\chi^2}$ minimum value of $\log{P(bar)} = -0.9$ ($P = 0.13$ bar) are presented in Table \ref{tab:plev}; the model CO (2--1) spectrum for three of these solutions is plotted against the scaled 1-mm data in Fig. \ref{fig:testplev}. The good match of all of these spectra to the data implies that there is a range of transition pressure levels that produce quality fits to the data.  This is also clear from the best-fit $\widehat{\chi^2}$ values; for example, the fit solution at $\log{P(bar)} =-0.8$ ($P = 0.16$ bar) also has a $\widehat{\chi^2}$ of 11.1, with a slightly lower best-fit CO abundance of 1.0 ppm in the upper atmosphere and 0.0 ppm CO in the lower atmosphere.  Therefore, the best-fit answer alone does not fully characterize the range of solutions allowed by our data. We also find that models using our nominal thermal profile produce solutions with lower values of $\widehat{\chi^2}$ than the best solutions from models using other thermal profiles, for a range of transition pressures (see Tables \ref{tab:2lev} and \ref{tab:plev}). These solutions have CO mole fractions of 0.0-0.3 ppm in the troposphere and 0.8-1.3 ppm in the stratosphere, with level transitions between 0.06 and 0.25 bar. This very shallow minimum in $\widehat{\chi^2}$ as a function of transition pressure is likely a result of the thermal profile being nearly isothermal in this region around the tropopause. In addition to the default assumption of a dry adiabat, we also test thermal profiles that used a wet adiabat to extrapolate to high pressures. We find that, in general, using a wet adiabat for the thermal profile does not alter the models significantly, and increases the values of $\widehat{\chi^2}$ slightly.

\begin{figure}
\begin{center}$
\includegraphics[width=0.45\textwidth]{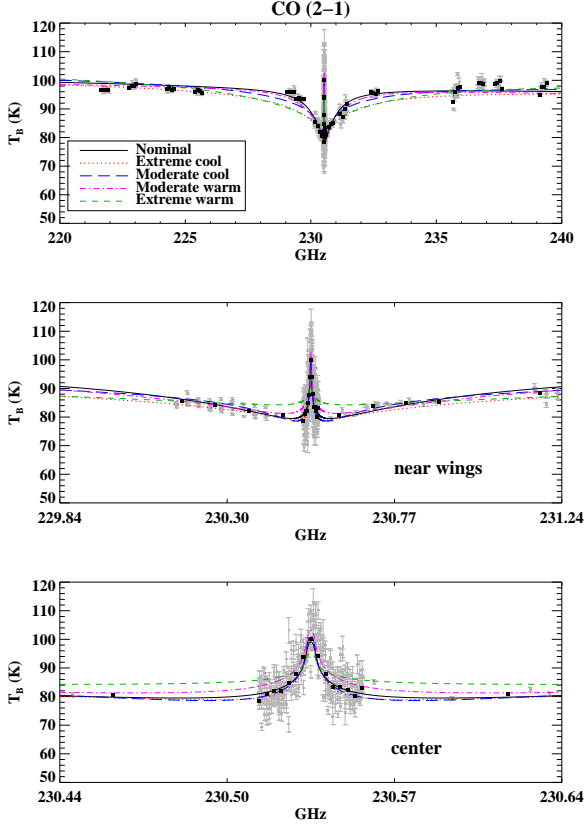}
$
\end{center}
\caption[Comparison of best-fit two-level models to the CO (2--1) data]{\label{fig:coalltp1mm}  \small Comparison of best-fit two-level models to the CO (2--1) data using different TP profiles.}
\end{figure}

\begin{figure}
\begin{center}$
\includegraphics[width=0.45\textwidth]{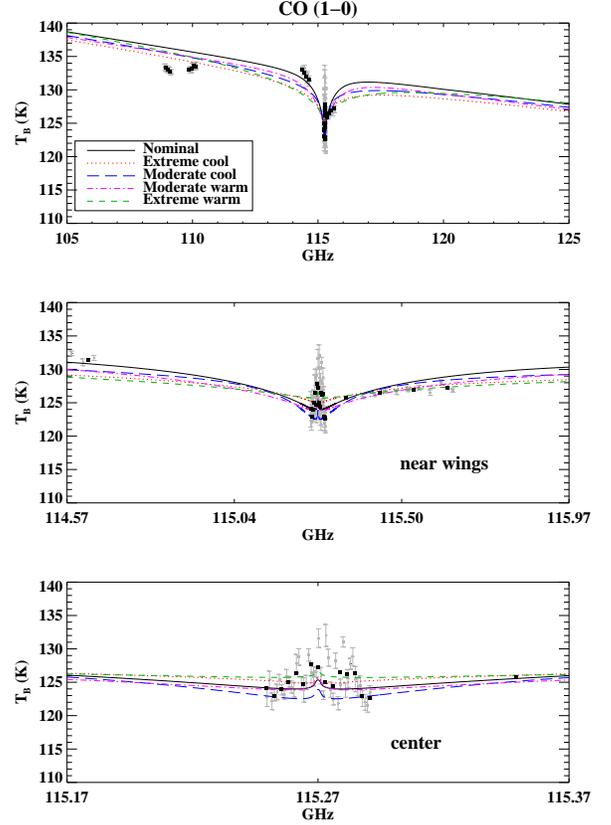}
$
\end{center}
\caption[Comparison of best-fit two-level models to the CO (1--0) data]{\label{fig:coalltp3mm} \small Comparison of best-fit two-level models to the CO (1--0) data using different TP profiles.}
\end{figure}

Of all of our test thermal profiles, models using the extreme warm profile agree least well with the data, in a $\widehat{\chi^2}$ sense.  This thermal profile is also the least consistent with the published TP profiles (see Fig. \ref{fig:tpdiff}). The best-fit transition pressure for this thermal profile is at  0.4 bar, which is below the tropopause, and therefore probably unphysical.   The best-fit moderate warm CO profile is qualitatively similar to the best model from the nominal TP profile, with 1.1 ppm of CO in the stratosphere and 0.0 ppm CO below the tropopause. Figures \ref{fig:coalltp1mm} and  \ref{fig:coalltp3mm} show that both warm TP profiles produce line profiles that are too high in the near wings of the line ($\leq 200$ MHz from center), particularly at 1 mm. The extreme warm profile is also too broad in the far wings of the CO line. 

Model fits using the cooler TP profiles favor two-level profiles with level transitions occurring well above the tropopause; the transition for the best-fit, moderate cool thermal profile case is at  25 mbar; and at   2.5 mbar for the extreme cool profile. Fits using cooler TP profiles  favor higher CO mole fractions in both levels of the atmosphere: the best-fit solutions give 0.3 ppm CO in the lower level of the atmosphere for both thermal profiles, and 1.4 and 1.9 ppm CO in the upper atmosphere for the moderate and extreme TP profiles, respectively. As Figs. \ref{fig:coalltp1mm} and  \ref{fig:coalltp3mm} show, models using the cooler TP profiles have the correct absorption depth and match the emission peak well;  but are too broad in the far wings, particularly at 1 mm.

Given these findings, we conclude that the data are best matched by the \cite{fletcher10} thermal profile, with $0.1^{+0.2}_{-0.1}$ ppm of CO in the troposphere and $1.1^{+0.2}_{-0.3}$ ppm of CO in the stratosphere, as indicated by the range of values in Table \ref{tab:plev}. Fits to the 1-mm and 3-mm lines separately, which are performed using the nominal TP profile, give consistent best-fit solutions. Visual inspection of all of the models generally indicates better agreement with the data at 1 mm; however since the 3-mm data have lower signal-to-noise, the $\widehat{\chi^2}$ value for the 3-mm data alone is actually lower than for the 1-mm data. Best-fit corrections to the flux density are 6-10\% at 1 mm and 9-10\% at 3 mm (Table \ref{tab:2lev}).

\subsection{Two-level models, H$_2$S included}\label{sec:resultsh2s}

\begin{figure}[]
\begin{center}$
\includegraphics[width=0.5\textwidth,clip,trim=.3in 0 0 0]{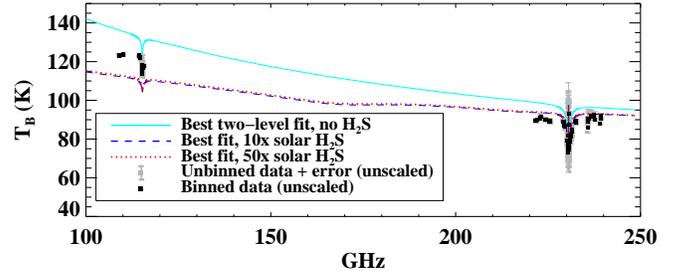}
$
\end{center}
\caption[Best-fit solutions, H$_2$S absorption]{\label{fig:coh2s} \small  Best-fit solutions for models containing no H$_2$S absorption (solid cyan), 10 times solar H$_2$S (blue dashed) and 50 times solar H$_2$S (red dotted). In this Figure, the data and models have \textit{not} been scaled by any flux density correction factor. Generally, all models are higher than our raw data at 1 mm. At 3 mm, the no-H$_2$S model is higher than the raw data, whereas the H$_2$S-enriched models are lower. The 10 and 50 times protosolar H$_2$S models match one another closely. }
\end{figure}

As discussed in Section \ref{sec:model}, we test the effect of including opacity due to a 10  and 50 times protosolar enrichment in H$_2$S. We find that the main effect of adding H$_2$S on the best-fit model parameters is to change the amplitude scaling factors for the 1- and 3-mm data. Most notably, without H$_2$S, the 3-mm data are lower than the model by $\sim 9\%$, whereas with 10$\times$ protosolar H$_2$S the data are higher than the model by $\sim 9\%$ (Fig. \ref{fig:coh2s}). Abundances of H$_2$S higher than 10$\times$ the protosolar value do not further change the millimeter spectrum, due to condensation into the H$_2$S ice cloud at pressures of a few bars \citep{depater91}.

\begin{figure}[]
\begin{center}$
\includegraphics[width=0.45\textwidth]{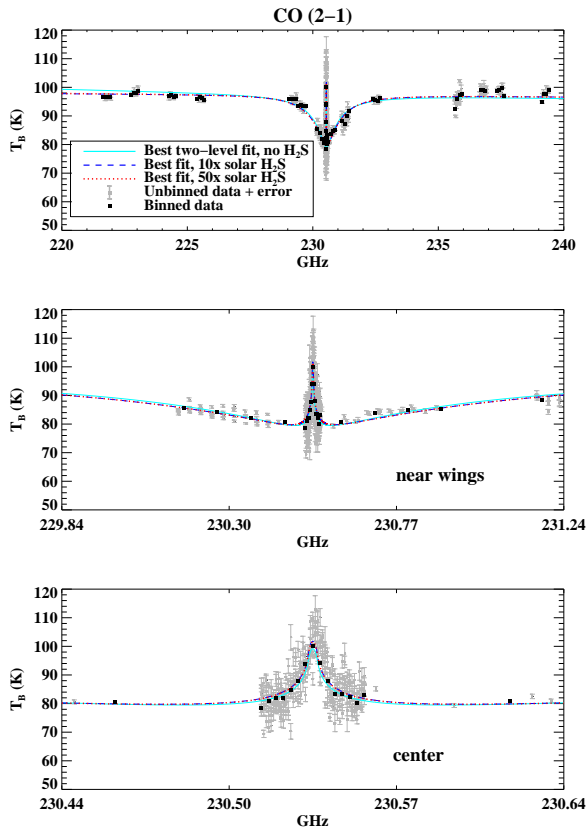}
$
\end{center}
\caption[Scaled H$_2$S-enriched model fits to the CO (2--1) data]{\label{fig:coh2s1mm} \small Model fits to the CO (2--1) line as in Fig. \ref{fig:coh2s}, except that the data and H$_2$S-enriched models are scaled to match the brightness temperature of the best-fit, no-H$_2$S model. }
\end{figure}

\begin{figure}[]
\begin{center}$
\includegraphics[width=0.45\textwidth]{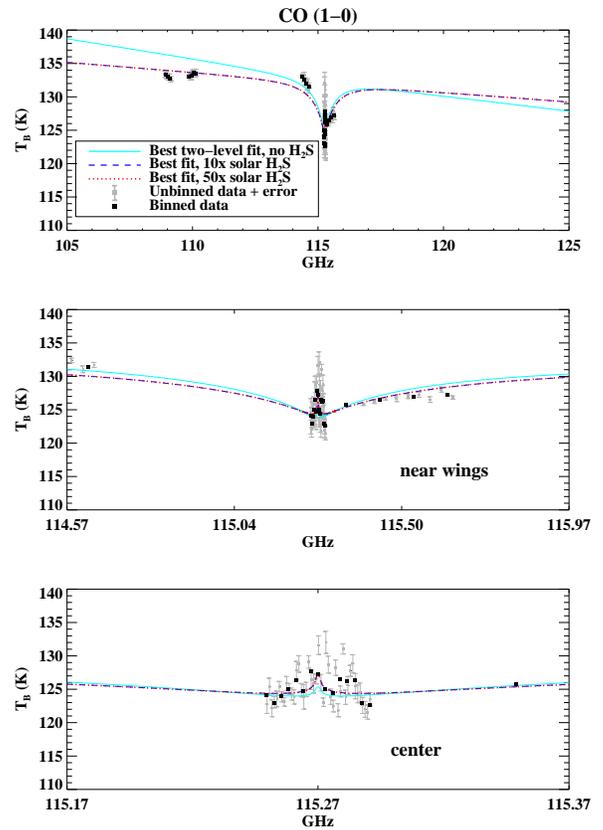}
$
\end{center}
\caption[Scaled H$_2$S-enriched model fits to the CO (1--0) data]{\label{fig:coh2s3mm}  \small Same as Fig. \ref{fig:coh2s1mm}, except for the CO (1--0) line.}
\end{figure}

Despite the significant effect of H$_2$S absorption on the 3-mm continuum, we find that the best-fit CO solution is not strongly affected by the inclusion of H$_2$S absorption: the best-fit CO vertical profile has 1.1 ppm CO in the stratosphere and 0.0 ppm of CO in the troposphere (Table \ref{tab:2lev}). The best-fit CO model spectra, scaled to match the best two-level fit for the no-H$_2$S case, are plotted with the (scaled) data in Figs. \ref{fig:coh2s1mm} and \ref{fig:coh2s3mm}. The  $\widehat{\chi^2}$ value for the fit improves with the addition of H$_2$S, which is primarily due to an improved fit to the shape of the far wings of the CO (1--0) line (Fig. \ref{fig:coh2s3mm}). There is also some improvement to the fit at the center of the CO (1--0) line. 

\subsection{Physical models}\label{sec:resultsphys}
 Physical CO profiles based on the diffusion models of \cite{moses92} and \cite{romani93} produce the best model spectra when cool thermal profiles are used.  The reason for this can be inferred from Fig. \ref{fig:cophys}: for all of these diffusion models, the mole fraction of CO in the atmosphere from an external source falls off at pressures less than $\sim$10 mbar, which is much higher in altitude than the transition level for the best-fit two-level model using our nominal thermal profile, but is consistent with the transition levels for the best-fit moderate and extreme cool TP profile models. The internal CO contribution for these fits is roughly independent of the eddy diffusion profile, with mole fractions of 0.3 and 0.4 ppm for the extreme and moderate cool thermal profiles, respectively. These values are very similar to the CO abundances found for the two-level fits using the same thermal profiles. The external CO flux is dependent on the choices for the eddy diffusion coefficient and thermal profiles: values range from $\phi_\CO=0.5-7\times10^8$ cm$^{-2}$s$^{-1}$. 

  \begin{figure}[]
\begin{center}$
\includegraphics[width=0.45\textwidth]{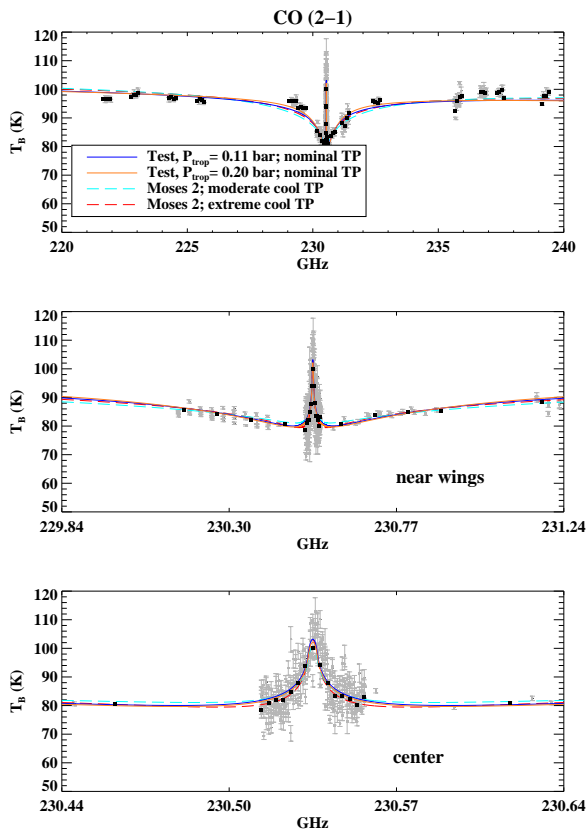}
$
\end{center}
\caption[Physical CO profile model fits to the CO (2--1) data]{\label{fig:cophys1mm} \small Model fits to the CO (2--1) line from a selection of best-fit physical CO profiles (see Table \ref{tab:physco}).}
\end{figure}

 \begin{figure}[]
\begin{center}$
\includegraphics[width=0.45\textwidth]{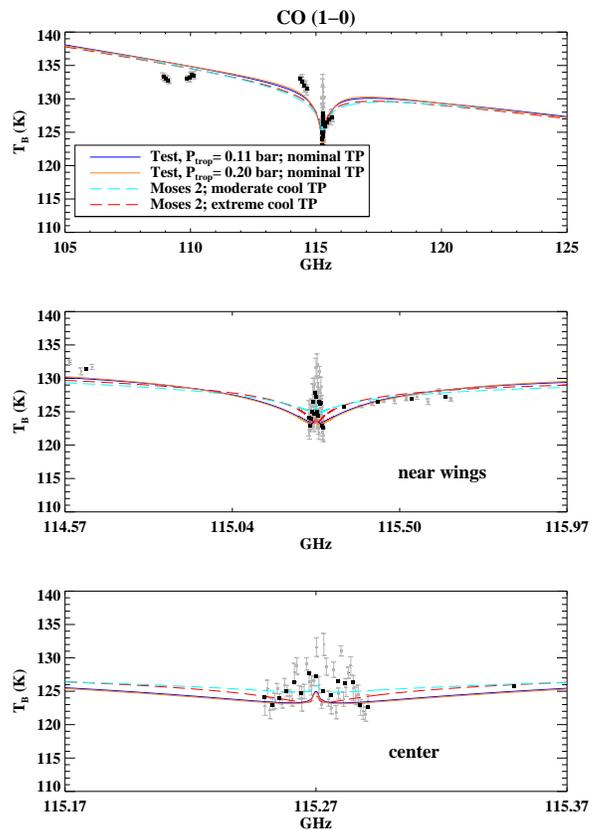}
$
\end{center}
\caption[Physical CO profile model fits to the CO (1--0) data]{\label{fig:cophys3mm} \small Same as Fig. \ref{fig:cophys1mm}, except for the CO (1--0) line. }
\end{figure}

To produce a physical CO profile that is more like our best-fit two-level CO distribution, we created two `test' eddy diffusion coefficient profiles (Fig. \ref{fig:eddy}). The profiles are designed to have a more rapid increase in the mixing from the diffusion rate minimum just above the tropopause, to allow high concentrations of externally supplied CO to reach deeper levels.  No attempt is made to evaluate the physical likelihood of such an eddy profile. The two test profiles are identical except for the location of the tropopause, which is defined by a sharp transition from fast to slow mixing. In addition to using a value for the tropopause of 0.11 bar as in the \cite{moses92} and \cite{romani93} cases, we also try locating the tropopause level at 0.20 bar. Using these test eddy profiles, we are able to produce physical CO models using the nominal TP profile that are in better agreement with the data. The best-fit parameters for the test profiles are $\phi=10-20\times10^8$ cm$^{-2}$s$^{-1}$ and CO$_{int}=0.2-0.3$ ppm. Unsurprisingly, the test eddy profiles, which are designed to be used with the nominal TP profile, do not do as well when used with the cool TP profiles.  A selection of the best model spectra based on physical CO distributions is shown in Figs. \ref{fig:cophys1mm} and \ref{fig:cophys3mm}.


\section{Discussion}\label{sec:discussion}	

Our analysis of the CARMA CO (2--1) and (1--0) data indicates a preference for the \cite{fletcher10} thermal profile in Neptune's upper atmosphere over warmer and cooler profiles. Good fits are characterized by CO mole fractions of 0.0-0.3 ppm below the tropopause and 0.8-1.3 ppm in the stratosphere. This is true for  the independent fits of the 1- and 3-mm lines, as well as the combined fit. The \cite{bezard99} thermal profile and our moderate warm profile give best-fit CO profiles that are consistent with the nominal thermal profile results. Cooler thermal profiles also produce acceptable fits to the data, but models using the cooler TP profiles are generally too broad in the far wings of the CO lines. The vertical CO profiles derived using our moderate and extreme cool thermal profiles are qualitatively different than the nominal TP profile best-fit solution: we find CO mole fractions of 0.3-0.4 ppm in the troposphere and lower stratosphere, and 1.4-1.8 ppm above 2.5-25 mbar.   Using a physical diffusion model, we determine that the stratospheric CO abundances for all thermal profiles correspond to an external CO flux of $\phi_\CO=0.5-20\times10^8$ cm$^{-2}$s$^{-1}$. Internal CO contributions for the physical CO profile solutions are typically 0.2-0.4 ppm.

Plots of the 3-mm spectrum (e.g. Figs. \ref{fig:co1vs2lay}, \ref{fig:coalltp3mm}, \ref{fig:coh2s3mm}, \ref{fig:cophys3mm}) show greater deviations  between the data and models than we find at 1 mm. We attribute this to the fact that, at 3 mm, we probe deeper levels of the atmosphere (Fig. \ref{fig:contrib}), where the atmospheric opacity is not well constrained. However, we are encouraged by the fact that the independent fits to our two CO lines give consistent CO profile solutions. Additionally, we find that the inclusion of 10-50$\times$ solar H$_2$S opacity improves the $\widehat{\chi^2}$ goodness-of-fit, but does not significantly affect the best-fit CO vertical profile.

 \subsection{Comparison with previous results}

\begin{figure*}[]
\begin{center}$
\begin{array}{cc}
\includegraphics[width=0.5\textwidth]{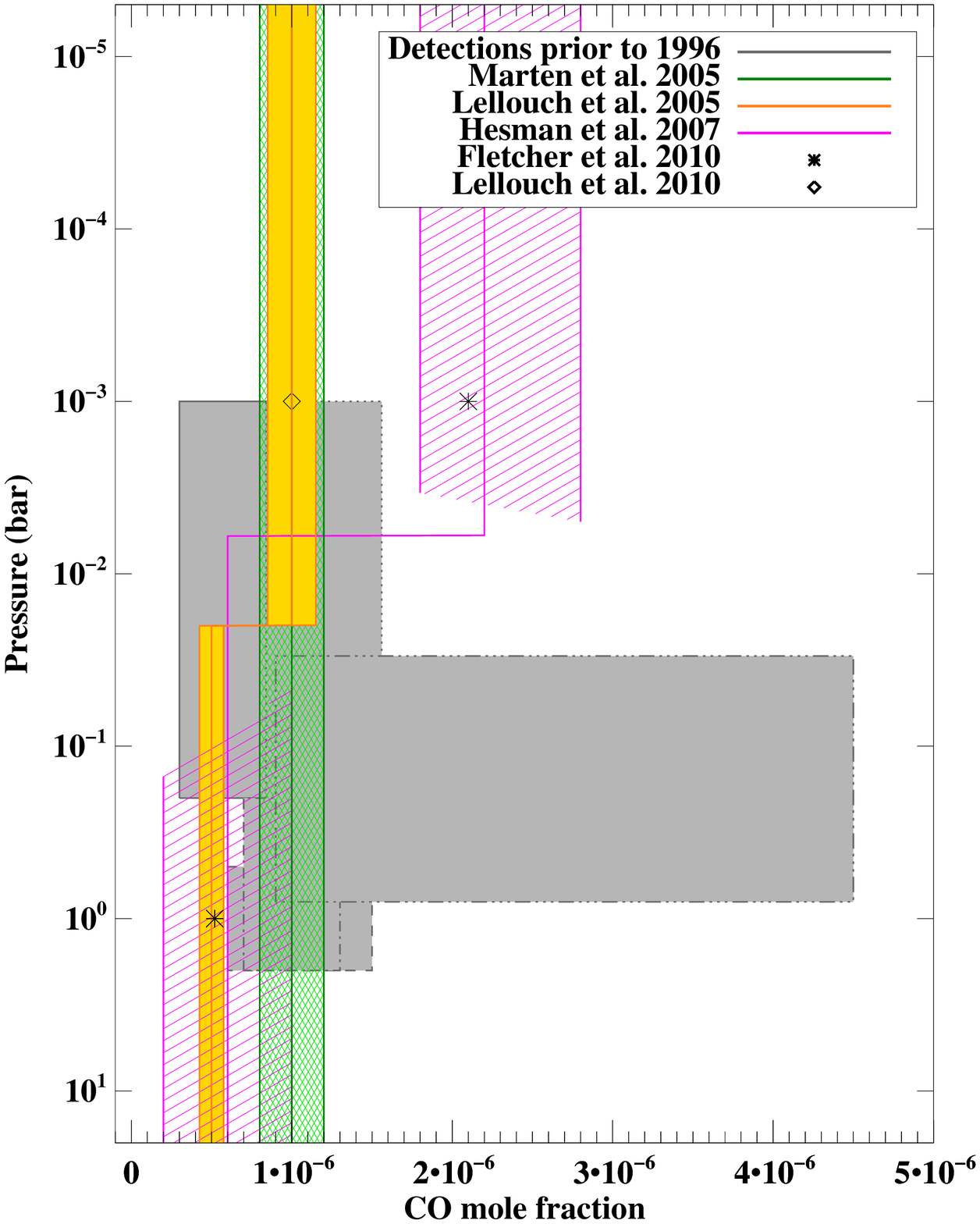}
\includegraphics[width=0.5\textwidth]{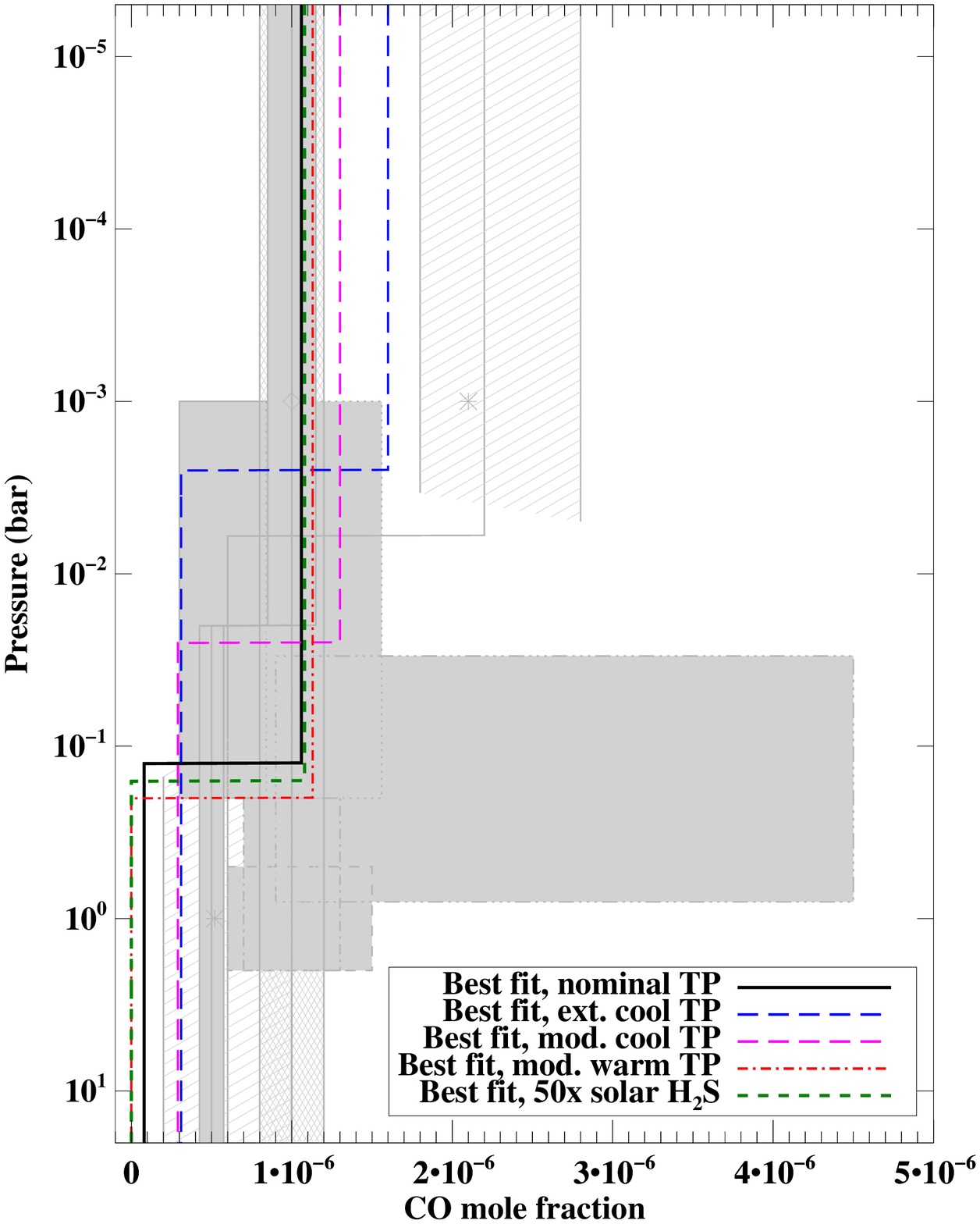}
\end{array}$
\end{center}
\caption[Comparison of our best-fit solutions with previously published CO profiles]{\label{fig:coprof}  \small Selected published CO profiles with errors (left), compared to our best-fit profiles (right). Grey boxes indicate all the published results from prior to 1996, as described in \cite{courtin96}. Colored regions indicate the best-fit CO profiles of \cite{marten05} (green, cross-hatched), \cite{lellouch05} (orange, shaded) and \cite{hesman07} (magenta, hatched); along with their uncertainties. The symbols indicate the values reported by \cite{fletcher10} (stars) and \cite{lellouch10} (diamond) for the stratospheric CO abundance. On the right, our best-fit two-level profiles are shown for several different thermal profiles, as well as for the case where absorption due to 50$\times$ solar H$_2$S is included.}

\end{figure*}

\begin{figure}[]
\begin{center}$
\includegraphics[width=0.45\textwidth]{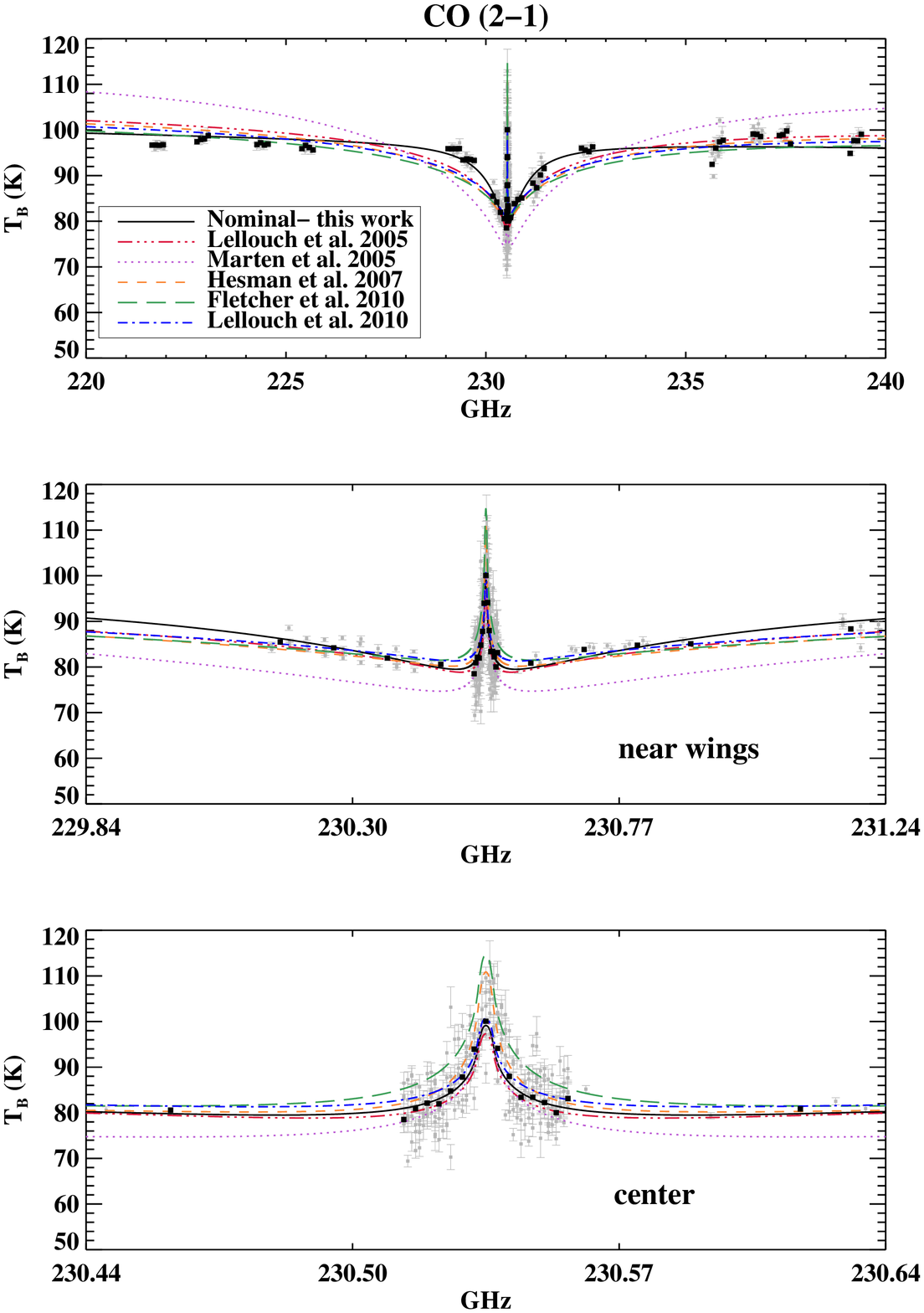}
$
\end{center}
\caption[Comparison of previously-published profiles to our CO (2--1) data]{\label{fig:coprev1mm}\small  Comparison of the CO (2--1) line data (this work) with a selection of best-fit CO profiles from previous authors (see Table \ref{tab:qual}). In each case, we use the reported CO and temperature profiles from the indicated study. Using our model code, we fit the 1- and 3- mm gain uncertainties to minimize $\widehat{\chi^2}$ and plot the scaled model. For reference, the best-fit two-level solution from this work is plotted as well (black). }
\end{figure}

\begin{figure}[]
\begin{center}$
\includegraphics[width=0.45\textwidth]{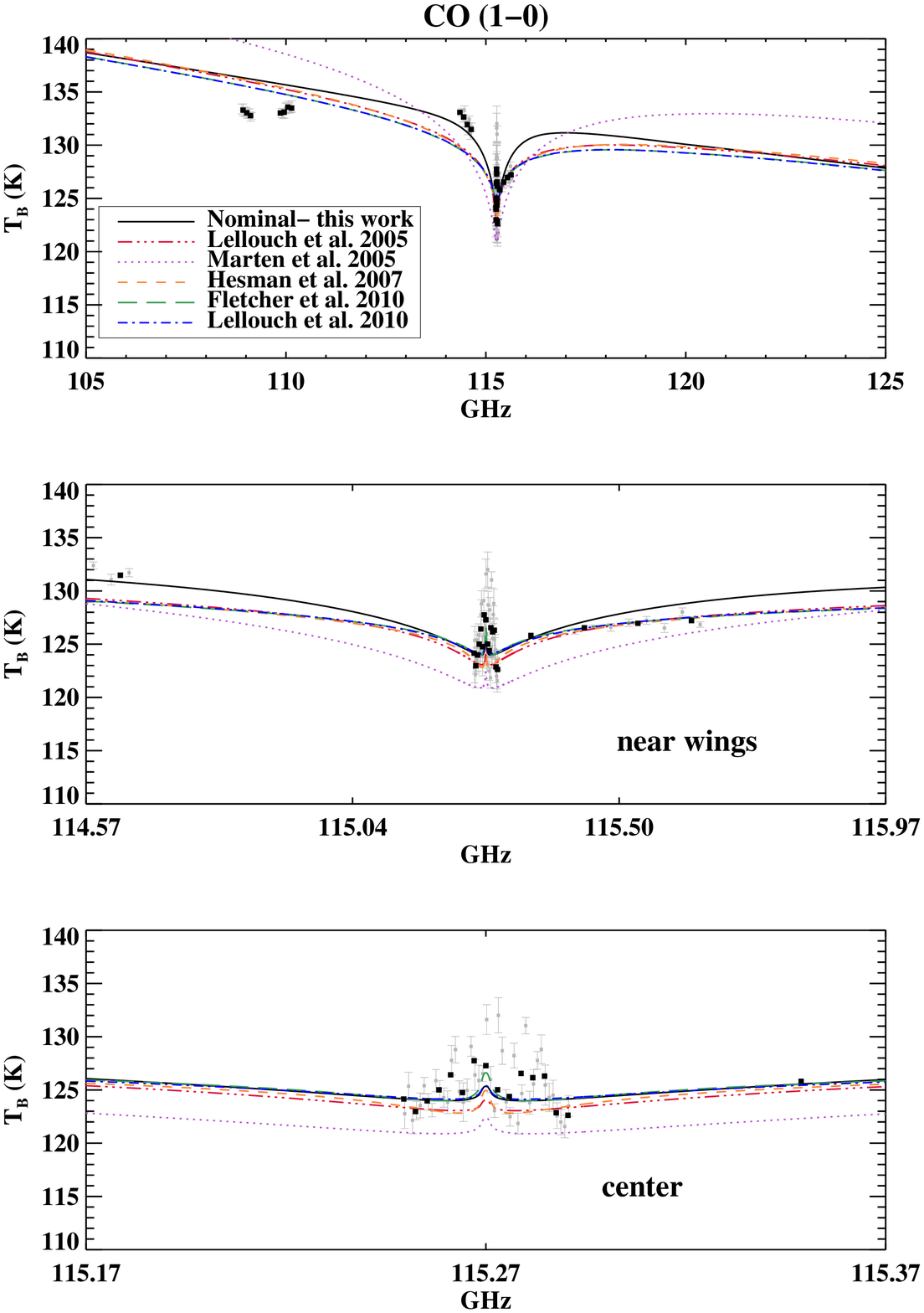}
$
\end{center}
\caption[Comparison of previously-published profiles to our CO (1--0) data]{\label{fig:coprev3mm}  \small Same as Fig. \ref{fig:coprev1mm}, except for the CO (1--0) line data. }
\end{figure}

 In Fig. \ref{fig:coprof} we compare our best-fit two-level CO profiles with previously reported values and profiles. Figures \ref{fig:coprev1mm} and \ref{fig:coprev3mm} use several of these recent results for the CO profile in conjunction with our radiative transfer code to produce model spectra. In each case we fit only the data amplitude scale factors at 1 and 3 mm; the lowest $\widehat{\chi^2}$ for these literature CO profiles and our data are presented in Table \ref{tab:qual}. 

Of the three most recent previous measurements of millimeter CO line shapes \citep{marten05,lellouch05,hesman07}, only \cite{marten05}  found a result consistent with a constant CO mole fraction. However, their data only cover wavelengths up to 50 MHz away from the center of the CO (4-3) line. From the CO line contribution functions of the first three rotational transitions (Fig. \ref{fig:contrib}), we estimate that their data are only sensitive down to pressures of a few tens of mbar, which implies that they would not detect a tropospheric decrease in the CO abundance. Our best-fit two-level, nominal TP profile model has an abundance of 1.1 ppm in the upper atmosphere, which agrees well with the \cite{marten05} value. As Figs. \ref{fig:coprev1mm} and \ref{fig:coprev3mm} and Table \ref{tab:qual} illustrate, this profile is the least successful of those tested at reproducing our data.
 
 Our two-level solutions, which have a smaller abundance of CO in the lower atmosphere, are qualitatively consistent with the results of the  two recent millimeter studies by \cite{lellouch05} and \cite{hesman07}. Quantitatively, there are important differences between these two previous works and with our own result. The \cite{hesman07} best-fit solution has a stratospheric CO abundance that is inconsistent with the \cite{lellouch05} result, as well as a lower pressure that defines the transition between the upper and lower levels of a two-level model. Our analysis demonstrates that there is a strong relationship between the stratospheric CO mole fraction and the pressure that defines the transition between the upper and lower levels of a two-level model. Furthermore, the goodness-of-fit, as measured by $\widehat{\chi^2}$,  tends to be relatively constant over a wide range of transition pressures, so that small variations in the data may lead to significant changes in the derived CO abundances. This leads to uncertainties in the CO abundances that are greater than the 20\% uncertainties adopted by \cite{lellouch05}, and may be a dominant source of inconsistency between the solutions of \cite{lellouch05} and \cite{hesman07}. In comparison with our results, \cite{lellouch05} and \cite{hesman07}  have higher tropospheric CO abundances ($0.5\pm 0.1$ and $0.6\pm0.4$ ppm, respectively) and transition altitudes (20 and 6 mbar, respectively) than our nominal best-fit solution of less than 0.1 ppm in the troposphere, with a transition near the tropopause. Interestingly, our best-fit solution using a moderate cool thermal profile, which is similar to the TP profiles used by  \cite{lellouch05} and \cite{hesman07}, produces a two-level CO profile that is roughly similar to the \cite{lellouch05} CO vertical profile (see Table \ref{tab:2lev}), although still outside of their quoted uncertainties; and a tropospheric CO abundance that is consistent with \cite{hesman07}. Using a cooler thermal profile also pushes the best-fit transition pressure to lower pressures (Fig. \ref{fig:coprof}).  As an additional test of the effect of the thermal profile on the CO best-fit solution, we used the plot from  \cite{hesman07} to approximate the  \cite{hesman07} data, and modeled the CO (3--2) line in a similar way to our own data. We note that we do not include the small leakage correction \citep{hesman07} when converting from antenna temperature to brightness temperature. We find that our best-fit model to the approximate \cite{hesman07} data using our nominal thermal profile has a CO mole fraction of 1.1 ppm in the upper atmosphere; this value increases to 1.7 ppm of CO in the upper atmosphere if we perform the fit using a cooler thermal profile, which is closer to the best-fit value found by \cite{hesman07}. This behavior is similar to what we see when fitting our own CO (2--1) and (1--0) line data.
 
 While we find better agreement with the CO profiles of \cite{lellouch05} and \cite{hesman07} when we use a cooler thermal profile, our results do favor the warmer nominal thermal profile. As Figs. \ref{fig:coprev1mm} and \ref{fig:coprev3mm} illustrate, model spectra produced using the \cite{lellouch05} and \cite{hesman07} CO and thermal profiles, but with our radiative transfer code, are less consistent with our data than our nominal fit; this is particularly true in the line wings, 0.5-2 GHz from line center. The \cite{hesman07} solution also appears to be too high in the CO (2--1) emission peak.  In addition to their use of a cooler thermal profile, the remaining cause of discrepancy between our result and that of  \cite{lellouch05} may be the more limited frequency coverage of their dataset: if we consider only our data within the frequency range covered by the \cite{lellouch05} dataset, we find a best-fit vertical CO profile with $1.00 \pm 0.08$ ppm in the stratosphere (above 3 mbar) and $0.62 \pm 0.05$ ppm in the troposphere.  We considered two additional factors that may contribute to the difference between our solution and that found by \cite{hesman07}: the fact that \cite{hesman07} observe a higher frequency line, which probes slightly higher altitudes in the atmosphere (Fig. \ref{fig:contrib}); and the uncertainty in the JCMT beam efficiency, which is used to convert the \cite{hesman07} data from antenna temperature to brightness temperature. By fitting  the approximate \cite{hesman07} data, we find that of the two, the uncertainty in the JCMT beam efficiency has a stronger effect on the best-fit CO profile solution. The authors allow for a 5\% uncertainty in the beam efficiency. With this constraint, our best fit to their data has a mole fraction of 0.5 ppm in the lower stratosphere and troposphere. If we allow for a beam efficiency uncertainty of $\pm 10\%$, our best-fit models to the approximate \cite{hesman07} data have mole fractions of only 0.0-0.1 ppm of CO in the lower atmosphere. We therefore conclude that an underestimate of the uncertainties in the beam efficiency could potentially account for the remaining difference between our best-fit solution and the best-fit solution of  \cite{hesman07}.

\begin {table*}[]
 \begin{minipage}{1.0\textwidth}
\begin{center}

 \small
 \caption{} {\small Goodness-of-fit  $\widehat{\chi^2}$ values obtained using the best-fit CO profiles from previous authors with our radiative transfer code and our data. See Section \ref{sec:discussion}.}
 
 \vspace{\baselineskip}
 
 \begin{tabular}{l l l l l l l l }
\hline
\hline
CO profile reference & TP profile  &$\widehat{\chi^2}$(1 mm)&  $\widehat{\chi^2} $(3 mm)&  $\widehat{\chi^2} $all  \\
\hline
this work\footnote{best-fit one-level model, from Table \ref{tab:1lev}}	&nominal&15.3 &14.6 &15.3\\
 \cite{marten05}\footnote{1.0 ppm CO everywhere}	&\cite{marten05}	&70.2 &71.5&70.3\\

\hline
this work\footnote{best-fit two-level model, no H$_2$S, from Table \ref{tab:2lev}}	&nominal&10.7 &9.3 &11.1\\
 \cite{lellouch05}\footnote{1.0 ppm CO at altitudes above 20 mbar; 0.5 ppm CO deeper than 20 mbar}	&\cite{bezard91}\footnote{same as used in \cite{lellouch05}}&15.8&      19.9&       16.3\\
 Hesman et al. 2007\footnote{2.2 ppm CO at altitudes above 6 mbar; 0.6 ppm deeper than 6 mbar}& \cite{hesman07} &18.0	&21.0&18.4 \\
 \cite{fletcher10}\footnote{2.1 ppm CO at altitudes above 10 mbar; 0.5 ppm deeper than 10 mbar} &nominal \footnote{from \cite{fletcher10}} &25.3&16.8&24.0\\
 \cite{lellouch10}\footnote{we assume the same CO profile as in \cite{lellouch05}; only the stratospheric CO was constrained from \cite{lellouch10}}&\cite{lellouch10}&14.2&16.6&14.5\\

\hline
\label{tab:qual}
\end{tabular}
\end{center}
\end{minipage}
\end{table*}

Two other recent papers look at the CO abundance on Neptune using (far)infrared data. \cite{fletcher10} observed fluorescent lines with AKARI, and fit two profiles to the CO (2--1) fluorescent line: in their Profile 1, CO is limited to altitudes above 10 mbar. Profile 2 is similar to the \cite{hesman07} best-fit profile: 2.1 ppm of CO is present in the stratosphere, decreasing by a factor of four at altitudes below 10 mbar.  They find that they require some CO below 10 mbar (their Profile 2) to reproduce their data. As with the \cite{hesman07} result, we find that the \cite{fletcher10} CO profile produces models that are a poor fit in the line core and far wings. We note that for a physical CO distribution, the CO abundance will not be constant with altitude, and the derived value for the stratospheric CO abundance will be dependent on the pressure levels one is sensitive to (see Fig. \ref{fig:cophys}).  Using Herschel measurements of CO lines at 153-187 $\mu$m, \cite{lellouch10} find a stratospheric CO abundance that is similar to the \cite{lellouch05} number; roughly 1 ppm. This value is consistent with our results (though inconsistent with the \cite{fletcher10} solution).

 \subsection{Implications: internal CO}
 
 The CO abundance that originates from Neptune's deep atmosphere acts as a probe of Neptune's global oxygen abundance: according to the net thermochemical reaction (Eq. 1) the equilibrium CO mole fraction is directly proportional to the equilibrium abundance of H$_2$O, and under the conditions of Neptune's deep atmosphere, nearly all the gas phase oxygen is contained in water. As first described by \cite{prinn77}, the observed tropospheric CO mole fraction (0.0-0.3 ppm from our analysis) represents the equilibrium abundance  at the CO `quench level', which is defined as the depth at which
 \begin{equation}\label{eq:timescales}
 \tau_{chem} = \tau_{mix}
 \end{equation}
  where $\tau_{chem}$ is the timescale for chemical conversion of CO into CH$_4$ and $\tau_{mix}$ is the atmospheric mixing timescale. Above the quench level, vertical mixing transports CO to higher altitudes before the constituents have a chance to equilibrate ($\tau_{chem}>\tau_{mix}$); and the CO mole fraction remains constant at a level which can be much higher than the equilibrium value. The H$_2$O mole fraction can then be determined from the CO mole fraction via the equilibrium relation
\begin{eqnarray}\label{eq:ratio}
\frac{q\CO\left(q\HH_2\right)^3\left(P_\mathrm{T}\right)^2}{q\CH_4\ q\HH_2\OO} = e^{-\frac{\Delta_fG^0(\CO)-\Delta_fG^0(\CH_4)-\Delta_fG^0(\HH_2\OO)}{RT}} 
\end{eqnarray}
where $R$ is the gas constant, $T$ is the temperature and $P_\mathrm{T}$ is the total pressure in bars at the quench level, $q$X is the mole fraction of species X, and $\Delta_fG^0($X$)$ is the Gibbs free energy of formation of species X at the temperature of the quench level.

Determining the CO quench level requires knowledge of the limiting reaction rate for converting CO to CH$_4$; this is set by the slowest reaction step in the fastest chemical pathway for CO $\rightarrow$ CH$_4$ conversion. The original chemical scheme analyzed by \cite{prinn77} for Jupiter had as the rate-limiting step
\begin{eqnarray}
\HH_2\CO + \HH_2 \rightarrow \CH_3  +\OH
\end{eqnarray}

This scheme was adopted by \cite{lodders94} to estimate the O/H ratio implied by the early detections of $\sim$1 ppm of CO on Neptune \citep{marten91,rosenqvist92,guilloteau93,naylor94}. They found that a 440 times solar oxygen abundance is required to produce a 1 ppm CO abundance; this is roughly a factor of 10 higher than the C/H enrichment suggested by CH$_4$ measurements.  Using updated laboratory measurements for the rate coefficients, \cite{griffith99} showed that  the \cite{prinn77} reaction is actually about 4 orders of magnitude slower that originally calculated, and is therefore too slow to play a role in CO quenching kinetics. \cite{griffith99} adopted instead a chemical scheme first advocated by \cite{yung88} for their analysis of Gliese 229B, which has as its rate-limiting step
\begin{eqnarray}
\HH+\HH_2\CO+\M\rightarrow \CH_3\OO+\M
\end{eqnarray}
where M represents any third body (atom or molecule). \cite{bezard02}  also use the \cite{yung88} scheme for their analysis of CO chemistry on Jupiter. More recently, \cite{visscher10}, \cite{moses11} and \cite{visscher11} incorporate further updates to reaction rate coefficients, and compare the rates of all relevant reactions to determine the dominant kinetic mechanism for CO $\rightarrow$ CH$_4$ conversion. \cite{visscher11} find that the dominant pathway for conditions in Jupiter, cool brown dwarfs and hot Jupiters is:
 \begin{subequations}
 \begin{align}
 \HH+\CO+\M&\rightarrow \HH\CO+\M\\
 \HH_2+\HH\CO&\rightarrow \HH_2\CO+\HH\\
 \HH+\HH_2\CO+\M&\rightarrow \CH_2\OH+\M\\
 \HH_2+\CH_2\OH&\rightarrow \CH_3\OH+\HH\\
 \CH_3\OH+\M&\rightarrow \CH_3+\OH+\M \label{eq:limit}\\ 
 \HH_2+\CH_3&\rightarrow \CH_4+\HH\\
 \HH+\OH+\M&\rightarrow \HH_2\OO+\M
 \end{align}
 \end{subequations}
in which Eq. (\ref{eq:limit}) is the rate-limiting step. (Note: \cite{visscher11} indicate a second reaction scheme which may be important under some conditions. We include this second reaction in our models as well, but find the rate is always more than two orders of magnitude slower than the above scheme. For simplicity, we omit this second reaction pathway here.)

 A second source of error in the \cite{lodders94} determination of Neptune's CO quench level was described by \cite{smith98}, who showed that  the typical estimate of the mixing time 
 \begin{eqnarray}\label{eq:mix}
\tau_{mix}= L^2/K
 \end{eqnarray}
is not correct when the pressure scale height $H$ is used for the characteristic mixing length scale $L$. According to the calculations of \cite{smith98}, effective mixing lengths are typically of order $0.1--0.2 H$ which means that assuming $L=H$ will lead to an overestimate of $\tau_{mix}$ by up to two orders of magnitude.  
 
 Using the  \cite{visscher11} rate-limiting step and the \cite{smith98}  recipe for estimating the effective mixing length, we evaluate the O/H enrichment implied by our observed deep CO mole fractions. 
 
\begin {table*}[]
\begin{center}
 \caption{}{\small Mole fractions in the deep atmosphere (below condensation levels) for different O enrichments. The enrichment factor for C is 48$\times$ solar, S is 50$\times$ solar, N is 1$\times$ solar, and the O enrichment varies. }
 
 \vspace{\baselineskip}
 
 \small
\begin{tabular}{l l l l l l}
\hline

Molecule	&mole fraction&&&&\\
		& O/H=50$\times$ solar& =100$\times$ solar& =400$\times$ solar&=600$\times$ solar&=700$\times$ solar \\
\hline
H$_2$		&0.785				&0.744				&0.483 				&0.292 &	0.191	\\
He			&0.141				&0.134				&0.0869				&0.0526&	0.0345	\\
CH$_4$		&0.0250				&0.0252				&0.0266				&0.0276& 0.0281		\\
NH$_3$		&$1.28\times10^{-4}$	&$1.29\times10^{-4}$	&$1.36\times10^{-4}$	&$1.41\times10^{-4}$&$1.44\times10^{-4}$		\\
H$_2$O		& 0.0473				&0.0954				&0.402				&0.626&	0.744	\\
H$_2$S		&$1.28\times10^{-3}$	&$1.29\times10^{-3}$	&$1.36\times10^{-3}$	&$1.41\times10^{-3}$&$1.44\times10^{-3}$		\\
\hline
\label{tab:deep}

\end{tabular}
\end{center}
\end{table*}
Pursuant to Eq. \ref{eq:ratio}, the equilibrium CO abundance depends on the H$_2$, CH$_4$ and H$_2$O mole fractions, temperature, and pressure. Therefore, in order to determine the CO quench level implied by our data, we extend our model of the thermal profile and composition to the deep atmosphere ($P$ of order 10$^5$ bar). As described in Section \ref{sec:composition},  the atmospheric C, N and S abundances are constrained by previous studies. We assume that in the deep atmosphere the C/H and S/H enrichments are $\sim$50 times the protosolar value, and that the CH$_4$ and H$_2$S abundances represent the total elemental abundances of C and S, respectively. For nitrogen, we assume  that the NH$_3$/H ratio is equal to the protosolar N/H value, as constrained by cm wavelength observations, noting that the total atmospheric N/H enrichment can be higher if much of the nitrogen is in N$_2$.  The O/H enrichment is the quantity we wish to determine from these calculations.  

 For large enrichments of H$_2$O, the approximation 
$ \OO_{gas}/\HH \approx q\HH_2\OO/\left(2\cdot q\HH_2\right)$, where $q$X is defined as the mole fraction of species X, does not hold, since a significant fraction of the hydrogen is contained in species other than H$_2$. Therefore to relate the solar oxygen enrichment to the various abundances we approximate 

 \begin{multline}\label{eq:oh}
 \OO_{gas}/\HH \approx \\  \frac{q\HH_2\OO}{2\cdot q\HH_2+2\cdot q\HH_2\OO+3\cdot q\mathrm{NH}_3+4\cdot q\CH_4+2\cdot q\HH_2\mathrm{S}}\
 \end{multline}
Similar equations relate the C/H, S/H and NH$_3$/H enrichments to the mole fractions of CH$_4$, H$_2$S and NH$_3$. Table \ref{tab:deep} presents the deep atmospheric mole fractions for relevant species, for several different values of the oxygen enrichment. In relating the global composition of Neptune to protosolar abundances, we have assumed that the atmospheric abundances of C, N, O and S represent Neptune's global C, N, O and S enrichments. In particular, we ignore the removal of oxygen from the atmosphere by the formation of rock: for protosolar composition gas, of order 20\% of the oxygen is trapped in rock \citep{lodders04,visscher11}. This means that Eq. \ref{eq:oh} is an underestimate of Neptune's total global O/H ratio.   The protosolar abundances are taken from \cite{asplund09}.  We note that the values for the protosolar abundances have been revised significantly over the past two decades; for example, the protosolar O/H ratio has been adjusted downwards by more than 10\%  from the \cite{grevesse93} values used by  \cite{lodders94}. 

\begin{figure}[]
\begin{center}$
\includegraphics[width=0.45\textwidth]{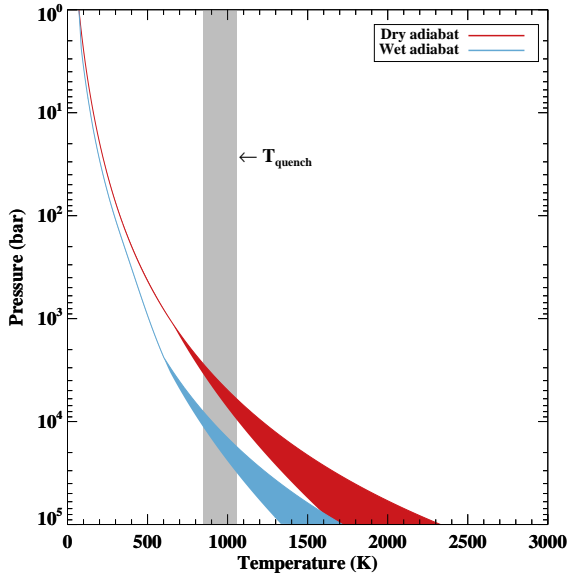}
$
\end{center}
\caption[Thermal profiles for the deep atmosphere]{\label{fig:tpdeep} \small Thermal profiles for the deep atmosphere, assuming an adiabatic extrapolation. The value at 1 bar is matched to the Voyager 2 measurement. The red profile assumes a dry adiabatic lapse rate throughout the atmosphere; the blue profile assumes the appropriate wet lapse rate at  expected cloud condensation levels. The range in the profiles indicates variation in composition, from 50 to 600 times the solar O/H ratio. The range of CO quench temperatures (T$_{quench}$) for $K=10^7-10^9$ cm$^2$ s$^{-1}$ is indicated by the vertical grey bar.}
\end{figure}

The model used for calculating the thermal profile down to high pressures is described in \cite{depater91} and references within; starting with the deep composition (Table \ref{tab:deep}) at an initial temperature and pressure, the model calculates the temperature at successive steps upwards in altitude by assuming the temperature follows either (1) a dry adiabat (regardless of the relative humidity of the atmospheric constituents); or (2) the appropriate wet adiabat (Fig. \ref{fig:tpdeep}). We expect the wet adiabat case to be more appropriate. Clouds condense when the partial pressure of a given species exceeds its saturation vapor pressure, and the atmospheric composition changes accordingly. No clouds form at temperatures above the critical temperature of water (647 K). To calculate the adiabatic TP profile, we use an ideal gas equation of state and ``intermediate'' hydrogen, i.e. the specific heat is near that of normal hydrogen, and the \textit{ortho} to \textit{para} ratio is close to the equilibrium value \citep{wallace80}. As in Section \ref{sec:TP}, we adjust the starting temperature at the deepest layer so that the thermal profile roughly matches the \cite{lindal92} value of 71.5 K at 1 bar. 

Utilizing the \cite{visscher11} rate-limiting reaction,  the chemical lifetime of CO is given by 
\begin{eqnarray}
\tau_{chem} = \frac{[\CO]}{d[\CO]/dt} = \frac{[\CO]}{k_{\ref{eq:limit}}[\CH_3\OH][\M]}
\end{eqnarray}
where [X] is the number density of species X and $k_{\ref{eq:limit}}$ is the reaction rate of step (\ref{eq:limit}).  Given the rate of the reverse reaction $k_{\ref{eq:limit}R}$ we can write $\tau_{chem}$ as

\begin{align}
\tau_{chem} &= \frac{[\CO]}{[\CH_3][\OH][\M]k_{\ref{eq:limit}R}} \\
		    &= \frac{nK_{eq}}{P_\mathrm{T}[\HH_2]^2 [M] k_{\ref{eq:limit}R}}\\
		    &= \frac{K_{eq}}{n^2P_\mathrm{T} q\HH_2^2 k_{\ref{eq:limit}R}}
\end{align}	    
where $P_\mathrm{T}$ is the total pressure in bars, $n$ is the total number density in cm$^{-3}$, and $K_{eq}$ is the equilibrium constant of $\OH+\CH_3\leftrightarrow \CO+2\HH_2$:
\begin{align}
K_{eq}=e^{-\frac{\Delta_fG^0(\CO)-\Delta_fG^0(\CH_3)-\Delta_fG^0(\OH)}{RT}}
\end{align}

The Gibbs free energies of formation as a function of temperature are taken from the NIST-JANAF tables \citep{chase98}. The reverse reaction rate $k_{\ref{eq:limit}R}$ is calculated as described in \cite{visscher11} using the expression

\begin{eqnarray}
k=\frac{k_0}{1+(k_0[\M]/k_\infty)}F_c^\beta
\end{eqnarray}
where
\begin{eqnarray}
\beta=\left(1+\left[\frac{\log_{10}(k_0[\M]/k_\infty)}{0.75-1.27\log_{10}F_c}\right]^2\right)^{-1}
\end{eqnarray}
and the parameters for calculating $k_{\ref{eq:limit}R}$ are given by \cite{jasper07}:
\begin{align}
k_0&=1.932\times10^3 T^{-9.88}e^{-7544/T}&\\
       & \ \ +5.109\times10^{-11}T^{-6.25}e^{-1433/T}\ \ \cm^6\ \s^{-1}\\
k_\infty&=1.031\times10^{-10}T^{-0.018}e^{16.74/T} \ \ \ \cm^3\ \s^{-1}\\
F_c&= 0.1855 e^{-T/155.8}+0.8145e^{-T/1675}+e^{-4531/T}
\end{align}

We calculate the atmospheric mixing timescale $\tau_{mix}$ according to Eq. (\ref{eq:mix}) using the \cite{smith98} recipe for determining the effective mixing length $L$. In general, we find that $L\approx0.12-0.18 H$, which is consistent with the \cite{smith98} findings. The eddy mixing coefficient in the deep atmosphere is constrained by the observed heat flux $\phi= 433\pm46$ erg cm$^{-2}$ s$^{-1}$ \citep{pearl91}. Using the scaling relationship for free dry convection \citep{stone76}

\begin{eqnarray}
K\sim\left(\frac{\phi R}{C_P \rho} \right)^{1/3}H\label{eq:conv}
\end{eqnarray}
we estimate that K is of order $2\times 10^8$ cm$^2$ s$^{-1}$. In Eq. (\ref{eq:conv}),  $\rho$ is the gas density, $R$ is the gas constant and $C_P$ is the specific heat at constant pressure. Following \cite{lodders94} we assume a plausible range for $K$ of $10^7-10^9$ cm$^2$ s$^{-1}$. 

\begin{figure*}[]
\begin{center}$
\begin{array}{cc}
\includegraphics[width=0.5\textwidth]{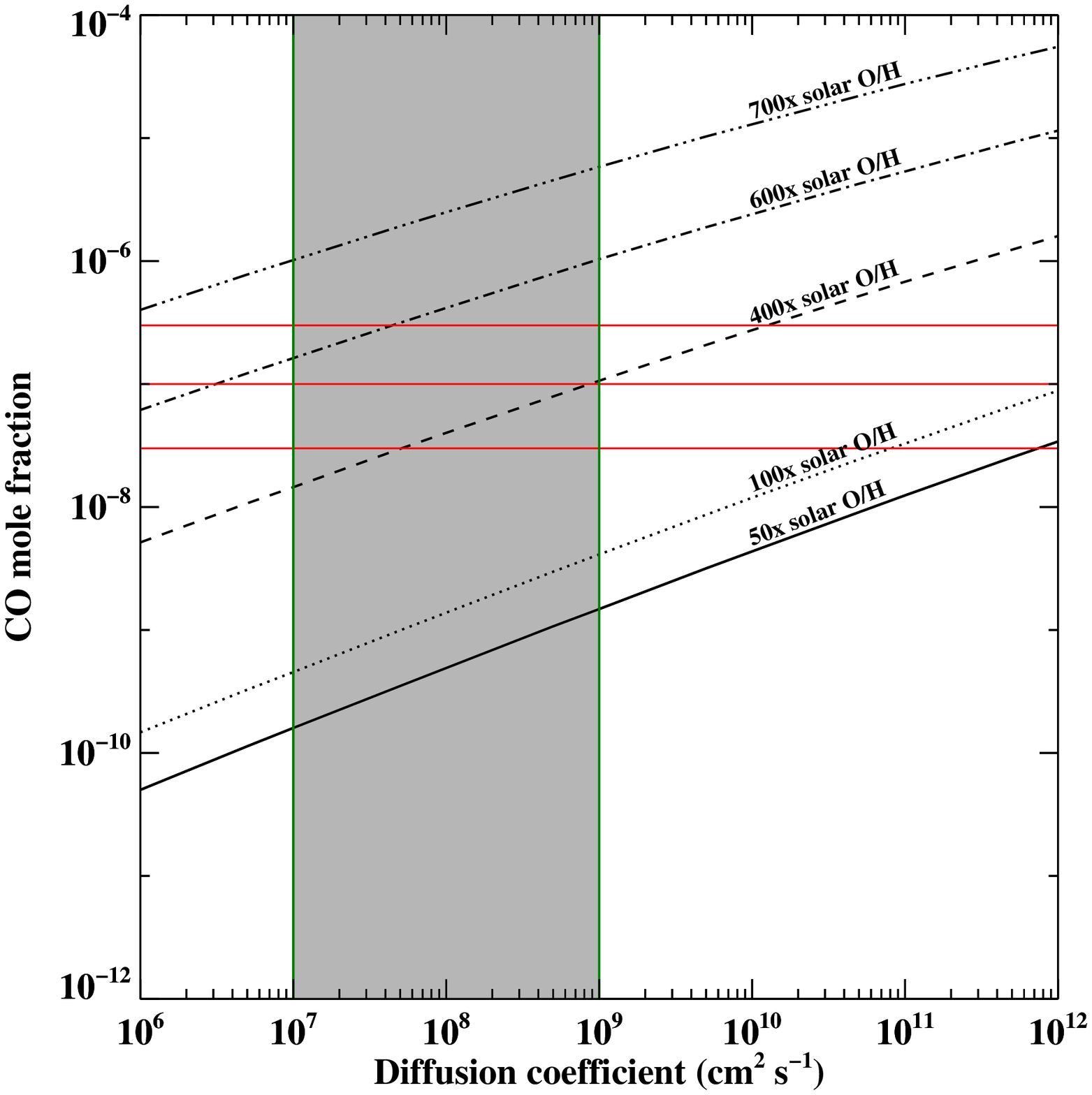}
\includegraphics[width=0.5\textwidth]{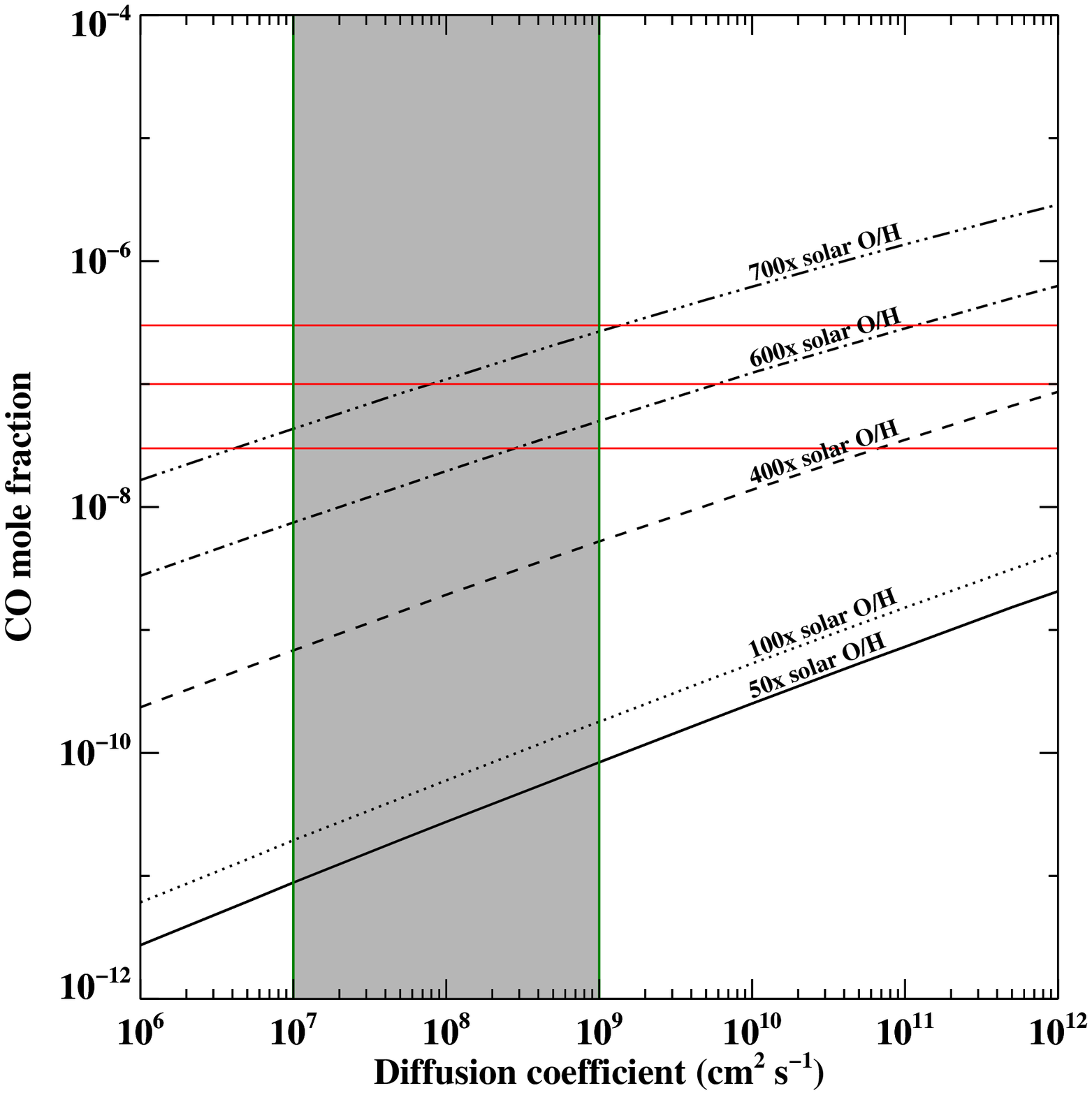}
\end{array}$
\end{center}
\caption[Predicted CO mole fractions due to upwelling]{\label{fig:coint} \small Predicted CO mole fraction in the visible atmosphere of Neptune due to upwelling from the deep atmosphere as a function of the eddy mixing coefficient $K$ and the O/H enrichment over solar, using a dry adiabatic extrapolation (left) and a wet adiabatic extrapolation (right) for the thermal profile. Horizontal lines indicating 0.3, 0.1 and 0.03 ppm of CO are shown (red); the shaded region indicates the plausible range of $K$ values. O/H enrichments of 50, 100, 400 and 600 times solar are shown. }
\end{figure*}

Using Eqs. (\ref{eq:timescales}) and (\ref{eq:ratio}), we now solve for the CO quench level and the predicted CO mole fraction, as a function of the eddy diffusion coefficient and the assumed value of the O/H ratio. We find that, for $K=10^7-10^9$ cm$^2$ s$^{-1}$, Eq. (\ref{eq:timescales}) becomes true at a temperature  $T_{quench} = 850-1100$ K; coincidentally this is quite similar to the \cite{lodders94} value of $T_{quench} =$ 998 K. The quench temperature, which is indicated in Fig. \ref{fig:tpdeep}, is insensitive to the thermal profile (dry or wet lapse rate) used. Since the quench level is below the condensation level of the deepest cloud, the abundances of H$_2$, CH$_4$, and H$_2$O are the same for the dry and wet adiabat cases; however the pressure at which the quench temperature is reached changes (Fig. \ref{fig:tpdeep}), which affects the CO mole fraction at the quench level (Eq. \ref{eq:ratio}). The quench level (and therefore, the predicted observable) CO mole fractions are plotted in Fig. \ref{fig:coint} for the dry and wet adiabat cases; for reference, we have indicated CO abundances of 0.03, 0.1 and 0.3 ppm with horizontal lines. We find that, for the dry adiabat case, 0.1 ppm of upwelled CO and $K\leq 10^9 $ cm$^2$ s$^{-1}$ implies a global oxygen enrichment of at least 400 times solar; this is roughly 8 times the C/H enrichment implied by Neptune's observed CH$_4$ abundance.  For a wet adiabat, 0.1 ppm of upwelled CO and $K\leq 10^9 $ cm$^2$ s$^{-1}$ implies a global oxygen enrichment of at least 650 times solar.

We might expect the planets to be uniformly enriched in heavy elements \citep{owen99}; if this is so, then for fast mixing (K$\sim 10^9$ cm$^2$ s$^{-1}$) we would expect an upwelled CO abundance of $1.5\times10^{-9}$ or $8.4\times10^{-11}$, for the dry and wet adiabat cases, respectively. We note that such low internal CO mole fractions are consistent with our data.

\subsection{Implications: external CO}
In addition to the abundance of CO that is vertically transported from Neptune's deep atmosphere, a significant external source ($\phi_\CO=0.5-20 \times10^8$ molecules cm$^{-2}$ s$^{-1}$ ) is implied by our analysis (see Table \ref{tab:physco}). These inferred CO production rates are in general agreement with the $10^8$ molecules cm$^{-2}$ s$^{-1}$ found by \cite{lellouch05} and lower than the $10^{10}$ molecules cm$^{-2}$ s$^{-1}$ estimated by \cite{hesman07}. The precise value of $\phi_\CO$ depends strongly on the eddy diffusion  and thermal profiles used in the model. 

The high rates of $\phi_\CO$ inferred by the stratospheric CO mole ratio, and the high CO/H$_2$O ratio on Neptune have led to the suggestion that comet impacts are the primary supply mechanism of CO to Neptune's upper stratosphere \citep{lellouch05,hesman07}. We explore effectiveness of producing these abundances of CO through cometary impacts by calculating the approximate CO production from estimates of impact rates, similar to the analysis performed by \cite{bezard02} for Jupiter. We use for the impact rates of comets at Jupiter the estimate from \cite{zahnle03}:

\begin{equation}
\dot{N}_J(D> 1.5 \text{ km}) = 0.005^{+0.006}_{-0.003}\ \  \text{ yr}^{-1}
\end{equation}
as well as their approximation that the rate for Neptune is roughly 

\begin{equation}
\dot{N}_N(D> 1.5 \mathrm{\ km}) \approx 0.25 \dot{N}_J (D > 1.5 \mathrm{\ km})
\end{equation}
where $D$ is the comet diameter. We test two possible comet size distributions: the first is the Shoemaker and Wolfe (1982) distribution:

\begin{equation}
\dot{N}_N(>D) =  \dot{N}_N(D> 1.5 \mathrm{\ km}) \left(\frac{D(\mathrm{ km})}{1.5}\right)^{-2}\ \text{ yr}^{-1}
\end{equation}
Secondly, we test the \cite{zahnle03} ``Case B''  impact distribution, which was determined from the size distribution of craters on Triton:

 \begin{align*} 
\dot{N}_N(>D) &=  2.62\ \dot{N}_N(D> 1.5 \mathrm{km}) \left(\frac{D(\mathrm{km})}{1.5}\right)^{-1.7}\ \  \text{ yr}^{-1}  \\
&\ \ \ \ \ \ \ \ \ \ \ \ \ \ \ \ \ \ \ \ \ \ \ \ \  \ \ \ \ \ \ \  \  (D<1.5 \ \mathrm{ km}) \tag{29a} \\ 
\dot{N}_N(>D) &=  0.129\ \dot{N}_N(D> 1.5 \mathrm{km}) \left(\frac{D(\mathrm{km})}{5}\right)^{-2.5}\ \text{ yr}^{-1}  \\	
&\ \ \ \ \ \ \ \ \ \ \ \ \ \ \ \ \ \ \ \ \ \ \ \ \  \ \ \ \ \ \ \  \ (D>1.5 \ \mathrm{ km}) \tag{29b} \\
 \end{align*}
\refstepcounter{equation}
Using these comet distributions, we calculate the mass rate of CO due to comet impacts as
\begin{equation}
\dot{M}= f_{\OO}f_{\CO} \int_{R_{min}}^{R_{max}} 4\pi R^2\rho \dot{N}(>R)\ dR \ \ \  \text{ g yr}^{-1}
\end{equation}
where $R$ is the comet radius, $f_\OO$ is the fraction of the comet mass in the form of oxygen and $f_{\CO}$ is the fraction of the oxygen that ends up as CO. Following \cite{bezard02} we adopt values of $f_\OO=0.5$ and $f_{\CO}=0.9$, and a typical comet nucleus density  $\rho= 0.55$ g cm$^{-3}$. 

To calculate the size range of comets to use, we follow \cite{bezard02} in choosing a minimum radius of R$_{min}=0.15 \text{ km}$, and calculate the maximum radius $R_{max}$ as the maximum size for a comet for which the impact timescale is equal to the diffusion timescale:

\begin{equation}
\frac{1}{ \dot{N}(>D_{max})} \sim \tau_K
\end{equation}
The diffusion timescale can be approximated as
\begin{equation}
\tau_K \approx \frac{2 H_0^2}{K_0}
\end{equation}
where H$_0$ and K$_0$ are the scale height and diffusion rate at the tropopause, where mixing is slowest. We consider three values of K$_0$: 200, 800 and 2000 cm$^2$ s$^{-1}$. The slowest mixing rate of 200 cm$^2$ s$^{-1}$  corresponds to the minimum eddy diffusion rate in the Romani `A'  profile and our `test' eddy profile; the Romani `B' and Moses `2' profiles have K$_0$ values near 2000 cm$^2$ s$^{-1}$ , and the Moses `1' profile has an intermediate K$_0\sim 800$ cm$^2$ s$^{-1}$.  The CO production rate $\phi_\CO$ is determined from the CO mass influx rate $\dot{M}$ for the range of $\dot{N}_N(D> 1.5 \mathrm{\ km})$ and for the two different comet size distributions described above. These calculated CO production rates are shown in Table \ref{tab:comet}. 

Using a CO flux to the upper atmosphere of $0.5-20\times 10^8$ cm$^{-2}$ s$^{-1}$, as inferred by our physical fits, we calculate the diameter $D_1$ of a single large comet impact required to reproduce the observed CO abundance, given $\tau_K$ and for each value of $K_0$. We also report the timescales for impacts of size $D_1$ -- which can be compared to the timescale estimates reported by \cite{lellouch05} and \cite{hesman07}. We note that, without detailed knowledge of the vertical eddy diffusion profile, the two-level CO profiles reported by \cite{lellouch05} and \cite{hesman07} do not preclude a constant CO influx by (sub)kilometer-sized comets (rather than the single event discussed by these authors.) Eddy diffusion profiles like the Romani `B' and Moses `2' profiles will produce CO profiles that, when mimicked by a two-level profile, will have transition pressures in the 10 mbar range (Section \ref{sec:resultsphys}). Since very large impacts are rare, a constant infall rate of smaller comets may be a preferable explanation. We find that, of our two comet size distributions, the \cite{zahnle03} Case B distribution gives more optimistic predictions for CO production: the rates of CO injection found for the small and intermediate K$_0$ values are $0.44^{+0.90}_{-0.33} \times 10^8$ cm$^{-2}$ s$^{-1}$ and $0.20^{+0.55}_{-0.17} \times 10^8$ cm$^{-2}$ s$^{-1}$  , respectively, suggesting that cometary impacts are in fact a feasible mechanism for supplying the observed abundance of CO to Neptune's upper atmosphere. However, \cite{zahnle03} suggest that their impact rate for 1.5 km comets at Neptune is likely an overestimate, meaning the true CO production rate due to comets could be smaller.

\begin{table*}[]
 \begin{minipage}{1.0\textwidth}
\begin{center}
\setlength{\tabcolsep}{0.05in}
 \footnotesize
 \caption[Production rate of CO from comets]{\small Production rate of CO from comets.} 
\begin{tabular}{lll lll llll}
 \hline
						&									&								&\multicolumn{3}{l }{Shoemaker and Wolfe} 									&\multicolumn{3}{l}{Zahnle et al. Case B} 			\\
 \hline
 K$_0$ (cm$^{2}$ s$^{-1}$)	&$\tau_K$ (yr)\footnote{diffusion time scale}	&D$_1$ (km) \footnote{diameter of single comet impact required to produce $\phi_\CO=0.5-20\times10^8$ cm$^{-2}$ s$^{-1}$}	& D$_{max}$ (km) 				& $\log{\phi\footnote{total CO flux expected based on impact rates}}$( cm$^{-2}$ s$^{-1}$)	& $\tau_{D_1}$ (yr)\footnote{1/$\dot{N}_N(>D_1)$} 	& D$_{max}$ (km) 			& $\log{\phi^c}$(cm$^{-2}$ s$^{-1}$) & $\tau_{D_1}$ (yr)\\
 \hline
 200 						& 930							& $3.4-12$		&$1.6^{+0.8}_{-0.6}$ 	&$7.1^{+0.5}_{-0.7}$&$1900-120000$&$2.3^{+0.9}_{-0.7}$ &$7.7^{+0.5}_{-0.6}$	&$1100-130000$\\
 800						& 230							&$2.2-7.4$		&$0.8^{+0.4}_{-0.3}$	& $6.7^{+0.6}_{-0.8}$&$750-49000$ &$1.3^{+0.6}_{-0.5}$ &$7.3^{+0.6}_{-0.8}$&$350 -41000 $\\
 2000 					&93								&$1.6-5.4$		&$0.5^{+0.2}_{-0.2}$	& $6.3^{+0.7}_{-1.4}$&$410-26000$ &$0.7^{+0.4}_{-0.3}$ &$6.9^{+0.7}_{-1.0}$&$160 -19000 $\\
 \hline

\label{tab:comet}
\end{tabular}
\end{center}
\end{minipage}
\end{table*}

 \section{Summary/conclusions}
We have observed Neptune in the CO (2--1) and (1--0) rotational lines with the Combined Array for Research in Millimeter-wave Astronomy.  Our radiative transfer analysis indicates a preference for the \cite{fletcher10} thermal profile in Neptune's upper atmosphere over warmer and cooler profiles, and we find that the best-fit solution for the CO vertical profile is strongly dependent on the atmospheric TP profile. Adopting the \cite{fletcher10} thermal profile, we find that good fits to the data (both to the individual lines and to the combined dataset) are characterized by CO mole fractions of $0.1^{+0.2}_{-0.1}$ parts per million (ppm) in the troposphere, and $1.1^{+0.2}_{-0.3}$ ppm in the stratosphere. Higher CO mole fractions, particularly in the stratosphere, are favored when cooler thermal profiles are used. If the CO mole fraction resulting from vertical mixing is greater than 0.1 ppm, our calculations imply an O/H abundance that is at least 400 times the protosolar value.  However, since we do not rule out a 0.0 ppm tropospheric CO mole fraction, we cannot place a lower limit on Neptune's global O/H ratio, and our data do not require that Neptune's oxygen enrichment exceeds its carbon enrichment. We find that the stratospheric deposition rate of CO is $0.5-20 \times 10^8$ CO molecules cm$^{-2}$ s$^{-1}$; such a high abundance could be supplied by impacts from (sub)kilometer-sized comets, as long as the eddy diffusion rate near the tropopause is small. 

This work will be followed up with spatially-resolved mapping of Neptune in the CO (2--1) line with CARMA, to look for latitudinal variations in the CO abundance. ALMA will play a key role in the three dimensional mapping of trace species on Neptune, as well as on the other Solar System giant planets. Such studies will provide further insight into the composition, atmospheric circulation, and environments of the giant planets.

\begin{acknowledgements}
The data presented in this work were obtained with CARMA. Support for CARMA construction was derived from the states of California, Illinois, and Maryland, the James S. McDonnell Foundation, the Gordon and Betty Moore Foundation, the Kenneth T. and Eileen L. Norris Foundation, the University of Chicago, the Associates of the California Institute of Technology, and the National Science Foundation. Ongoing CARMA development and operations are supported by the National Science Foundation under a cooperative agreement, and by the CARMA partner universities. This work was supported by NASA Headquarters under the NASA Earth and Space Science Fellowship program - Grant NNX10AT17H; and by NSF Grant AST-0908575. The authors would like to thank L. Fletcher for providing his temperature and CH$_4$ profiles, and G. Orton for providing his H$_2$ CIA absorption coefficients. The authors would also like to acknowledge R.L. Plambeck, M. Wright and A. Bauermeister for many helpful discussions.
\end{acknowledgements}

\end{document}